%% file: MeSH_EtSH_12.tex
  \documentclass[structabstract]{aa}
\usepackage{graphicx}
\usepackage{txfonts}
\usepackage{natbib}
\usepackage{ulem}

\bibpunct[]{(}{)}{;}{a}{}{,}

\begin{document}
   \title{Exploring molecular complexity with ALMA (EMoCA): Alkanethiols and alkanols 
          in Sagittarius~B2(N2)}

   \author{Holger S.~P. M{\"u}ller\inst{1}
           \and
           Arnaud Belloche\inst{2}
           \and
           Li-Hong Xu \textbf{()}\inst{3}
           \and
           Ronald M. Lees\inst{3}
           \and
           Robin T. Garrod\inst{4}
           \and
           Adam Walters\inst{5,6}
           \and
           Jennifer van Wijngaarden\inst{7}
           \and
           Frank Lewen\inst{1}
           \and
           Stephan Schlemmer\inst{1}
           \and
           Karl M. Menten\inst{2}
           }

   \institute{I.~Physikalisches Institut, Universit{\"a}t zu K{\"o}ln,
              Z{\"u}lpicher Str. 77, 50937 K{\"o}ln, Germany\\
              \email{hspm@ph1.uni-koeln.de}
              \and
              Max-Planck-Institut f\"ur Radioastronomie, Auf dem H\"ugel 69, 
              53121 Bonn, Germany
              \and
              Centre for Laser, Atomic and Molecular Sciences (CLAMS), Department of Physics, 
              University of New Brunswick, Saint John, NB E2L 4L5, Canada
              \and
              Departments of Chemistry and Astronomy, University of Virginia, Charlottesville, 
              VA 22904, USA
              \and
              Universit{\'e} de Toulouse, UPS-OMP, IRAP, Toulouse, France
              \and
              CNRS, IRAP, 9 Av. colonel Roche, BP 44346, 31028 Toulouse cedex 4, France
              \and
              Department of Chemistry, University of Manitoba, Winnipeg, 
              MB R3T 2N2, Canada
              }

   \date{Received 29 September 2015 / Accepted 24 November 2015}

  \abstract
{Over the past five decades, radio astronomy has shown that  molecular complexity is a 
natural outcome of interstellar chemistry, in particular in star forming regions. However, 
the pathways that lead to the formation of complex molecules are not completely understood 
and the depth of chemical complexity has not been entirely revealed. In addition, the sulfur 
chemistry in the dense interstellar medium is not well understood.}
{We want to know the relative abundances of alkanethiols and alkanols in the Galactic Center 
source Sagittarius~B2(N2), the northern hot molecular core in Sgr~B2(N), whose relatively 
small line widths are favorable for studying the molecular complexity in space.}
{We investigated spectroscopic parameter sets that were able to reproduce published laboratory 
rotational spectra of ethanethiol and studied effects that modify intensities in the 
predicted rotational spectrum of ethanol. We used the Atacama Large Millimeter Array (ALMA) 
in its Cycles~0 and 1 for a spectral line survey of Sagittarius~B2(N) between 84 and 114.4~GHz. 
These data were analyzed by assuming local thermodynamic equilibrium (LTE) for each molecule. 
Our observations are supplemented by astrochemical modeling; a new network is used for the first 
time that includes reaction pathways for alkanethiols.}
{We detected methanol and ethanol in their parent $^{12}$C species and those with one $^{12}$C 
atom substituted by $^{13}$C; the latter were detected for the first time unambiguously in the 
case of ethanol. The $^{12}$C/$^{13}$C ratio is $\sim$25 for both molecules. In addition, 
we identified CH$_3^{18}$OH with a $^{16}$O/$^{18}$O ratio of $\sim$180 and a 
$^{13}$CH$_3$OH/CH$_3^{18}$OH ratio of $\sim$7.3. Upper limits were derived for the next 
larger alkanols \textit{normal}- and \textit{iso}-propanol. We observed methanethiol, CH$_3$SH, 
also known as methyl mercaptan, including torsionally excited transitions for the first time. 
We also identified transitions of ethanethiol (or ethyl mercaptan), though not enough to claim 
a secure detection in this source. The ratios CH$_3$SH to C$_2$H$_5$SH and C$_2$H$_5$OH to 
C$_2$H$_5$SH are $\gtrsim 21$ and $\gtrsim 125$, respectively. In the process of our study, 
we noted severe discrepancies in the intensities of observed and predicted ethanol transitions 
and propose a change in the relative signs of the dipole moment components. In addition, we 
determined alternative sets of spectroscopic parameters for ethanethiol. The astrochemical models 
indicate that substantial quantities of both CH$_3$SH and C$_2$H$_5$SH may be produced on the 
surfaces of dust grains, to be later released into the gas phase. The modeled ratio 
CH$_3$SH/C$_2$H$_5$SH = 3.1 is lower than the observed value of $\gtrsim 21$; the model value 
appears to be affected most by the underprediction of CH$_3$SH relative to CH$_3$OH and 
C$_2$H$_5$OH, as judged by a very high CH$_3$OH/CH$_3$SH ratio.}
{The column density ratios involving methanol, ethanol, and methanethiol in Sgr~B2(N2) are 
similar to values reported for Orion~KL, but those involving ethanethiol are significantly 
different and suggest that the detection of ethanethiol reported toward Orion~KL is uncertain. 
Our chemical model presently does not permit the prediction of sufficiently accurate column 
densities of alkanethiols or their ratios among alkanethiols and alkanols. 
Therefore, additional observational results are required to establish the level of 
C$_2$H$_5$SH in the dense and warm interstellar medium with certainty.}

\authorrunning{H.~S.~P. M{\"u}ller et al.}
\titlerunning{Alkanethiols and alkanols in Sgr~B2(N2)}

\maketitle
\hyphenation{For-schungs-ge-mein-schaft}
\hyphenation{vi-nyl-a-mine}
\hyphenation{al-kane-thi-ole}

\section{Introduction}
\label{intro}

The degree of chemical complexity observed toward star-forming regions known as ``hot cores'' 
is greater than that found thus far in any other portion of the interstellar medium (ISM), 
encompassing a range of organic molecules of up to 12 atoms \citep{i-PrCN_det_2014}. 
The significance of these detections may go beyond the immediate chemistry of hot cores 
themselves; the apparent similarity between the abundances of complex interstellar molecules 
and the chemical composition of comets suggests that interstellar material may have been 
incorporated directly into cometary and/or planetary solids during the planet formation process. 
It is therefore critical to understand the depth and breadth of chemical complexity that 
may develop in the early stages of star formation, and which may influence the chemistry 
of later planetary systems and perhaps the ultimate emergence of life.

Sagittarius (Sgr for short) B2 is an especially well-suited source for studying saturated 
or nearly saturated organic molecules, and most of the larger ones have been detected 
toward this source for the first time \citep{SgrB2_3mm_2013}. Sgr~B2(N) and Sgr~B2(M) 
are the two main sites of star formation in the Sgr~B2 molecular cloud complex, one of 
the most massive star-forming regions in our Galaxy located close to the Galactic Center. 
Sgr~B2(N) contains two dense, compact hot cores separated by about $5''$ 
\citep{det_AAN,cont_SMA_2011}, the more prominent one is the so-called Large Molecule 
Heimat Sgr~B2(N-LMH), also called Sgr~B2(N)-SMA1 \citep{cont_SMA_2011} or P1 \citep{det_AAN}; 
the other is called Sgr~B2(N)-SMA2 \citep{cont_SMA_2011} or P2 \citep{det_AAN} and is to 
the north of the first. We propose that Sgr~B2(N1) and Sgr~B2(N2) be adopted because they 
are short and unambiguous names. Single-dish observations of Sgr~B2(N) usually cover both 
sources, but differences in the local standard of rest (lsr) velocities provide the 
means to distinguish between the two sources.

In order to study the molecular complexity in Sgr~B2, we used the IRAM 30~m telescope 
to carry out molecular line surveys toward Sgr~B2(N) and Sgr~B2(M) at 3~mm wavelengths 
with additional observations at 2 and 1.3~mm \citep{SgrB2_3mm_2013}. Even though each 
frequency setup was observed for a modest amount of time, line confusion was reached 
for a considerable part of the surveys, in particular toward Sgr~B2(N). Nevertheless, 
in this study we detected aminoacetonitrile, a possible precursor of glycine (the simplest 
amino acid), for the first time in space \citep{det_AAN}, as well as ethyl formate 
and \textit{n}-propyl cyanide \citep{det-PrCN_EtFo}, the $^{13}$C isotopologs of 
vinyl cyanide \citep{13C-VyCN_2008}, and several other minor isotopic species and 
transitions in excited vibrational states \citep{SgrB2_3mm_2013}.

Other recent investigations into the molecular complexity of Sgr~B2(N) include a line 
survey with the HIFI high-resolution spectrometer on board the \textit{Herschel} 
satellite covering its entire frequency region (most of $\sim$500 to $\sim$1900~GHz) 
\citep{SgrB2(N)_HIFI_2014} or the PRIMOS Legacy Project (up to about 50~GHz) employing 
the 100~m GBT dish, which led to the detection of ethanimine \citep{ethanimine_det_2013} 
and \textit{E}-cyanomethanimine \citep{E-Cyanomethanimine_rot_det_2013}, among others. 
Among the less recent studies of Sgr~B2, we note the 1.3~mm line survey of 
\citet{SgrB2_Nummelin_1998,SgrB2_Nummelin_2000} who studied three positions (M, N, 
and NW) using the SEST 15~m dish.

The Orion Molecular Cloud, and in particular the Kleinmann-Low Nebula (Orion~KL), have also 
been studied extensively in the context of molecular complexity \citep{Orion-KL_HIFI_2014}. 
An IRAM 30~m molecular line survey led to the detection of several molecules including methyl 
acetate \citep{MeAc_det_2013} (an isomer of ethyl formate), and numerous minor isotopic 
species, such as methyl-deuterated methyl formate, HCO(O)CH$_2$D \citep{HCOOCH2D_det_2013}. 
The detection of methyl formate in its second excited torsional state was reported through 
different observations \citep{MeFo_vt2_det_2012}. In recent years, considerable levels of 
molecular complexity have also been investigated in other star-forming regions. 
Ethanediol, for example, which is also known as ethylene glycol, is a molecule with ten atoms. 
It was detected in the hot corinos associated with the Class~0 protostars NGC 1333-IRAS2A 
\citep{NGC1333-IRAS2A_survey_2014} and, tentatively, IRAS 16293-2422B 
\citep{IRAS-16293_glycald_2012} after the first detection toward Sgr~B2(N) 
\citep{det-eglyc_2002}.

Following up on our observations with the IRAM 30~m telescope, we used the Atacama Large 
Millimeter Array (ALMA) in its Cycle~0 for a spectral line survey of Sagittarius~B2(N) 
between 84 and 111~GHz with the aim of Exploring Molecular Complexity with ALMA (EMoCA). 
An additional setup in Cycle~1 extended the frequency range to 114.4~GHz. 
We expected a reduction in the line confusion through the spatial resolution of Sgr~B2(N1) 
and Sgr~B2(N2). In fact, we observed reduced line confusion toward the weaker source 
Sgr~B2(N2) caused not only by the spatial separation of the two sources, but also by 
narrower lines. As a first result, we were able to detect the first branched alkyl molecule 
in space, \textit{iso}-propyl cyanide and to determine a 1~:~2.5 ratio relative to the more 
abundant \textit{n}-propyl cyanide in Sgr~B2(N2) \citep{i-PrCN_det_2014}. 
The line confusion is high in Sgr~B2(N1), therefore, we focus our initial analyses on 
Sgr~B2(N2). In the present work, we report on alkanols and alkanethiols in this source.

Methanol, CH$_3$OH, is the lightest alkanol and one of the most abundant organic 
molecules in the ISM. It was among the first molecules to be detected in space, 
toward Sgr~A and Sgr~B2, by means of radio astronomy \citep{det-MeOH_1970}. 
In fact, $^{13}$CH$_3$OH and CH$_3$OD were detected in Sgr~B2 shortly thereafter 
\citep{det-13C-MeOH_1979}. We note, however, that the identification of CH$_3$OD in 
that work has been questioned recently \citep{deuterated_SgrB2N2_2015}. A few years 
later, \citet{CH3O-18-H_det_1989} reported on the detection of CH$_3^{18}$OH in Sgr~B2. 
The great diagnostic value of methanol lines has been pointed out several times 
\citep{MeOH_diag_2004,MeOH_diag_2007,MeOH_maser_dark-cloud_2004,MeOH_diag_Orion_2011}. 
Rotational lines of $^{13}$CH$_3$OH have been used frequently to derive proper 
column densities because those of the main isotopic species are often saturated 
in dense molecular clouds \citep{SgrB2_3mm_2013,D_in_KL_2013}.

Ethanol, C$_2$H$_5$OH, the next heaviest alkanol, was also among the first molecules 
to be detected in space employing radio astronomy \citep{a-EtOH_det_1975}. 
The detection was also made toward Sgr~B2. Initially, it was observed only 
via its \textit{anti} (or \textit{trans}) low-energy conformer, the higher lying 
\textit{gauche} conformer was first detected in Orion~KL several years later 
\citep{g-EtOH_det_1997}. Ethanol was not only detected in several hot cores, but also 
in kinetically moderately warm Galactic Center clouds \citep{cold-GC-clouds_30m_2006}. 
In addition, there is evidence for the molecule in the molecular clouds surrounding 
the Class~0 protostars IRAS 16293-2422A and B \citep{16293_SMA_2008}. Ethanol with one 
$^{13}$C has not yet been detected with certainty, but we reported on a tentative 
detection in the course of the 30~m line survey of Sgr~B2(N) \citep{SgrB2_3mm_2013}.

In the case of propanol, C$_3$H$_7$OH, there are two isomers, the straight chain 
\textit{normal}- or \textit{n}-propanol and the branched \textit{iso}- or 
\textit{i}-propanol. We are aware of two reports attempting to detect 
\textit{n}-propanol in space \citep{cold-GC-clouds_GBT_2008,search_EME_n-PrOH_2015}. 
There appears to be no published report on the search for \textit{i}-propanol in space.

Replacing an O atom in an alkanol with an S atom yields an alkanethiol. 
The simplest one is methanethiol, CH$_3$SH, also known as methyl mercaptan. 
It was among the earliest molecules detected in space by radio astronomical means, 
first tentatively \citep{CH3SH_tent_1977} and confirmed subsequently \citep{MeSH_det_1979}, 
again toward Sgr~B2. It was also observed toward the hot core G327.3-0.6 
\citep{CH3SH_etc_G327_2000}, the cold core B1 \citep{det-CH3O_2012}, and the Orion~KL 
hot core \citep{EtSH_lab_Orion-KL_2014}. The last study also reported the detection 
of ethanethiol, C$_2$H$_5$SH, also known as ethyl mercaptan. Even though many spectral 
features were ascribed to the molecule, most of them are partially or completely blended 
with emission from other species. In addition, the column density ratio of ethanethiol 
with respect to methanethiol or ethanol appears somewhat higher than the column density 
ratios of the latter two molecules with respect to methanol. 
The puzzlingly high apparent abundance of C$_2$H$_5$SH in Orion~KL motivates 
probing the abundances of alkanols and alkanethiols in other hot cores, in particular 
Sgr~B2(N), in order to understand better the interstellar chemistry of these two families 
of organic molecules.

\section{Laboratory spectroscopy}
\label{lab-spec}

In this section, we present background information on the spectroscopy of the molecules in 
the present study. One general aspect concerns the partition function of organic molecules 
and the related abundances. These molecules often have low-lying vibrations, such as OH-, 
SH-, or CH$_3$-torsions, which are not always considered in the derivation of the partition 
function. We comment on necessary corrections of the partition function and, equivalently, 
the column densities under local thermodynamic equilibrium (LTE). Similar corrections 
may be required if not all of the thermally populated conformations were considered 
in the derivation of the partition function.

The rotational spectrum of a saturated or nearly saturated organic molecule may differ 
considerably from that of a simple asymmetric top rotor because the torsion or inversion 
of functional groups such as CH$_3$, NH$_2$, OH, or SH may be hindered only slightly. 
These large amplitude motions may lead to additional splitting in the rotational 
transitions, which sometimes can be resolved in radio astronomical spectra.

\citet{RAS_1972} proposed a reduced axis system (RAS) for the treatment of inversion 
problems, i.e., tunneling between two equivalent minima. Its main off-diagonal term of 
\textit{a}-symmetry is $F_{bc} \{J_b,J_c\}$ with $\{J_b,J_c\} = J_bJ_c + J_cJ_b$. 
Distortion corrections, such as 
$\{\{J_b,J_c\}, F_{bc,K}J_a^2 + F_{bc,J}J^2 + F_{2bc}(J_+^2 + J_-^2) + ...\}$ 
with $J_{\pm} = J_b \pm iJ_c$, may also be required. Terms of \textit{b}- and 
\textit{c}-symmetry are defined equivalently. The symmetry of the allowed terms in the 
Hamiltonian depends on the symmetry of the inversion motion. The designation of the 
parameter as $F_{bc}$ is fairly common, but other designations also exist (e.g., $D_{bc}$). 
Parameters such as $F_{2bc}$ may also be called $F_{bc}^{\pm}$.

Reduced axis system Hamiltonians were employed successfully for a variety of molecules 
of interest for astrochemistry, including NH$_2$D \citep{NH2D_rot_1982}, vinylamine 
\citep{VyNH2_rot_1999}, \textit{aGg'}-ethanediol \citep{aGg-eglyc_2003}, NHD$_2$ 
\citep{NHD2_rot_2006}, H$_2$DO$^+$ \citep{H2DO+_anal_2010}, phenol \citep{PhOH_rot_2013}, 
and 1,3-propanediol \citep{1_3_PD_rot_2013}.

A RAS Hamiltonian still appears to be well suited to describe the tunneling between 
two equivalent conformations if a third minimum is sufficiently higher in energy. 
This third minimum usually has a dihedral angle of $\sim$0$^{\circ}$ (\textit{syn}) 
or $\sim$180$^{\circ}$ (\textit{anti}); the two equivalent conformations have dihedral 
angles of about $\pm$120$^{\circ}$ or of about $\pm$60$^{\circ}$, respectively. 
In either case, the conformations are frequently called \textit{gauche}. 
Examples are peroxynitric acid \citep{HOONO2_rot_1986}, fluoromethanol 
\citep{CH2FOH_rot_1986}, and propargyl alcohol \citep{Propargyl-OH_rot_2005}. 
If a third minimum is closer in energy to the \textit{gauche} states, it may be necessary 
to include parameters of odd order into the Hamiltonian to describe the tunneling between 
the \textit{gauche} states. The lower order parameters of \textit{a}-symmetry are 
$\{iJ_a, G_a + G_{a,K}J_a^2 + G_{a,J}J^2 + G_2(J_+^2 + J_-^2) + ... \}$, those of 
\textit{b}- and \textit{c}-symmetry are defined equivalently. This approach has been 
used, for example, in the treatment of the rotation-tunneling spectrum of ethanol 
\citep{g-EtOH_rot_1996,EtOH_rot_2008}.

In the case of tunneling between two equivalent conformations if a third minimum is 
sufficiently higher in energy, the RAS interaction Hamiltonian with even order parameters, 
such as $F_{ab}$, may be replaced by an interaction Hamiltonian with odd order parameters, 
such as $G_{c}$. \citet{Propargyl-OH_rot_2005} have shown that equivalent fits can be 
obtained with only even or with only odd order interaction parameters in the case of 
propargyl alcohol.

We note that the sign of the first off-diagonal interaction parameter of a given symmetry 
and between specific conformations or vibrational states is not determinable in a fit 
of field-free rest frequencies. This parameter is usually the one of lowest order, i.e., 
$G_i$ or $F_{jk}$. However, the sign of any additional interaction parameter of the same 
symmetry and between those specific conformations or vibrational states are determinable 
with respect to that of the first parameter. Moreover, in the absence of off-diagonal 
interaction parameters, the signs of the dipole moment components are not determinable 
and do not matter. However, in the presence of off-diagonal interaction parameters 
and with their signs fixed, the relative signs of the dipole moment components may matter; 
in fact, it is the sign of the dipole moment component times the sign of the overlap 
integral that is determined. In that case, they may be determinable by Stark 
spectroscopy \citep{NH2D_rot_1982,HOONO2_rot_1986,CH2FOH_rot_1986}. 
The relative signs of the dipole moment components may also lead to changes 
in the intensities of some transitions, and conversely, relative intensity 
measurements may lead to the determination of the relative signs of the dipole 
moment components \citep{aGg-eglyc_2003,NHD2_rot_2006,H2DO+_anal_2010}.

The treatment of torsional large amplitude motions (e.g., of CH$_3$ groups) often 
requires more specialized Hamiltonian models (see, e.g., \citet{Herbst-MeOH_1984} or 
\citet{LAM_review_2010} for a review) because the regular $K$ energy level structure 
of an asymmetric top rotor could be modified significantly by torsional contributions 
that oscillate as a function of $K$. The amplitude of this oscillation increases rapidly 
with excitation to torsional states of successively higher energy relative to the 
torsional barrier. The frequency of the $K$-variation depends on the ratio of the moment 
of inertia of the rotating top to that of the whole molecule, hence torsional contributions 
to the energies oscillate faster with $K$ for heavy tops and light frames.

\subsection{Methanol, CH$_3$OH}
\label{lab_MeOH}

The torsional energy for methanol varies rapidly with $K$ because the moment of inertia of the 
CH$_3$ group dominates that of the light OH frame. In addition, the intermediate-size barrier 
to rotation of the CH$_3$ group leads to torsional contributions to the $K$ energy level structure 
in the ground vibrational state that are of similar size to the $K$ level splitting at low $K$. 
As a consequence, the origins of the $b$-type rotational transitions (with $\Delta K = 1$) 
differ considerably from those expected for an asymmetric top rotor. In contrast, the patterns of 
the  $a$-type rotational transitions (with $\Delta K = 0$) resemble those of an asymmetric top 
in the absence of perturbations; this applies even to excited torsional states. The first excited 
torsional state straddles the barrier to internal rotation, and even higher torsional states 
display a $K$-level structure that increasingly approaches that of a free internal rotor.

We have used rotational transitions of $^{13}$CH$_3$OH initially to infer the column density of 
methanol because of frequent saturation of lines pertaining to the main isotopic species. 
The spectroscopic line data were taken from the CDMS 
catalog\footnote{http://www.astro.uni-koeln.de/cdms/catalog; 
http://cdms.ph1.uni-koeln.de/cdms/portal/} \citep{CDMS_1,CDMS_2}. The first version 
is based on \citet{13C-MeOH_rot_1996} and \citet{12-13C-MeOH_1997} with experimental 
transition frequencies in the range of our ALMA data from \citet{13C-MeOH_rot_1986} 
and from the privately communicated methanol atlas of the Toyama University 
\citep{MeOH_atlas_1995}. An update of the CDMS entry is in preparation. 
It is based on extensive measurements taken in Cologne that extend up to 
$\sim$1500~GHz and include data in the range of our survey \citep{13C-MeOH_rot_2014}. 
The magnitudes of the dipole moment components employed in the calculation are 
$|\mu_a| = 0.899$~D and $|\mu_b| = 1.44$~D from initial measurements on the main 
isotopic species \citep{MeOH_dip_1953}. \citet{13C-MeOH_dip_1994} determined slightly 
different dipole moment components of $^{13}$CH$_3$OH. The vibrational contributions 
to the partition function at 160~K are essentially converged by inclusion of 
$\varv _{\rm t} = 2$, a correction of the column density because of the population 
of vibrational states is not required.

We used the CDMS entry to evaluate the column density of CH$_3^{18}$OH. The entry is based on 
\citet{18O-MeOH_rot_2007} with extensive far-infrared data from that work. Most of the rest 
frequencies determined with microwave accuracy are from \citet{18O-MeOH_rot_1996} and from 
\citet{18O-MeOH_rot_1998}. The dipole moment components are slightly different from those 
of the main species and were taken from \citet{18O-MeOH_dip_1996}. As for $^{13}$CH$_3$OH, 
vibrational corrections to the partition function at 160~K are not needed.

We used the JPL catalog\footnote{http://spec.jpl.nasa.gov/ftp/pub/catalog/catdir.html} 
\citep{JPL-catalog_1998} entry to estimate the CH$_3$OH column density because we expect 
to observe highly rotationally or torsionally excited CH$_3$OH transitions with line widths 
much wider than in dark clouds. The entry is based on the extensive study of \citet{MeOH_rot_2008} 
which extends to $\varv _{\rm t} = 2$ and high $J$ and $K$ quantum numbers. Rest frequencies in 
the range of our study were taken mainly from the methanol atlas of the Toyama 
University\footnote{http://www.sci.u-toyama.ac.jp/phys/4ken/atlas/} \citep{MeOH_atlas_1995} with 
additional data from \citet{MeOH_rot_1968,MeOH_rot_1984,MeOH_maser_dark-cloud_2004,MeOH_rot_2008}. 
An older entry based on \citet{12-13C-MeOH_1997} is available in the CDMS. Even though it 
covers only $\varv _{\rm t} = 0$ and 1 and has weaknesses for transitions with higher 
$J$ and $K$ values, it is frequently more accurate for lower energy transitions that may be 
important for dark cloud or maser observations, which require high frequency accuracy. 
This is achieved in part by merging corresponding laboratory data \citep{MeOH_maser_dark-cloud_2004}. 
The dipole moments were taken from \citet{MeOH_dip_1953}. Very recently, determinations of the 
dipole moment components were reported by \citet{MeOH_dip_2015} with very similar values. 
The small change in the $a$-dipole moment component upon torsional excitation probably 
has negligible effects for the strongest $a$-type transitions even in a perfect fit of 
astronomical data. There may be non-negligible effects in transitions in which contributions 
of the $a$- and $b$-dipole moment components almost cancel out. Vibrational corrections to 
the partition function at 160~K are again not needed.

We suspected that transitions pertaining to $\varv _{\rm t} = 3$ may be observable in our data. 
We took $J = 2 - 1$ $a$-type transition frequencies from \citet{MeOH_atlas_1995}, the lower state 
energies from \citet{Methanol-Atlas_1995}, and estimated the line strengths from lower torsional 
states \citep{MeOH_rot_2008}. It is worthwhile mentioning that substantial progress has been 
made recently in analyzing $\varv _{\rm t} = 3$ and higher vibrational states of methanol 
\citep{MeOH_vt3_2009,MeOH_high-v_2015}.

\subsection{Methanethiol, CH$_3$SH}
\label{lab_MeSH}

Replacing the O atom in methanol with an S atom yields methanethiol, reducing the moment of 
inertia ratio between the CH$_3$ group and the SH frame compared to the OH frame in methanol. 
Overall, the torsional splitting in CH$_3$SH is smaller than that in CH$_3$OH. The splittings 
in the  first excited torsional state of CH$_3$SH are similar to the ones in the ground 
vibrational state of CH$_3$OH.

Information on rest frequencies were taken from the CDMS. The entry is based on 
\citet{MeSH_rot_2012}. Extensive terahertz and far-IR data were combined with previous 
data, which include data in the range of our survey from \citet{MeSH_rot_1980} with 
additional data mainly from \citet{MeSH_rot_1986} and \citet{MeSH_rot_1999}.
Dipole moment components of $|\mu_a| = 1.312$~D and $|\mu_b| = 0.758$~D were cited by 
\citet{MeSH_dip_1989}. The vibrational contributions to the partition function are 
quite well accounted for at 180~K, so no correction to the column density needs to be made.

\subsection{Ethanol, C$_2$H$_5$OH}
\label{lab_EtOH}






\begin{figure}

\centering

\includegraphics[angle=0,width=8.8cm]{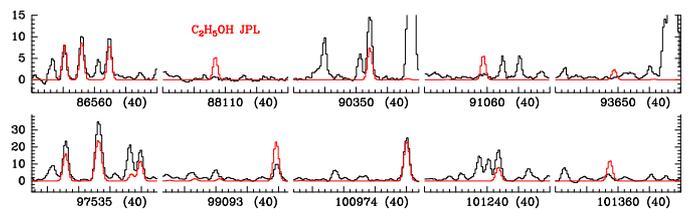}

  \caption{Sections of the continuum-subtracted spectrum recorded with 
           ALMA toward Sgr~B2(N2) in black.  
           The synthetic spectrum of ethanol using the official JPL catalog 
           entry is shown in red. Some lines are predicted to be much stronger 
           than they appear in the spectrum, e.g., near 88107~MHz and 91059~MHz. 
           A few other lines, e.g., near 90354~MHz and 101244~MHz, are stronger than 
           predicted. Compare with Fig.~\ref{EtOH_new_cat}. The central frequency 
           of each panel is indicated below the $x$-axis in MHz; its width is given 
           also in MHz in parentheses. The $y$-axis is labeled in brightness 
           temperature units (K).}
\label{EtOH_old_cat}

\end{figure}

\begin{figure}

\centering

\includegraphics[angle=0,width=8.8cm]{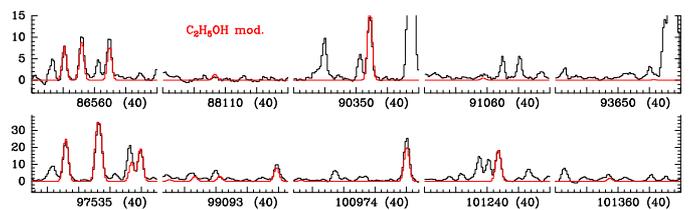}

  \caption{Same as in Fig.~\ref{EtOH_old_cat} with the same LTE parameters, 
           but the modeling was carried out with a modified catalog entry 
           in which the sign of one dipole moment component was altered, 
           see Table~\ref{EtOH_dip}.}

\label{EtOH_new_cat}

\end{figure}


There are two large amplitude motions in ethanol, the internal rotation of the methyl group, 
whose rather small splitting is only rarely resolved in the laboratory and is frequently 
neglected, and the internal rotation of the OH group, which causes more prominent splitting. 
The \textit{anti} conformer (also known as \textit{trans}) with a dihedral HOCC angle of 
180$^{\circ}$ is lowest in energy. The doubly degenerate \textit{gauche} conformer with 
dihedral angles of about $\pm60^{\circ}$ is considerably higher in energy. Tunneling between 
the two equivalent \textit{gauche} conformers causes them to split into the symmetric 
\textit{gauche}$^+$ and the antisymmetric \textit{gauche}$^-$ states. The former is 56.82~K 
higher than the \textit{anti} conformer and the latter 61.53~K \citep{EtOH_rot_2008}. 
Torsion-rotation interaction between the \textit{gauche} conformers occurs even at low 
quantum numbers \citep{g-EtOH_det_1997}. In contrast, such interaction can be neglected 
for the \textit{anti} conformer up to moderate quantum numbers ($J + 2K_a \lesssim 30$).

\citet{EtOH_rot_2008} reported extensively on the rotational spectra of the C$_2$H$_5$OH 
conformers and their mutual torsion-rotation interaction. Data associated with this work 
as well as resulting predictions of the ground state rotational spectrum are available in 
the JPL catalog. Transition frequencies in the range of our ALMA survey are from that work. 
Vibrational contributions to the partition function at 150~K are dominated by the two 
torsional modes \citep{EtOH_FIR_1990}. Combining \textit{anti} and \textit{gauche} conformers, 
the ground state partition function needs to be increased by a factor of 1.24; slight 
differences in the OH torsional energies of the two conformers are taken into account.

The entries of ethanol isotopologs with one $^{13}$C atom were taken from the CDMS catalog. 
They were based on \citet{13C-EtOH_rot_2012}, which restricted the investigations 
to the \textit{anti} conformers with $J$ and $K_a$ ranges in which interaction with 
the \textit{gauche} conformers can be neglected. The contribution of the \textit{gauche} 
conformers to the partition function increases the value at 150~K based on the \textit{anti} 
conformer only by a factor of $\sim$2.39 as estimated in the CDMS documentation from 
the main isotopic species. Vibrational contributions to the partition function were also 
derived from the main isotopic species.


\begin{table}
\begin{center}
\caption{Experimental dipole moment components $\mu_i$ (D) of ethanol$^a$ with proposed signs.}
\label{EtOH_dip}
\renewcommand{\arraystretch}{1.10}
\begin{tabular}[t]{lr@{}lcc}
\hline \hline
Component                                & \multicolumn{2}{c}{Value$^a$} & \multicolumn{2}{c}{Sign$^b$} \\ 
\cline{4-5}
                                         &   &                           &  Initial    & Present \\
\hline
$\mu_a(a \leftrightarrow a)$             &  0&.046~(14)                  &  +          &  (+)    \\
$\mu_b(a \leftrightarrow a)$             &  1&.438~(7)                   &  +          &  +      \\
$\mu_c(a \leftrightarrow a)$             &  0&.$^c$                      &             &         \\
$\mu_a(g^{\pm} \leftrightarrow g^{\pm})$ &  1&.246~(10)                  &  $-$        &  +      \\
$\mu_b(g^{\pm} \leftrightarrow g^{\pm})$ &  0&.104~(8)                   &  +          &  (+)    \\
$\mu_c(g^{\pm} \leftrightarrow g^{\mp})$ &  1&.101~(16)                  &  $-$        &  $-$    \\
\hline
\end{tabular}
\end{center}
\tablefoot{
$^a$ From \citet{a-EtOH_dip_rot_1968} and \citet{g-EtOH_dip_rot_1980} for \textit{anti} and \textit{gauche} 
     conformers of ethanol, respectively. Numbers in parentheses after the values refer to the reported 
     uncertainties in units of the least significant figures. $a$, $g^+$, and $g^-$ refer to the \textit{anti} 
     and to the symmetric and antisymmetric states of the \textit{gauche} conformers of ethanol, respectively. 
$^b$ With respect to the signs of the interaction parameters in the Hamiltonian model from \citet{EtOH_rot_2008}. 
     Initial values are from that work. Signs in parentheses are not certain. See also the second to last paragraph 
     in section~\ref{lab-spec}. 
$^c$ By symmetry.  
}
\end{table}


\begin{table}
\begin{center}
\caption{Quantum numbers, conformers Conf.$^a$, frequencies (MHz) of ethanol transitions 
         appearing too strong in initial models, and Einstein $A$ values (10$^{-7}$~s$^{-1}$) 
         from the JPL catalog and new values proposed in the present study.}
\label{EtOH_problematic_lines}
\renewcommand{\arraystretch}{1.10}
\begin{tabular}[t]{ccr@{}lr@{}lr@{}l}
\hline \hline
$J',K'_a,K'_c - J'',K''_a,K''_c$ & Conf. & \multicolumn{2}{c}{Frequency} &  \multicolumn{2}{c}{$A$(JPL)} &  \multicolumn{2}{c}{$A$(new)} \\ 
\hline
 28,10,18 $-$ 29, 9,20  & $g\mp$ &  88106&.42 &  5&.55 &  6&.26 \\
 28,10,19 $-$ 29, 9,21  & $g\mp$ &  88106&.57 &  5&.55 &  6&.26 \\
 26, 0,26 $-$ 25, 1,24  & $g\pm$ &  88106&.96 & 22&.76 &  2&.82 \\
 25, 2,24 $-$ 24, 3,21  & $a$    &  91058&.80 & 18&.64 &  1&.52 \\
 27, 0,27 $-$ 26, 1,25  & $g\pm$ &  93054&.89 & 26&.87 &  3&.20 \\
 31, 2,29 $-$ 32, 1,32  & $a$    &  93648&.78 & 23&.57 &  0&.18 \\
 28, 0,28 $-$ 27, 1,26  & $g\pm$ &  98097&.68 & 31&.55 &  3&.60 \\
 13, 0,13 $-$ 13, 1,13  & $g\mp$ &  99109&.25 & 68&.67 & 16&.37 \\
 13, 4, 9 $-$ 12, 5, 7  & $g\mp$ &  99109&.69 &  7&.04 &  7&.11 \\
 19, 3,16 $-$ 19, 2,18  & $g\pm$ &  99537&.22 & 34&.23 & 11&.36 \\
 24, 2,23 $-$ 23, 3,20  & $a$    &  99896&.28 & 24&.36 &  2&.40 \\
 32, 2,30 $-$ 33, 1,33  & $a$    & 101134&.52 & 29&.75 &  0&.21 \\
 15, 2,13 $-$ 15, 1,15  & $g\pm$ & 101357&.44 & 36&.42 &  5&.16 \\
 29, 1,29 $-$ 28, 2,27  & $g\pm$ & 102326&.79 & 16&.70 &  3&.93 \\
 29, 0,29 $-$ 28, 1,27  & $g\pm$ & 103215&.29 & 36&.82 &  4&.03 \\
 30, 0,30 $-$ 29, 1,28  & $g\pm$ & 108392&.97 & 42&.71 &  4&.29 \\
 33, 2,31 $-$ 34, 1,34  & $a$    & 108442&.67 & 36&.77 &  0&.24 \\
  8, 1, 8 $-$  8, 0, 8  & $g\mp$ & 110368&.52 & 94&.84 & 62&.45 \\
  7, 2, 6 $-$  6, 1, 6  & $g\pm$ & 110545&.85 & 32&.68 & 14&.58 \\
 22, 2,21 $-$ 21, 3,18  & $a$    & 113309&.00 & 34&.73 &  5&.15 \\
\hline
\end{tabular} 
\end{center}
\tablefoot{
$^a$ Transitions within the \textit{anti} or \textit{trans} conformer are indicated by $a$, 
  $g\mp$ and $g\pm$ indicate transitions with \textit{gauche}$^+$ as lower and upper state, 
  respectively, and \textit{gauche}$^-$ as upper and lower state, respectively.
}
\end{table}


The experimental dipole moment components of the ethanol conformers are given in 
Table~\ref{EtOH_dip} together with the proposed signs from \citet{EtOH_rot_2008}, 
which are important because of the extensive torsion-rotation interaction. 
\citet{EtOH_rot_2008} discussed line strength issues in section~5.2. 
They pointed out that intensities for a number of transitions are not correctly modeled 
by their Hamiltonian model and attributed this short-coming by the neglect of torsional 
corrections (sine and cosine functions) to the dipole moment components.

As shown in Fig.~\ref{EtOH_old_cat}, we noticed that several rather weak C$_2$H$_5$OH 
transitions in our line survey were predicted to be significantly too strong; these 
transitions are summarized in Table~\ref{EtOH_problematic_lines}. We suspected that 
the sign choices of the dipole moment components relative to the interaction parameters 
could be responsible for the observed intensity issues. We changed the signs of the dipole 
moment components from \citet{EtOH_rot_2008} one at a time relative to $\mu_b$ of the 
\textit{anti} conformer. The resulting predictions were compared with our ALMA observations. 
In addition, we inspected a laboratory spectrum of ethanol in the 3~mm wavelength 
region taken at the Universit{\"a}t zu K{\"o}ln for educational purposes. 
Sign changes of either $\mu_a$ or $\mu_c$ of the \textit{gauche} conformer yielded 
predictions that were compatible with our ALMA observations with a slight preference 
for a change in $\mu_a$. 
The intensities in the laboratory spectrum were in support of this view; e.g., the 
\textit{gauche}$^-$ transition $13_{1,12} -  13_{1,13}$ near 96716.5~MHz was about 
as strong as the $g^+ - g^-$ transition $39_{9,31} -  38_{10,29}$ near 96746.9~MHz, 
and the \textit{gauche}$^+$ transition $13_{1,12} -  13_{0,13}$ near 93279.0~MHz 
was about twice as strong. Retaining the sign of $\mu_a$ and changing it in $\mu_c$ 
results in intensities of the two transitions with $J = 13$ lower by a factor of about 2. 
Changing the signs of the dipole moment components that are small in magnitude, $\mu_a$ 
of the \textit{anti} conformer and $\mu_b$ of the \textit{gauche} conformer, had very 
little effect on the predicted intensities, as is expected. Therefore, we refrain 
from any statement concerning their signs. Figure~\ref{EtOH_new_cat} shows the modeling 
of ethanol with the altered sign of the dipole moment component. Not only are there 
now no weak transitions predicted to be much stronger than observed, in some cases 
the lines, for which initially only part of the observed signal was assigned to 
C$_2$H$_5$OH, are now ascribed entirely to this molecule. Further details on the model 
are given in in Sect.~\ref{ss:c2h5oh}; the entire model is presented in the online material.

We should also point out that the choice of spectroscopic parameters may also affect 
the intensities. The intensities of $\Delta K = 3$ transitions in the rotational spectrum 
of PH$_3$ were discussed as a recent example \citep{PH3_v=0_2013}. 
Inclusion or omission of the $\Delta K = 6$ term $h_3$ or fixing it to an 
estimated value has profound effects on the value of the $\Delta K = 3$ term $\epsilon$ 
in the fit which, in turn, affects the intensities of the weak $\Delta K = 3$ transitions. 
\citet{EtOH_rot_2008} discussed the determinability of odd order parameters (e.g., $G_a$) 
simultaneously with even order parameters (e.g., $F_{bc}$) although not explicitly 
in relation to intensities in the rotational spectrum of C$_2$H$_5$OH.

We should emphasize that neglecting torsional corrections to the dipole moment components 
may also account for part of the intensity issues. Even though our dipole moment model 
was only studied in the 3~mm wavelength region, intensities may also be improved outside 
of this range. As indicated in this and the previous paragraphs, the proposed sign change 
does not necessarily resolve all intensity issues, even at 3~mm. Inspection of the 
existing laboratory spectral recordings, in particular outside the 3~mm range, or 
temperature dependent intensity measurements, such as those carried out for methanol 
\citep{MeOH_int_600GHz_2014,MeOH_int_250GHz_2014}, may help resolve the intensity issues 
in the rotational spectrum of ethanol overall.

\subsection{Ethanethiol, C$_2$H$_5$SH: redetermination of spectroscopic parameters}
\label{lab_EtSH}

\citet{EtSH_rot_dip_1975} studied the microwave spectrum of ethanethiol, determined its 
dipole moment components, and estimated the \textit{anti} conformer to have a higher energy 
than the \textit{gauche} conformer by $204 \pm 22$~K. The spectroscopic analysis revealed 
\textit{gauche}$^-$ to be higher than \textit{gauche}$^+$ by only 1754~MHz (or 0.084~K). 
The dipole moment components are $\mu_a = 1.48 \pm 0.02$~D and $\mu_b = 0.19 \pm 0.10$~D 
within the \textit{gauche} conformers and $\mu_c = 0.59 \pm 0.02$~D between the 
\textit{gauche} conformers. Values for the \textit{anti} conformer are $\mu_a = 1.06 \pm 
0.03$~D and $\mu_b = 1.17 \pm 0.03$~D.

Far-infrared and Raman spectroscopy were also employed to evaluate the energy separation 
between the \textit{gauche} and the \textit{anti} conformer. Values of 144~K 
\citep{EtSH_i-PrSH_FIR_egy_1973}, 121~K \citep{i-PrOH_RSH_FIR_egy_1973}, 177~K 
\citep{EtSH_FIR_egy_1973}, and 160~K \citep{EtSH_EtOH_Ra-g_1975} were obtained. 
The last value may be a reasonable compromise among these values and the one from 
microwave measurements, but it should be viewed with caution because the uncertainty 
is at least 30~K, possibly more than 50~K. An energy difference of 160~K yields a 
population of 17\,\% for \textit{anti}-C$_2$H$_5$SH at 180~K, considering the ground 
vibrational states only. Combining the energy difference with the vibrational data from 
\citet{EtSH_EtOH_Ra-g_1975}, including in particular the differences in the SH torsional 
energies, the combined conformational and vibrational correction is $\sim$1.95 at 180~K.


\begin{table}
\begin{center}
\caption{Spectroscopic parameters$^a$ (MHz) of the \textit{gauche} conformers of 
         C$_2$H$_5$SH and C$_2$H$_5^{34}$SH determined in the present study.}
\label{gauche-spec-param}
\renewcommand{\arraystretch}{1.10}
\begin{tabular}[t]{lr@{}lr@{}l}
\hline \hline
Parameter & \multicolumn{2}{c}{C$_2$H$_5$SH} & \multicolumn{2}{c}{C$_2$H$_5^{34}$SH} \\
\hline
$A$                         & 28~746&.912~8(58)    & 28~708&.867~5(189)   \\
$B$                         &  5~294&.877~932(99)  &  5~175&.221~38(36)   \\
$C$                         &  4~846&.575~3(58)    &  4~745&.419~1(72)    \\
$D_K \times 10^3$           &    203&.821~84(264)  &    202&.21(114)      \\
$D_{JK} \times 10^3$        &  $-$18&.152~94(89)   &  $-$17&.715~64(185)  \\
$D_J \times 10^3$           &      3&.287~853(196) &      3&.147~245(116) \\
$d_1 \times 10^6$           & $-$519&.999(130)     & $-$485&.551(133)     \\
$d_2 \times 10^6$           &  $-$16&.383(53)      &  $-$14&.889(87)      \\
$H_K \times 10^6$           &      5&.417~4(50)    &      5&.40           \\
$H_{KJ} \times 10^6$        &   $-$1&.061~41(106)  &   $-$1&.029~0(72)    \\
$H_{JK} \times 10^9$        &   $-$3&.18(35)       &   $-$3&.0            \\
$H_J \times 10^9$           &      2&.661~63(252)  &      2&.491          \\
$h_1 \times 10^9$           &      1&.209~1(90)    &      1&.073~3(281)   \\
$h_2 \times 10^9$           &      0&.155~98(200)  &      0&.140          \\
$E$                         &    876&.994~80(247)  &    865&.478~2(62)    \\
$E_K \times 10^3$           &     71&.894(308)     &     64&.73(238)      \\
$E_J \times 10^3$           &  $-$13&.007(215)     &  $-$10&.926(248)     \\
$E_2 \times 10^3$           &  $-$13&.120(104)     &  $-$11&.948(118)     \\
$E_{KK} \times 10^6$        &    106&.9(42)        &     90&.             \\
$E_{JK} \times 10^6$        &   $-$6&.395(235)     &   $-$5&.4            \\
$E_{JJ} \times 10^9$        &     84&.4(84)        &     69&.             \\
$E_{KKK} \times 10^9$       & $-$540&.6(242)       & $-$420&.             \\
$E_{JKK} \times 10^9$       &   $-$8&.31(185)      &   $-$6&.4            \\
$E_{JJK} \times 10^9$       &      0&.469(41)      &      0&.36           \\
$E_{KKKK} \times 10^{12}$   &    844&.(41)         &    610&.             \\
$E_{JKKK} \times 10^{12}$   &     12&.85(303)      &      9&.1            \\
$F_{bc}$                    &     12&.200~46(194)  &     11&.935~60(118)  \\
$F_{bc,K} \times 10^3$      &     13&.28(74)       &     13&.0            \\
$F_{bc,J} \times 10^6$      &     93&.0(47)        &     91&.             \\
$F_{bc,KK} \times 10^6$     &      6&.70(43)       &      6&.6            \\
$F_{bc,JJ} \times 10^9$     &   $-$1&.565(304)     &   $-$1&.53           \\
$F_{ac}$                    &    100&.95(69)       &    100&.33(86)       \\
$F_{ac,J} \times 10^3$      &   $-$0&.951(39)      &   $-$0&.944          \\
$F_{{\rm 2}ac} \times 10^3$ &      1&.468~3(100)   &      1&.458          \\
\hline
\end{tabular}
\end{center}
\tablefoot{
$^a$ Watson's $S$ reduction was used in the representation $I^r$. 
Numbers in parentheses are one standard deviation in units of the least 
significant figures. Parameters of C$_2$H$_5^{34}$SH without uncertainties 
were taken from the main species and kept fixed in the fit.
}
\end{table}


Recently, \citet{EtSH_lab_Orion-KL_2014} reported on the extension of the rotational spectrum 
up to 660 and 880~GHz for the \textit{anti} and \textit{gauche} conformers of the main isotopic 
species, respectively, and on measurements of the \textit{gauche} conformer of the $^{34}$S 
isotopic species up to 300~GHz. Their Hamiltonian employed for the \textit{gauche} conformers 
was very similar to the one used earlier for the \textit{gauche} conformers of ethanol 
\citep{g-EtOH_rot_1996}. These data were used to create entries for the CDMS. We noted that 
the use of odd and even order interaction parameters resulted in very unstable fits for the 
\textit{gauche} conformers. Therefore, we tried to fit the data using a RAS, as described 
at the beginning of Sect.~\ref{lab-spec}, i.e., we tried to avoid using odd order parameters, 
such as $G_a$ and $G_b$, in the fit. We note that these parameters were called $Q$ and $N$, 
respectively, by \citet{EtSH_lab_Orion-KL_2014}. The magnitudes of 15.0~MHz and 6.3~MHz 
determined for the parent species are much smaller than the theoretically limiting values 
of $\sim$2$A$ (57494~MHz) and $\sim$2$B$ (10590~MHz), respectively, suggesting that both 
parameters have only fairly small effects in the fit. 
It is worthwhile mentioning that \citet{EtSH_rot_dip_1975} did not use these parameters 
in their fits. In fact, they used $F_{bc}$ and $F_{ac}$ for their more limited data set, 
which they named $D_{+-}$ and $E_{+-}$, respectively.

We determined spectroscopic parameters as averages between both states and expressed 
the differences as rotational corrections of the energy difference, as described above 
and as was done for ethanediol, for example \citep{aGg-eglyc_2003}. The advantage is 
that either parameter can be included in the fit without the other, whereas otherwise 
two distortion parameters are usually included in the fit, one for each state.
We used Watson's $S$ reduction of the rotational Hamiltonian because it is more 
appropriate for asymmetric top rotors so close to the prolate case; 
$\kappa = (2B - A - C)/(A - C) \approx -0.95$, close to the limiting case of $-$1. 
We employed in our first fits only the rotational and quartic distortion parameters, 
as well as $E$ and $F_{bc}$. The last was expected to be very important as it connects 
the upper asymmetry component of \textit{gauche}$^+$ with the lower asymmetry component 
of \textit{gauche}$^-$ with the same $J$ and $K_a$. The largest perturbations are expected 
if the asymmetry splitting is similar to the energy splitting between \textit{gauche}$^+$ 
and \textit{gauche}$^-$. We note that the energy difference is called $\Delta E$ in 
\citet{EtSH_lab_Orion-KL_2014} and equals $2E$ here by definition. At each stage of 
the fit, we searched among all reasonable parameters for the one that reduces the root 
mean square (rms) error as a measure of the quality of the fit most and which was determined 
with high significance. Neither $G_a$ nor $G_b$ was the parameter reducing the rms error 
most at any stage of the fit. Whenever one or the other was tried out in the fit, any 
appreciable reduction of the rms error made the fit highly unstable. This finding 
is hardly surprising as we do not have any indication that the transition frequencies of 
\textit{gauche}-C$_2$H$_5$SH in the fit are affected by perturbation with the 
\textit{anti} conformer. The final resulting parameters are given in 
Table~\ref{gauche-spec-param}.

We also determined new spectroscopic parameters for the \textit{gauche} conformer of 
C$_2$H$_5^{34}$SH. To this end, we took the parameters of the parent species and kept 
them fixed initially. Similarly to the fit of the parent species, we searched for the 
reasonable parameter whose floating in the fit caused the greatest reduction in the rms 
error until inclusion of new parameters caused only a minute reduction of the rms error. 
As some parameters showed larger isotopic changes than expected, we  scaled the parameters, 
which were kept fixed, with appropriate powers of $B + C$, $B - C$, and $E$. The final set 
of spectroscopic parameters for C$_2$H$_5^{34}$SH are also given in Table~\ref{gauche-spec-param}.


\begin{table}
\begin{center}
\caption{Spectroscopic parameters$^a$ (MHz) of \textit{anti}-C$_2$H$_5$SH 
         determined in the present study.}
\label{anti-spec-param}
\renewcommand{\arraystretch}{1.10}
\begin{tabular}[t]{lr@{}l}
\hline \hline
Parameter                 & \multicolumn{2}{c}{Value} \\
\hline
$A$                         & 28~416&.768~03(214)   \\
$B$                         &  5~485&.766~323(169)  \\
$C$                         &  4~881&.832~621(127)  \\
$D_K \times 10^3$           &    211&.166(149)      \\
$D_{JK} \times 10^3$        &  $-$22&.185~3(58)     \\
$D_J \times 10^3$           &      3&.787~500(119)  \\
$d_1 \times 10^6$           & $-$658&.795(45)       \\
$d_2 \times 10^6$           &  $-$23&.555(46)       \\
$H_K \times 10^6$           &    108&.19(267)       \\
$H_{JK} \times 10^9$        & $-$300&.6(63)         \\
$h_2 \times 10^9$           &      0&.774~0(234)    \\
$h_3 \times 10^9$           &      0&.253~8(158)    \\
\hline
\end{tabular}
\end{center}
\tablefoot{
$^a$ Watson's $S$ reduction was used in the representation $I^r$.
}
\end{table}


In addition, we redetermined the spectroscopic parameters of \textit{anti}-C$_2$H$_5$SH. 
\citet{EtSH_lab_Orion-KL_2014} used more centrifugal distortion parameters of eighth order 
for the \textit{anti} conformer than for the \textit{gauche} conformer even though fewer 
transition frequencies were determined extending to lower $J$ and, in particular, $K_a$ 
quantum numbers. We started with rotational and quartic centrifugal distortion parameters 
and weighted out lines that were difficult to reproduce. In the end, inclusion of only four 
additional sextic distortion parameters reproduced most of the transition frequencies well. 
The experimental line list is fairly sparse at higher $K_a$ values (9 to 15), and they 
were difficult to fit at somewhat higher $J$ (above 20). The residuals between observed 
and calculated frequencies showed only in some cases regular patterns. 
These residuals may be caused by perturbation of the rotational spectrum of the 
\textit{anti} conformer or by blending of the assigned lines with other, unidentified lines. 
Therefore, we omitted these lines together with few lines having lower values of $K_a$ 
which were also weak for the most part. Predictions beyond $K_a = 8$ or 9 should be viewed 
with caution because of these omissions. Transitions with such $K_a$ values may not be 
so important for astronomical observations because the \textit{anti} conformer is 
considerably higher in energy than the \textit{gauche} conformer. 
Our resulting spectroscopic parameters are presented in Table~\ref{anti-spec-param}. 
Associated line, parameter, and fit files along with additional auxiliary files will be 
available in the Fitting Spectra section of the 
CDMS\footnote{http://www.astro.uni-koeln.de/site/vorhersagen/pickett/beispiele/EtSH/}.


\begin{table}
\begin{center}
\caption{Spectroscopic parameters (MHz) of \textit{anti}-C$_2$H$_5$SH in Watson's $A$ 
         reduction from the present study in comparison to previous experimental values$^a$ 
         and those from quantum-chemical (QC) calculations$^b$.}
\label{anti-comp-param}
\renewcommand{\arraystretch}{1.10}
\begin{tabular}[t]{lr@{}lr@{}lr@{}l}
\hline \hline
Parameter                       & \multicolumn{2}{c}{Present exptl.} & \multicolumn{2}{c}{Previous exptl.} & \multicolumn{2}{c}{QC values}  \\
\hline
$A$                             &   28~416&.776~9(21)                &   28~416&.760~4(18)                 &   28~451&.975                  \\
$B$                             &    5~485&.778~73(28)               &    5~485&.779~01(15)                &    5~489&.912                  \\
$C$                             &    4~881&.817~82(17)               &    4~881&.817~62(12)                &    4~885&.345                  \\
${\it \Delta}_K \times 10^3$    &      211&.35(15)                   &      210&.03(18)                    &      198&.873                  \\
${\it \Delta}_{JK} \times 10^3$ &    $-$22&.395~1(48)                &    $-$22&.454~9(58)                 &    $-$23&.107                  \\
${\it \Delta}_J \times 10^3$    &        3&.830~86(36)               &        3&.832~17(27)                &        3&.767                  \\
$\delta_K \times 10^3$          &        7&.246(15)                  &        7&.342(20)                   &        5&.967                  \\
$\delta_J \times 10^3$          &        0&.657~25(19)               &        0&.656~64(11)                &        0&.555                  \\
${\it \Phi}_K \times 10^6$      &      103&.7(27)                    &       26&.7(80)                     &     $-$5&.464~7                \\
${\it \Phi}_{KJ} \times 10^6$   &         &                          &     $-$2&.717(106)                  &        1&.060~7                \\
${\it \Phi}_{JK} \times 10^9$   &         &                          &      367&.(18)                      &    $-$42&.3                    \\
${\it \Phi}_J \times 10^9$      &     $-$1&.36(28)                   &     $-$0&.86(17)                    &    $-$13&.0                    \\
$\phi_K \times 10^6$            &       13&.72(31)                   &       28&.89(97)                    &     $-$1&.345~0                \\
$\phi_{JK} \times 10^9$         &       80&.(14)                     &   $-$164&.(26)                      &   $-$226&.5                    \\
$\phi_J \times 10^9$            &     $-$1&.16(14)                   &     $-$0&.935(91)                   &     $-$6&.4                    \\
$L_K \times 10^6$               &         &                          &        1&.374(110)                  &         &                      \\
$L_{KKJ} \times 10^9$           &         &                          &        2&.11(40)                    &         &                      \\
$L_J \times 10^{15}$            &         &                          &       60&.(11)                      &         &                      \\
$l_{KJ} \times 10^9$            &         &                          &        4&.55(50)                    &         &                      \\
\hline
\end{tabular} 
\end{center}
\tablefoot{
$^a$ \citet{EtSH_lab_Orion-KL_2014}. 
$^b$ \citet{EtSH_DMS_ai_2014}; highest level CCSD(T)/cc-pVTZ values. 
}
\end{table}


The experimental transition frequencies of the \textit{gauche} conformers of the parent 
and $^{34}$S species of ethanethiol have been reproduced to 39.6~kHz and 38.1~kHz, 
respectively, about the same as in \citet{EtSH_lab_Orion-KL_2014}, where the rms values 
were 39~kHz and 41~kHz. However, we only required 34 parameters for the parent species 
whereas \citet{EtSH_lab_Orion-KL_2014} used 41. In the case of the $^{34}$S isotopic species, 
16 versus 19 parameters were floated. Therefore, we expect our parameters to describe 
the rotational spectrum better. Predictions much beyond the covered quantum number range 
($J \le 88$, $K_a \le 25$ for the parent species) should be viewed with caution. 
Such transitions are likely well beyond the scope of astronomical observations.

We looked into intensity alterations dependent on sign changes of the dipole moment 
components and found these alterations to be much smaller and much rarer than in ethanol 
(see Sect.~\ref{lab_EtOH}). Nevertheless, predicted intensities should be viewed with 
some caution.

Our values of $E_{KK}$, $E_{KKK}$, and $E_{KKKK}$ are fairly large in magnitude, possibly 
indicating perturbations of the spectrum by its lowest vibrational mode or by the 
\textit{anti} conformer. \citet{EtSH_rot_dip_1975} determined $F_{bc} = 12.20$~MHz and 
$F_{ac} = 139.34$~MHz. The former agrees perfectly with our value, and the agreement 
for the latter is reasonable if we take into account that in our case the uncertainty is 
0.69~MHz and in the case of \citet{EtSH_rot_dip_1975} it is at least several megahertz.

We have also determined spectroscopic parameters of the \textit{anti} conformer using Watson's 
$A$ reduction for comparison purposes with parameters of \citet{EtSH_lab_Orion-KL_2014} and 
with values from quantum chemical calculations \citep{EtSH_DMS_ai_2014}. 
These parameters are given in Table~\ref{anti-comp-param}.

The two sets of experimental rotational and quartic centrifugal distortion parameters 
agree well, the small, but significant differences in ${\it \Delta}_K$ and $\delta_K$ 
are caused by larger differences between the higher order parameters. 
The agreement is also good for values calculated by quantum chemical means. 
The rotational parameters include first-order vibrational corrections and are commonly 
marginally larger than the experimental ones at that level of computation. 
Quantum chemically calculated distortion parameters usually do not include vibrational 
corrections. The deviations between calculated quartic centrifugal distortion parameters 
and the experimental values are largely caused by this neglect.
The agreement between the two sets of experimental parameters is moderate to poor for 
the few sextic distortion parameters used in the present fit. The very large $L_K$ 
value in \citet{EtSH_lab_Orion-KL_2014} is largely responsible for the much smaller value 
of ${\it \Phi}_K$. In fact, contributions of $L_K$ are larger than those of ${\it \Phi}_K$ 
already at $K_a = 5$. The agreement is also poor between either set of sextic distortion 
parameters and those derived from quantum chemical calculations. \citet{EtSH_DMS_ai_2014} 
attributed these deviations entirely to errors in the experimental values. However, 
all three values (\textit{gauche}- and \textit{anti}-ethanethiol and dimethyl sulfide) of 
${\it \Phi}_K \approx H_K$ were reported as negative in that work. This is very unusual for 
an asymmetric top rotor close to the prolate limit, which usually has a positive value.  
For example, $\sim$23 and $\sim$18~Hz were obtained for the \textit{anti} conformers 
of the two isotopomers of ethanol with one $^{13}$C \citep{13C-EtOH_rot_2012}, 
13.73~Hz for dimethyl ether \citep{DME_rot_2009}, and 93~Hz for propargyl alcohol 
\citep{Propargyl-OH_rot_2005}. Therefore, we suspect that there are errors in the 
quantum chemical calculations of sextic distortion parameters in \citet{EtSH_DMS_ai_2014}.

\subsection{normal-Propanol, n-C$_3$H$_7$OH}
\label{lab_n-PrOH}

Replacing one H atom at the terminal C atom in ethanol with a CH$_3$ group yields the 
unbranched \textit{n}-propanol. There are three possibilities for doing so with CCCO 
dihedral angles of 180$^{\circ}$ or $\pm60^{\circ}$. As in the case of the HOCC dihedral, 
the conformations are called \textit{anti} (or \textit{trans}) and \textit{gauche}, 
the latter being doubly degenerate. Orientations of the methyl group are designated by 
upper-case characters, those of the OH group by lower-case characters. The \textit{Ga} 
conformer is lowest in energy, and the two non-equivalent \textit{Gg} and \textit{Gg'} 
conformers are higher by 68.81~K and 73.17~K, respectively, as determined from perturbations 
in rotational transitions of \textit{Ga} at high $K_a$ quantum numbers and from 
perturbations between \textit{Gg} and \textit{Gg'} \citep{n-PrOH_rot_dip-Ga_2010}. 
The energy of the \textit{Aa} conformer relative to \textit{Ga} is known only 
approximately, but it appears to be lower than that of either \textit{Gg} and \textit{Gg'} 
\citep{energies_ROH_2005}. The doubly degenerate \textit{Ag} conformer is also quite 
close in energy to the other conformers \citep{energies_ROH_2005}, and tunneling 
between the two equivalent conformers will likely lead to two distinguishable 
\textit{Ag}$^+$ and \textit{Ag}$^-$ states. Uncertainties of the relative energies 
of the \textit{A} conformers are difficult to estimate, but are probably at least 10~K 
if not a multiple thereof.

Predictions of the rotational spectrum of \textit{Ga-n}-propanol were taken from the CDMS. 
The entry is based on the extensive data from \citet{n-PrOH_rot_2006} with experimental 
data almost exclusively from that work. Frequencies of perturbed transitions were 
omitted. These transitions all have rather high values of $K_a$ and will be of importance 
for astronomical observations only if lower energy transitions have been detected with 
very high signal-tonoise ratios and at high rotational temperatures. Dipole moment 
components of $\mu_a = 0.4914~(4)$~D, $\mu_b = 0.9705~(13)$~D, and $\mu_c = 0.9042~(12)$~D 
were determined by \citet{n-PrOH_rot_dip-Ga_2010} from a single state fit. Based on 
the relative energies of the \textit{n}-propanol conformers, their total column density 
in the ground vibrational state at 150~K is about a factor of 3.3 higher than that 
determined for the \textit{Ga} conformer; inclusion of thermal population of vibrational 
states \citep{n-PrOH_FF_1968} yields a combined factor of about 5.2.

\subsection{iso-Propanol, i-C$_3$H$_7$OH}
\label{lab_i-PrOH}

Replacing one H atom at the central C atom in ethanol by a CH$_3$ group yields the 
branched \textit{i}-propanol. The two positions are equivalent. There are again three 
orientations for the OH group: \textit{anti} (or \textit{trans}) and the doubly degenerate 
\textit{gauche}. Tunneling between the two equivalent \textit{gauche} conformers leads 
again to \textit{gauche}$^+$ and \textit{gauche}$^-$, which are separated by 2.236~K 
\citep{i-PrOH_rot_2006}. There is consensus that the \textit{anti} conformer is higher 
in energy than the \textit{gauche} conformers, but the amount differs considerably. 
Intensity measurements at millimeter wavelengths \citep{i-PrOH_rot_2006} yielded 
a difference of $\sim$120~K with an estimated uncertainty of 15~K. This value is in 
reasonable agreement with 81~K \citep{energies_ROH_2005} and with $227 \pm 104$~K 
from intensity measurements in the microwave region \citep{g-i-PrOH_rot_dip_1979}. 
A value of 12.5~K \citep{i-PrOH_RSH_FIR_egy_1973} appears to be too small. We have 
employed an energy difference of 100~K as the average of the two determinations judged 
to be most reliable. At 150~K rotational temperature, this leads to a column density 
correction factor of 1.26 from thermal population of the \textit{anti} conformer. 
Taking into account vibrational data from \citet{i-PrOH_RSH_FIR_egy_1973} and 
\citet{i-PrOH_IR_2008} yields a combined conformational and vibrational correction 
factor to the partition function of $\sim$1.86.

Predictions were generated from data summarized in \citet{i-PrOH_rot_2006}, which are 
based largely on measurements from that work covering most of the 115$-$360~GHz region. 
\citet{g-i-PrOH_rot_dip_1979} determined $\mu_a = 1.114~(15)$~D between the \textit{gauche} 
substates and $\mu_b = 0.737~(25)$~D and $\mu_c = 0.8129~(49)$~D within the \textit{gauche} 
substates.

\section{Observations and data reduction}
\label{observations}

Part of the observations used in this article have been briefly described in 
\citet{i-PrCN_det_2014}. A detailed account of the observations, reduction, 
and analysis method of the full data set is reported in a companion paper 
\citep{deuterated_SgrB2N2_2015}. 

\section{Results}
\label{results}
\subsection{Methanol CH$_3$OH}
\label{ss:ch3oh}

Methanol is well detected toward Sgr~B2(N2) in its vibrational ground state and in its 
first torsionally excited state $\varv _{\rm t} = 1$ (Figs.~\ref{f:spec_ch3oh_ve0} and 
\ref{f:spec_ch3oh_ve1}). The second torsionally excited state is relatively well detected 
as well, with about three detected lines and seven additional lines that contribute 
significantly to the detected signal (Fig.~\ref{f:spec_ch3oh_ve2}). The six lines between 
96.19 and 96.56~GHz are the $a$-type ($\Delta K = 0$) transitions with $J = 2 - 1$. 
The remaining features are $b$-type transitions with $J \le 21$. The third torsionally 
excited state ($\varv _{\rm t} = 3$) cannot be considered as unambiguously detected; 
however, since it contributes significantly to the flux detected at 96268~MHz ($2_1 - 1_1$ 
of $E$ symmetry) according to our model that uses the same parameters as for the lower 
states, we include it in the full model (Fig.~\ref{f:spec_ch3oh_ve3}).

\onlfig{
\clearpage
\begin{figure*}
\centerline{\resizebox{0.9\hsize}{!}{\includegraphics[angle=0]{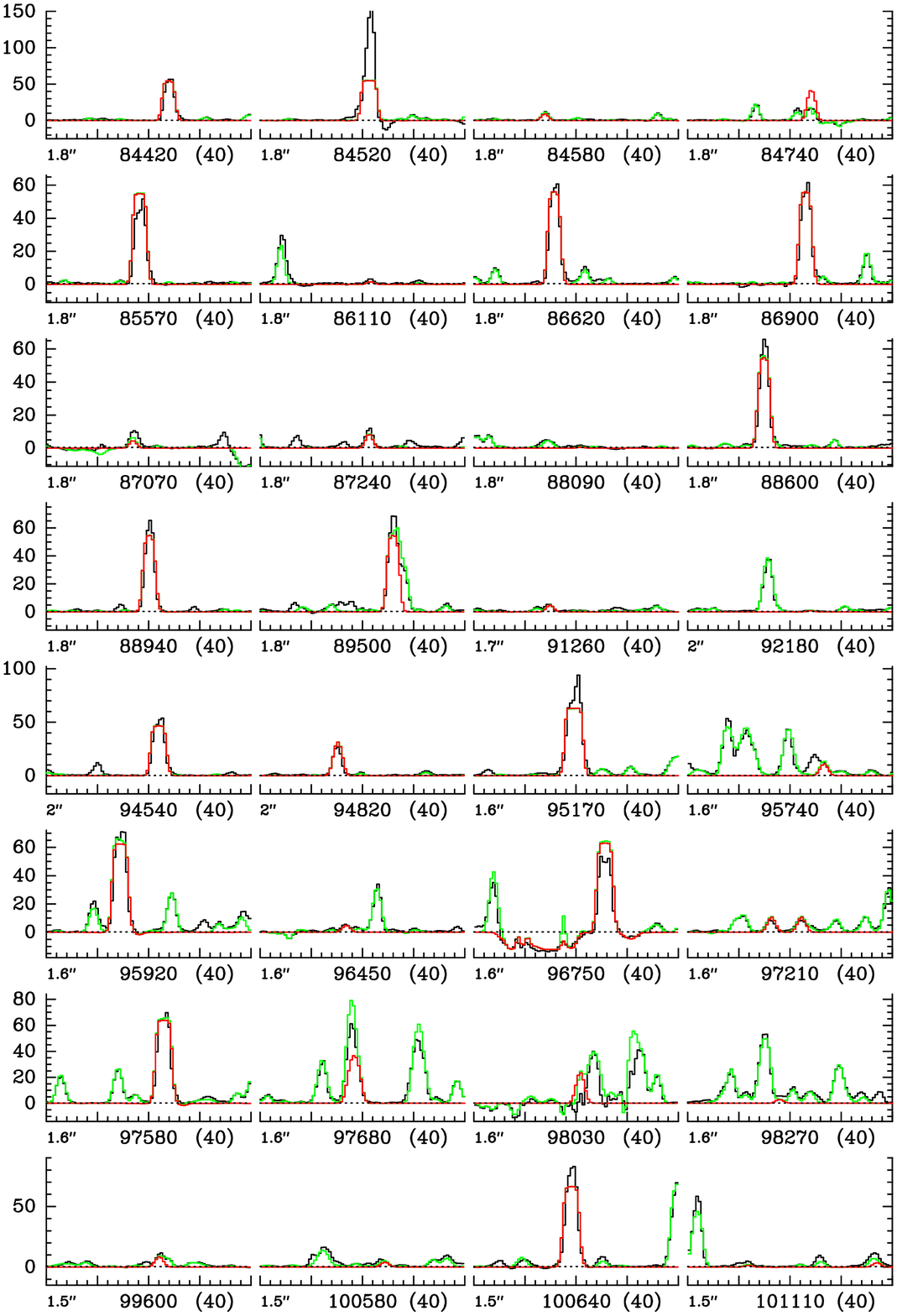}}}
\caption{Transitions of CH$_3$OH, $\varv = 0$ covered by our ALMA survey. The best-fit LTE 
synthetic spectrum of CH$_3$OH is displayed in red and overlaid on the observed spectrum 
of Sgr~B2(N2) shown in black. The green synthetic spectrum contains the contributions 
of all molecules identified in our survey so far including the species shown in red. 
The central frequency of each panel is indicated below the $x$-axis in MHz; its width 
is given also in MHz in parentheses. The angular resolution is also indicated. 
The $y$-axis is labeled in brightness temperature units (K). The dotted line marks 
the $3\sigma$ noise level.
}
\label{f:spec_ch3oh_ve0}
\end{figure*}
}

\onlfig{
\clearpage
\begin{figure*}
\addtocounter{figure}{-1}
\centerline{\resizebox{0.9\hsize}{!}{\includegraphics[angle=0]{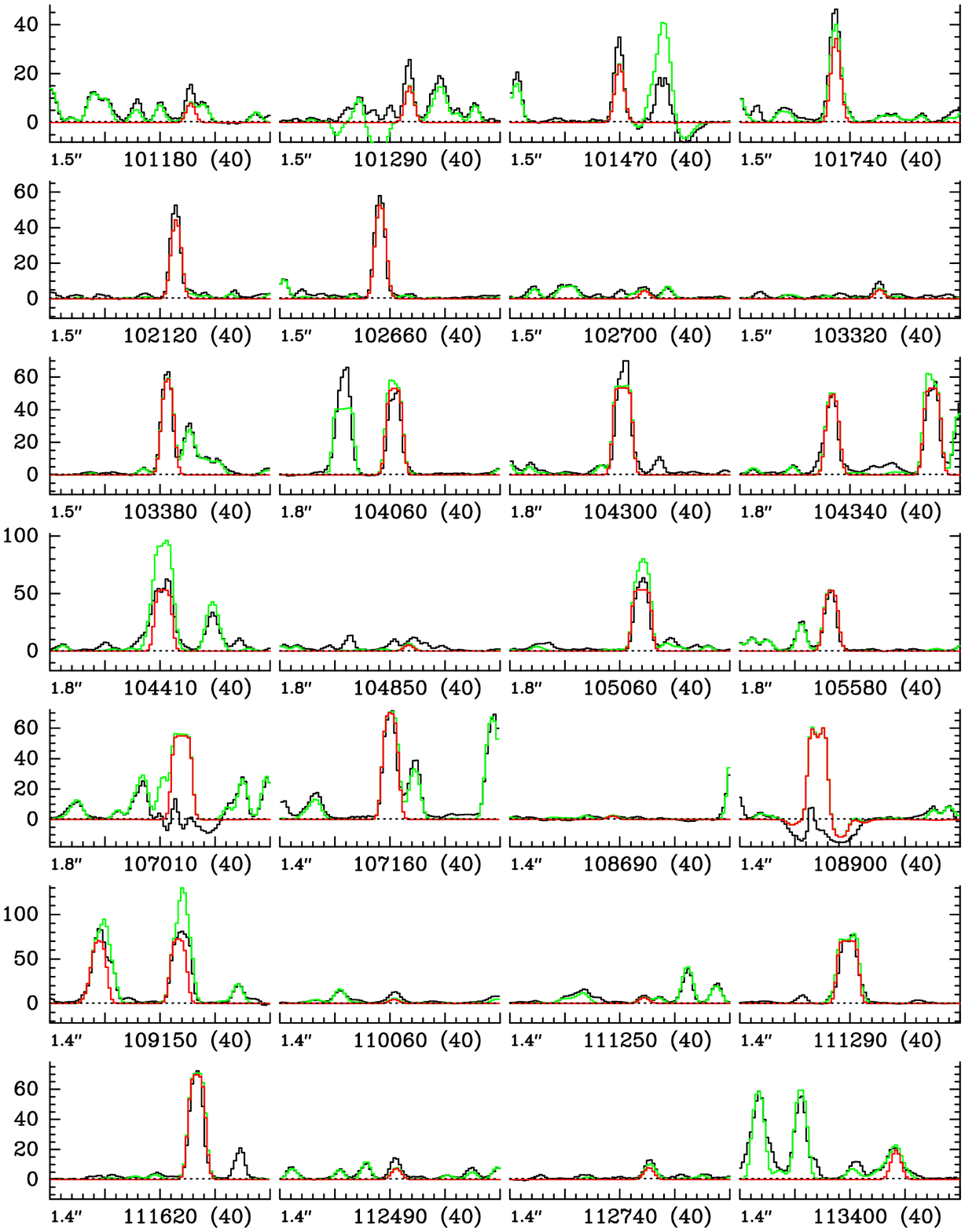}}}
\caption{continued.}
\end{figure*}
}
\addtocounter{figure}{-1}

\onlfig{
\clearpage
\begin{figure*}
\centerline{\resizebox{0.9\hsize}{!}{\includegraphics[angle=0]{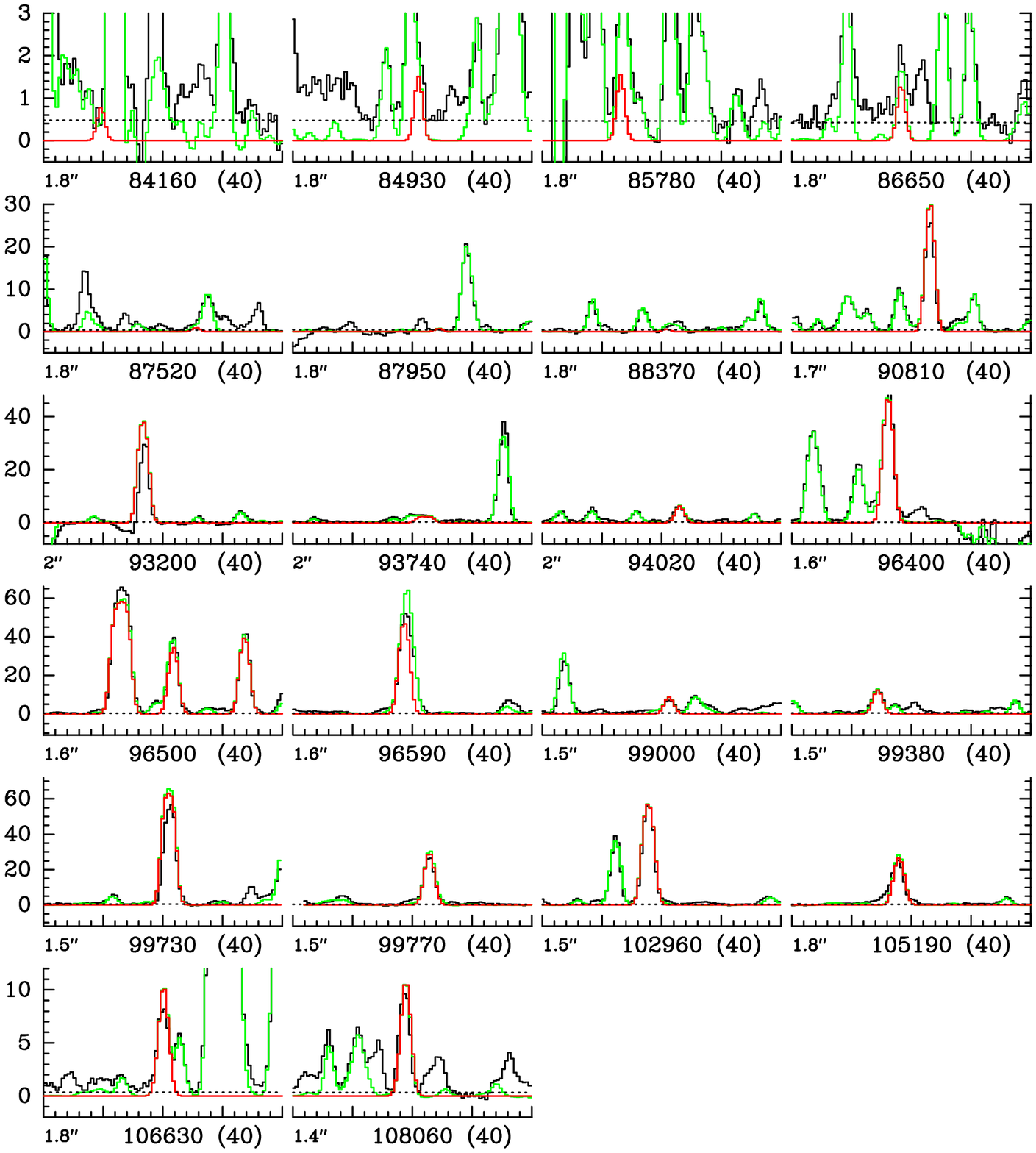}}}
\caption{Same as Fig.~\ref{f:spec_ch3oh_ve0} for CH$_3$OH, $\varv_{\rm t}=1$.
}
\label{f:spec_ch3oh_ve1}
\end{figure*}
}

\onlfig{
\clearpage
\begin{figure*}
\centerline{\resizebox{0.9\hsize}{!}{\includegraphics[angle=0]{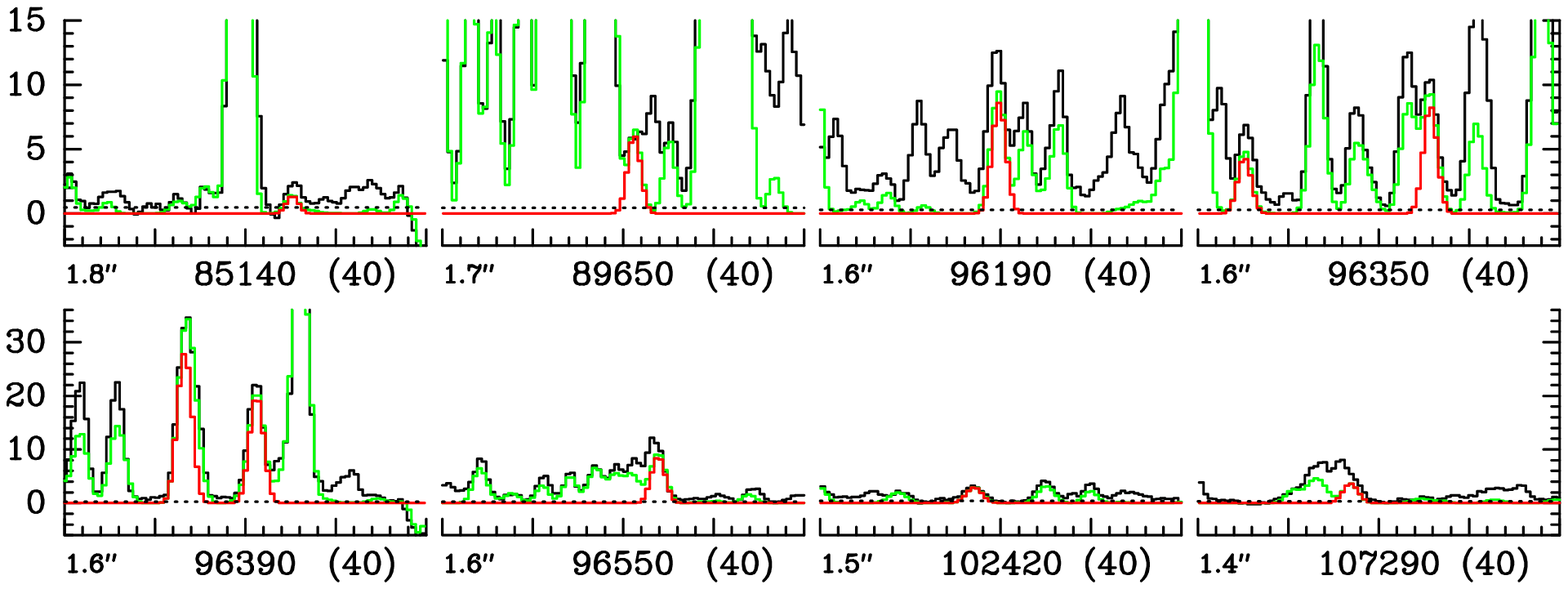}}}
\caption{Same as Fig.~\ref{f:spec_ch3oh_ve0} for CH$_3$OH, $\varv_{\rm t}=2$.
}
\label{f:spec_ch3oh_ve2}
\end{figure*}
}

\onlfig{
\begin{figure*}
\centerline{\resizebox{0.45\hsize}{!}{\includegraphics[angle=0]{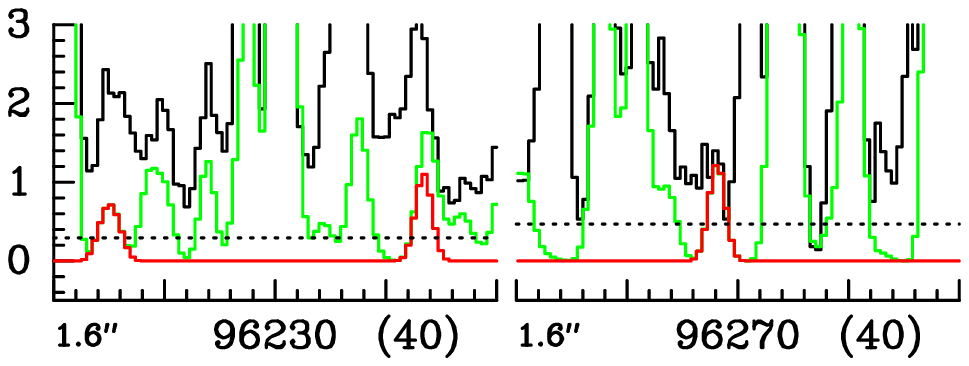}}}
\caption{Same as Fig.~\ref{f:spec_ch3oh_ve0} for CH$_3$OH, $\varv_{\rm t}=3$.
}
\label{f:spec_ch3oh_ve3}
\end{figure*}
}


The $^{13}$C isotopolog of methanol is also clearly detected, both in its ground state 
and in $\varv _{\rm t} = 1$ (Figs.~\ref{f:spec_ch3oh_13c_ve0} and \ref{f:spec_ch3oh_13c_ve1}). 
Some lines in the ground state are marginally optically thick ($\tau_{\rm max} = 1.3$). 
About eight lines of the $^{18}$O isotopolog are detected in its vibrational ground state, 
which makes the assignment secure (Fig.~\ref{f:spec_ch3oh_18o_ve0}). 
The $J = 2 - 1$ $a$-type transitions occur at 91958~MHz (not shown, strongly blended), 
around 92730~MHz, and at 93506~MHz. The remainder are $b$-type transitions with low values 
of $K$ and low to moderate values of $J$. Their emission is optically thin ($\tau_{\rm max} 
= 0.14$). Transitions from within $\varv _{\rm t} = 1$ are not unambiguously detected for 
this isotopolog, but its emission contributes significantly to the line detected at 
97069~MHz (Fig.~\ref{f:spec_ch3oh_18o_ve1}) which is why we included it in the full model. 
Our model allows us to derive the following isotopic ratios for methanol: $^{12}$C/$^{13}$C 
$\approx 25$ and $^{16}$O/$^{18}$O $\approx 180$. The $^{13}$CH$_3$OH/CH$_3^{18}$OH ratio 
is $\sim$7.3.


\onlfig{
\clearpage
\begin{figure*}
\centerline{\resizebox{0.9\hsize}{!}{\includegraphics[angle=0]{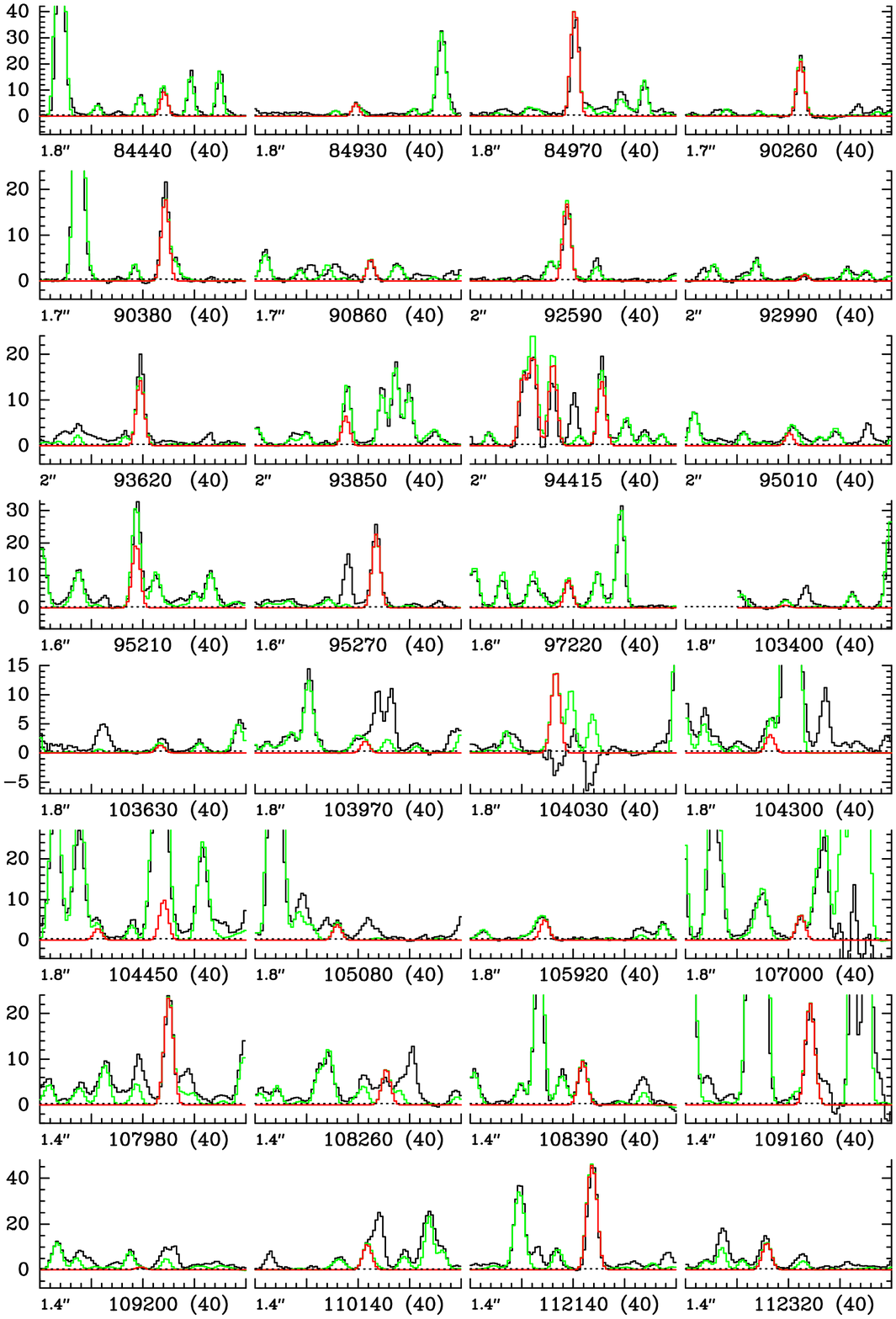}}}
\caption{Same as Fig.~\ref{f:spec_ch3oh_ve0} for $^{13}$CH$_3$OH, $\varv=0$.
}
\label{f:spec_ch3oh_13c_ve0}
\end{figure*}
}

\onlfig{
\clearpage
\begin{figure*}
\addtocounter{figure}{-1}
\centerline{\resizebox{0.45\hsize}{!}{\includegraphics[angle=0]{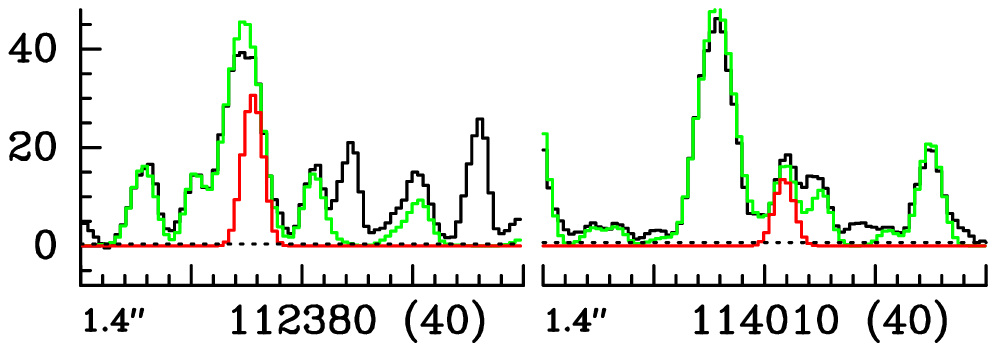}}}
\caption{continued.}
\end{figure*}
}
\addtocounter{figure}{-1}

\onlfig{
\begin{figure*}
\centerline{\resizebox{0.9\hsize}{!}{\includegraphics[angle=0]{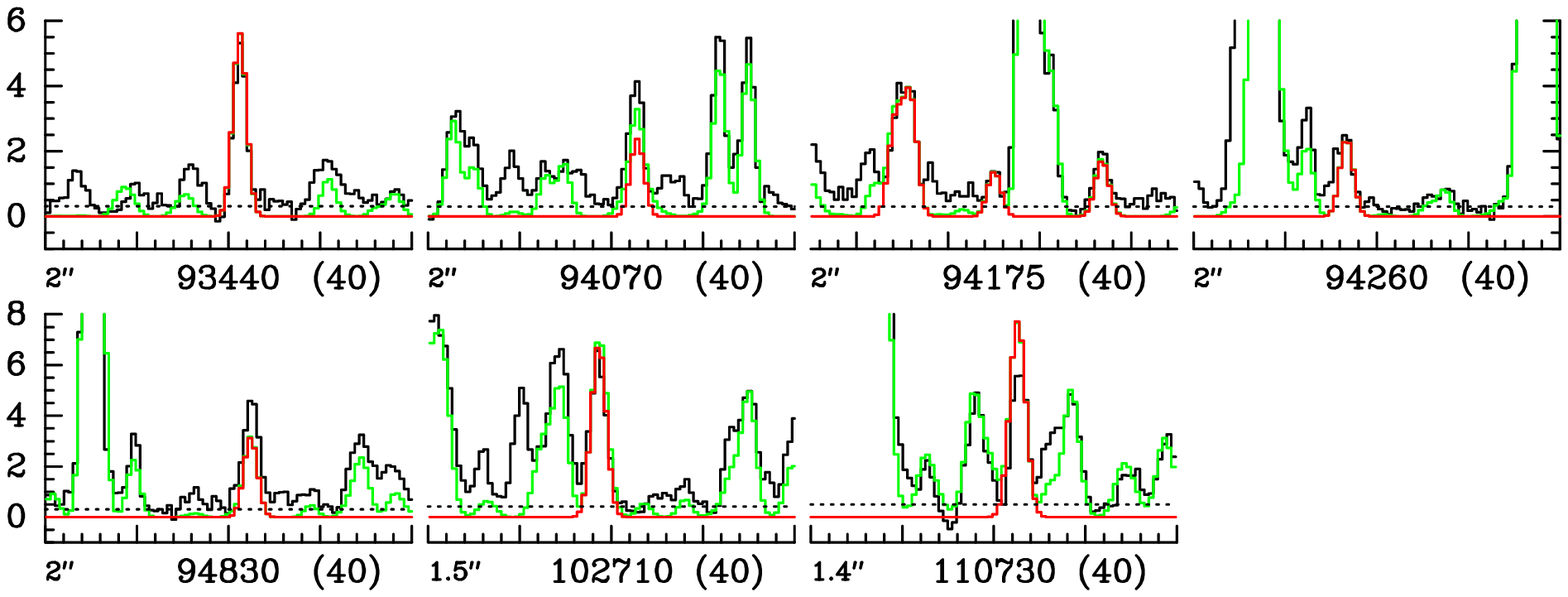}}}
\caption{Same as Fig.~\ref{f:spec_ch3oh_ve0} for $^{13}$CH$_3$OH, $\varv_{\rm t}=1$.
}
\label{f:spec_ch3oh_13c_ve1}
\end{figure*}
}

\onlfig{
\clearpage
\begin{figure*}
\centerline{\resizebox{0.9\hsize}{!}{\includegraphics[angle=0]{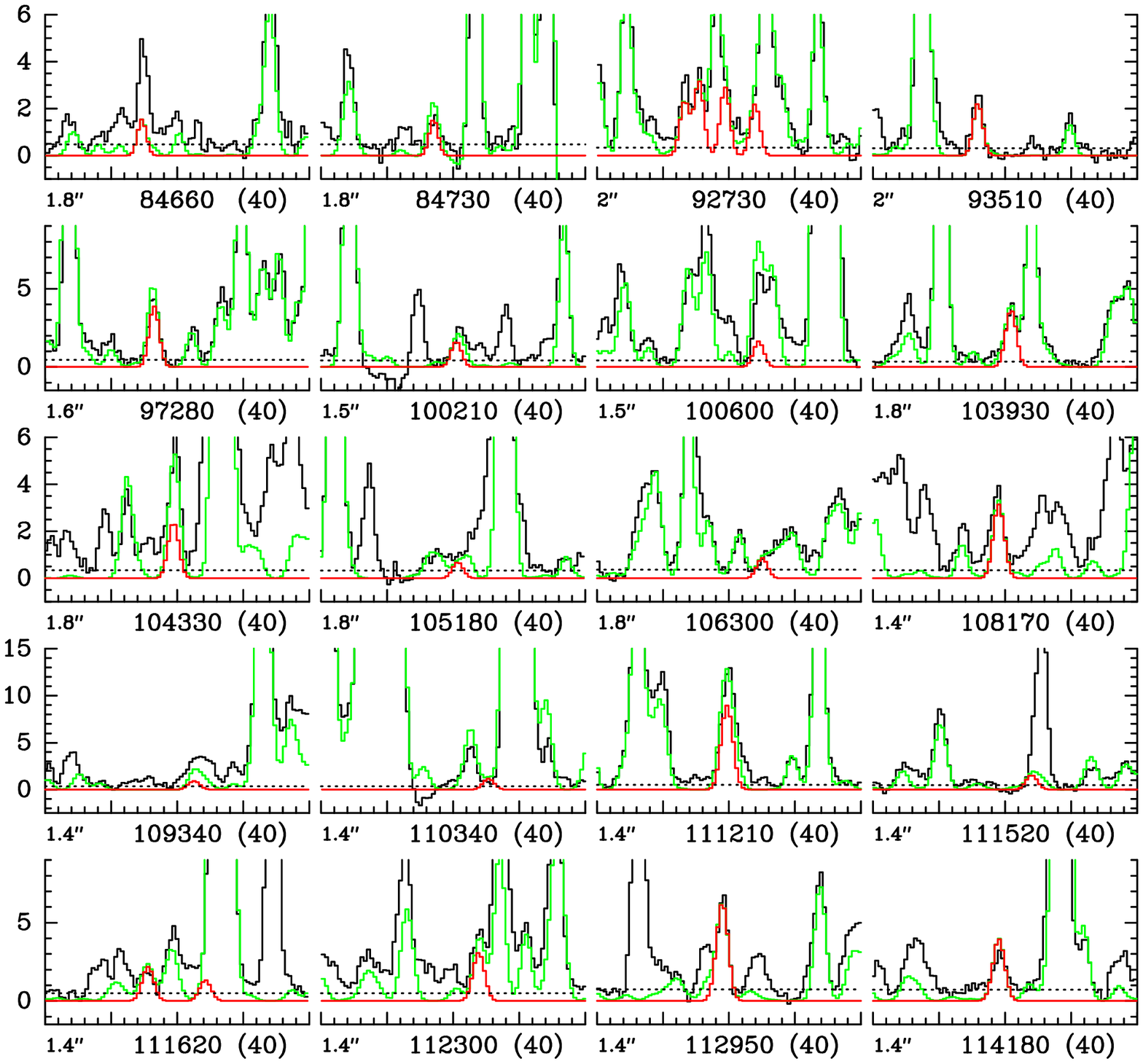}}}
\caption{Same as Fig.~\ref{f:spec_ch3oh_ve0} for CH$_3$$^{18}$OH, $\varv=0$.
}
\label{f:spec_ch3oh_18o_ve0}
\end{figure*}
}

\onlfig{
\begin{figure*}
\centerline{\resizebox{0.9\hsize}{!}{\includegraphics[angle=0]{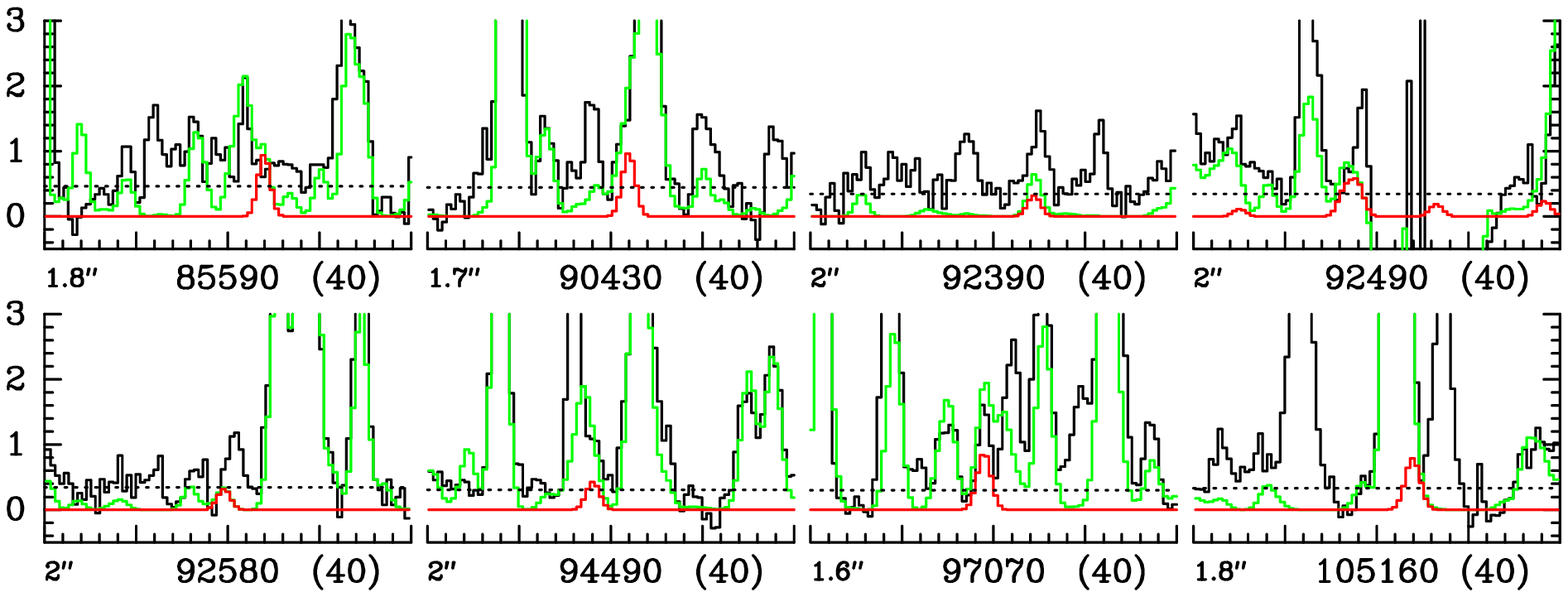}}}
\caption{Same as Fig.~\ref{f:spec_ch3oh_ve0} for CH$_3$$^{18}$OH, $\varv_{\rm t}=1$.
}
\label{f:spec_ch3oh_18o_ve1}
\end{figure*}
}


The fits to the integrated intensity maps of the detected methanol lines yield a source 
size that decreases with increasing energy level, from $\sim$1.4$\arcsec$ at low energy to 
$\sim$1.0$\arcsec$ at $E_{\rm up} \approx 1000$~K. A similar behavior is found for the $^{13}$C 
isotopolog over a narrower energy range. The population diagrams of methanol (restricted 
to its transitions with an opacity lower than 2) and its isotopologs are very well fitted 
with a rotation temperature of about 140$-$160~K (Table~\ref{t:popfit} and 
Figs.~\ref{f:popdiag_ch3oh}, \ref{f:popdiag_ch3oh_13c}, and \ref{f:popdiag_ch3oh_18o}). 
The rotation temperature of the CH$_3$OH ground vibrational state is constrained at low to 
moderate energies, for example, by the $J_{-2} - J_1$ transitions of $E$ symmetry. 
The lines are very weak at low $J$ (2$-$4 near 101.1~GHz), but have substantial opacities 
at higher $J$; $J = 17$ at 111626.5~MHz is the last line in the region of our survey. 
On the higher energy side are the two asymmetry split $26_5 - 25_6$ transitions of $A$ 
symmetry around 97210~MHz. The spectra of the optically thick lines of methanol require 
a large size and high temperature to be well fitted. Our LTE model thus assumes a source 
size of $1.4\arcsec$ and a rotational temperature of 160~K. The fit is good for most lines 
of all three isotopologs, except for the shape of the very optically thick lines of methanol 
(with $\tau$ up to 19) and the likely masing lines at 84521 and 95169~MHz ($5_{-1} - 4_0$ 
of $E$ symmetry and $8_0 - 7_1$ of $A^+$ symmetry, respectively; see, e.g., 
\citealt{MeOH_maser_dark-cloud_2004}).


\onlfig{
\clearpage
\begin{figure}
\centerline{\resizebox{1.0\hsize}{!}{\includegraphics[angle=0]{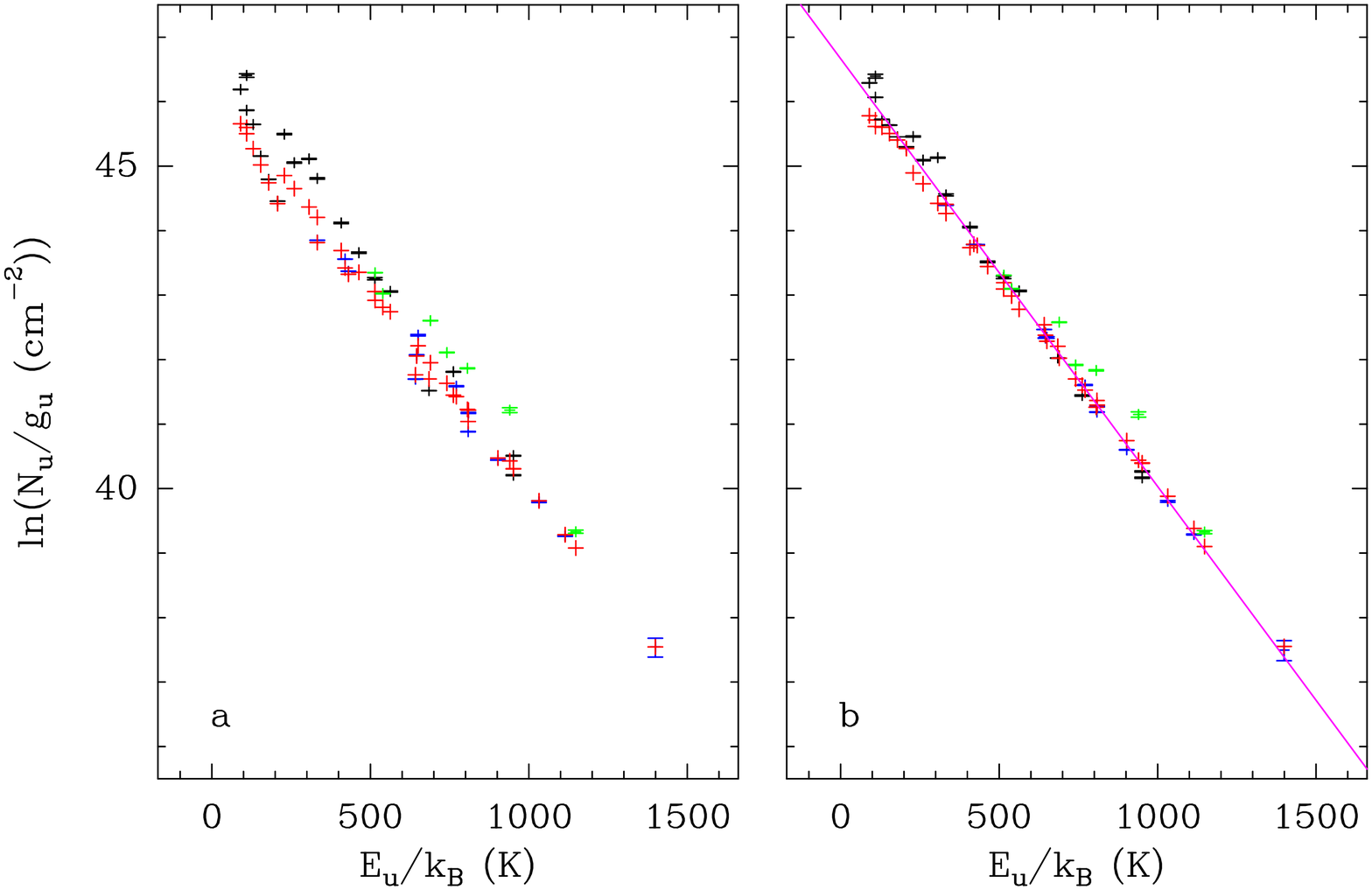}}}
\caption{Population diagram of CH$_3$OH, $\varv=0$, $\varv_{\rm t}=1$, and $\varv_{\rm t}=2$
toward Sgr~B2(N2). Only the lines that are clearly detected, do not suffer too much from 
contamination from other species, and have an opacity below 2 are displayed. The observed 
data points are shown in black while the synthetic populations are shown in red. No correction 
is applied in panel \textbf{a}. In panel \textbf{b}, the optical depth correction has been 
applied to both the observed and synthetic populations and the contamination from all other 
species included in the full model has been removed from the observed data points. The straight 
line is a linear fit to the observed populations (in linear-logarithmic space). 
The rotational temperature derived in this way is reported in Table~\ref{t:popfit}.}
\label{f:popdiag_ch3oh}
\end{figure}
}

\onlfig{
\begin{figure}
\centerline{\resizebox{1.0\hsize}{!}{\includegraphics[angle=0]{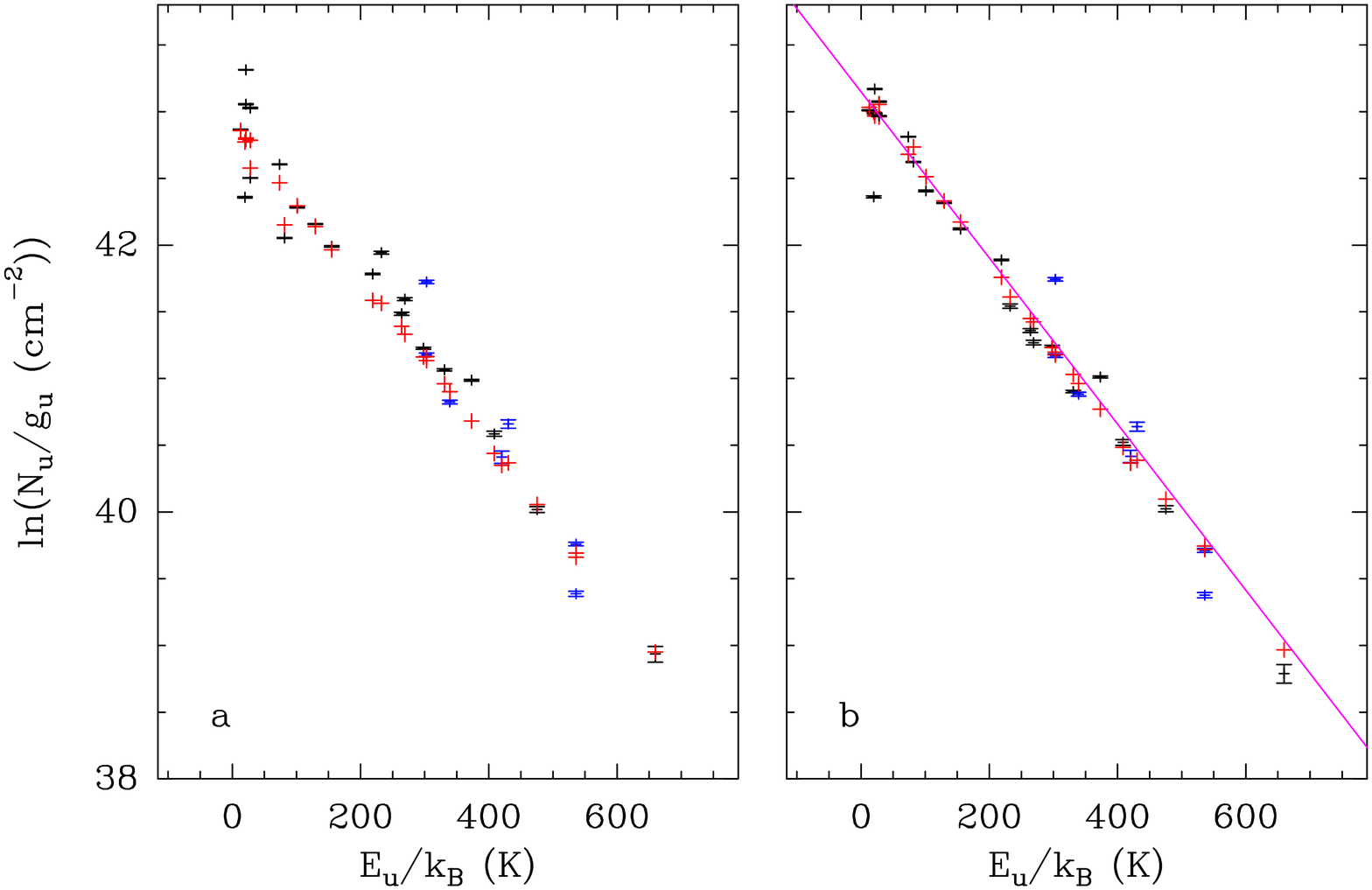}}}
\caption{Same as Fig.~\ref{f:popdiag_ch3oh} for $^{13}$CH$_3$OH, $\varv=0$ and
$\varv_{\rm t}=1$.}
\label{f:popdiag_ch3oh_13c}
\end{figure}
}

\onlfig{
\begin{figure}
\centerline{\resizebox{1.0\hsize}{!}{\includegraphics[angle=0]{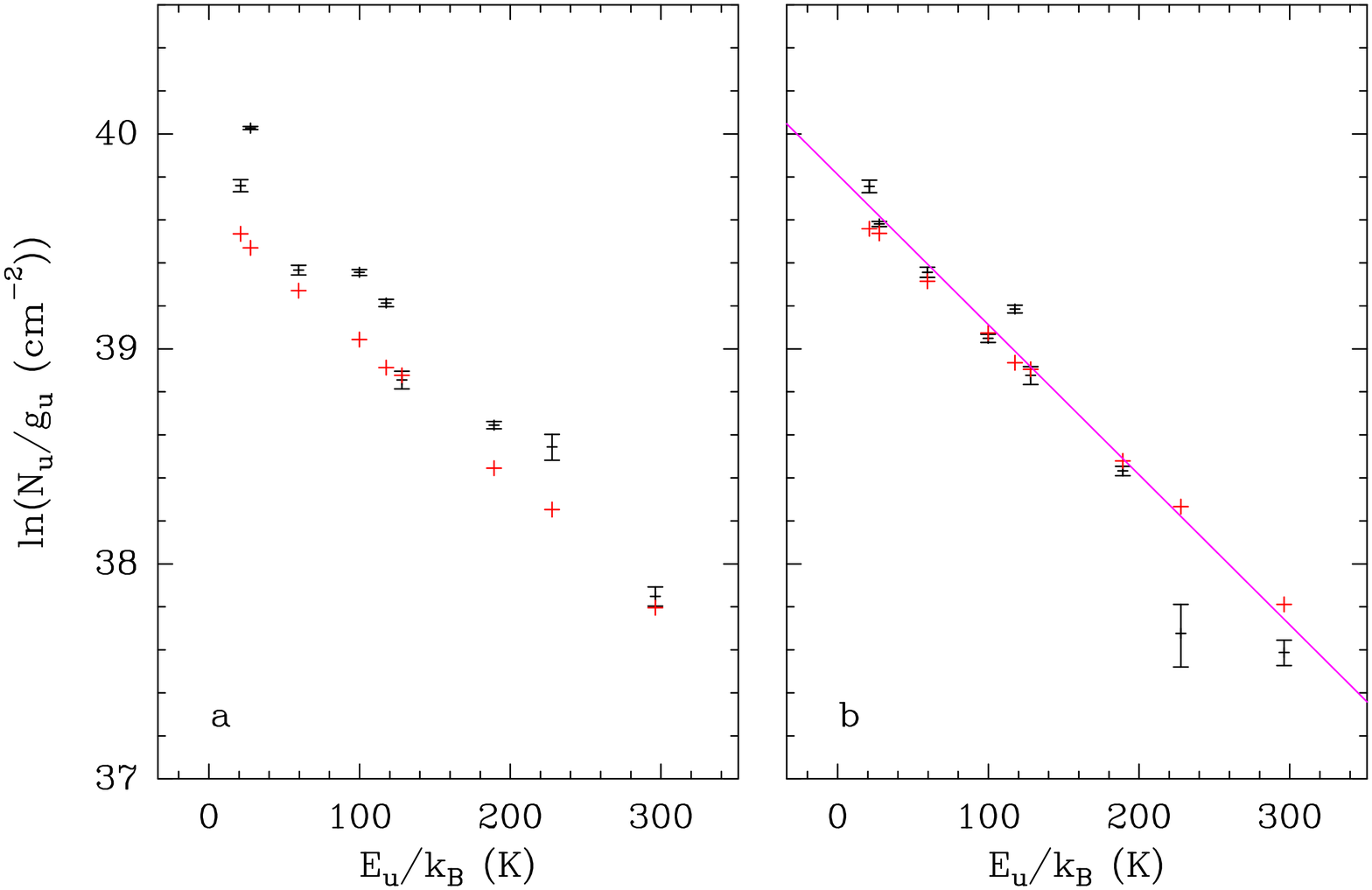}}}
\caption{Same as Fig.~\ref{f:popdiag_ch3oh} for CH$_3$$^{18}$OH, $\varv=0$.}
\label{f:popdiag_ch3oh_18o}
\end{figure}
}

\input{abb/tab_r-sh_popfit.tex}

\input{abb/tab_r-sh_weedsmodel.tex}
\subsection{Ethanol C$_2$H$_5$OH}
\label{ss:c2h5oh}

About 170 lines of ethanol are detected toward Sgr~B2(N2) in its vibrational 
ground state (Fig.~\ref{f:spec_c2h5oh_ve0}). The fits to the integrated 
intensity maps of these lines yield a median source size of $\sim$1.3$''$, 
with no clear trend as a function of upper-level energy. The population diagram 
shown in Fig.~\ref{f:popdiag_c2h5oh} yields a well-constrained rotational 
temperature (Table~\ref{t:popfit}). We use a slightly higher temperature 
(150~K versus 140~K), still fully consistent with the intensities of the 
detected lines. With this source size and temperature, some lines turn out to 
be marginally optically thick ($\tau_{\rm max} = 1.2$).

\onlfig{
\clearpage
\begin{figure*}
\centerline{\resizebox{0.9\hsize}{!}{\includegraphics[angle=0]{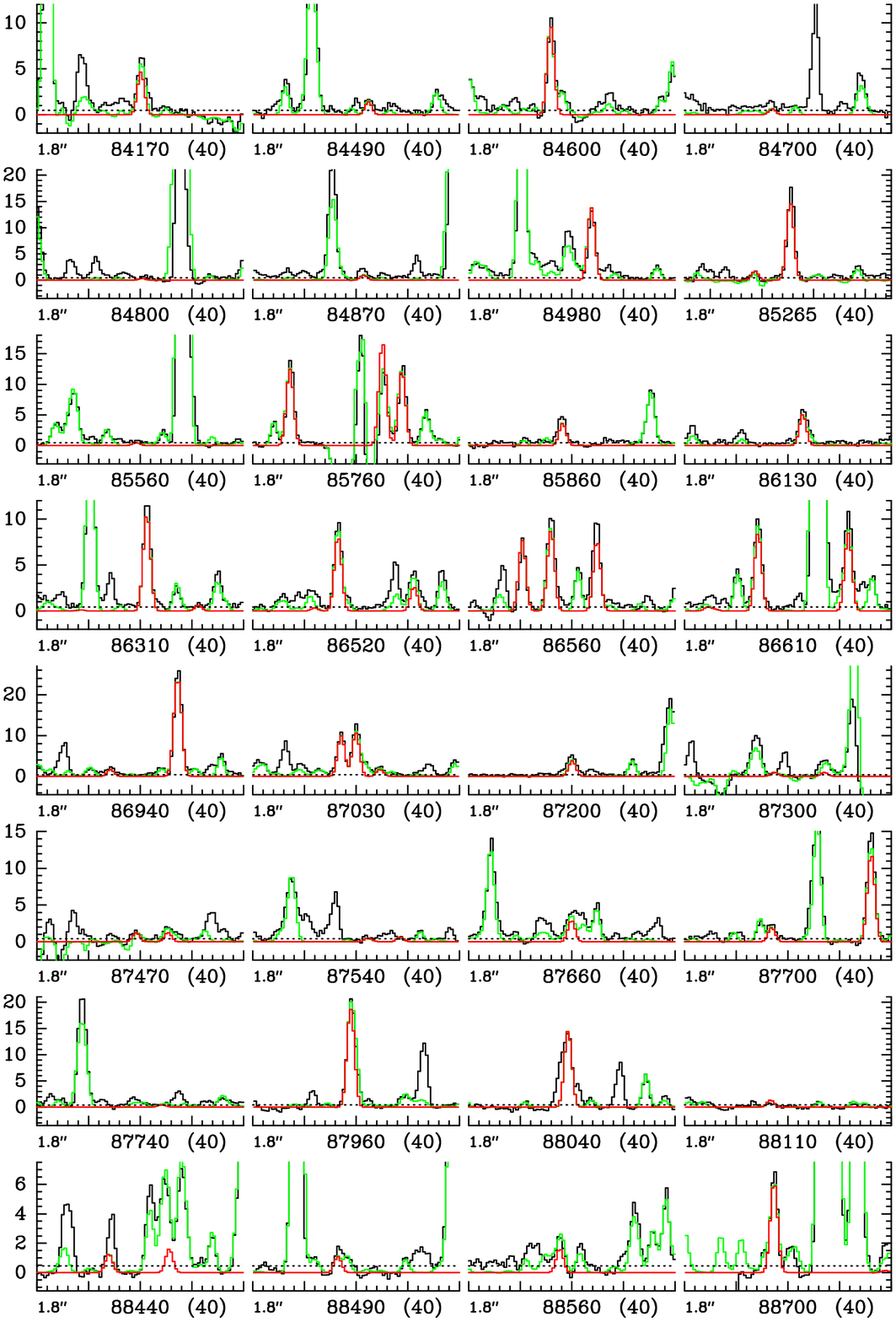}}}
\caption{Same as Fig.~\ref{f:spec_ch3oh_ve0} for C$_2$H$_5$OH, $\varv=0$.
}
\label{f:spec_c2h5oh_ve0}
\end{figure*}
}

\onlfig{
\begin{figure*}
\addtocounter{figure}{-1}
\centerline{\resizebox{0.9\hsize}{!}{\includegraphics[angle=0]{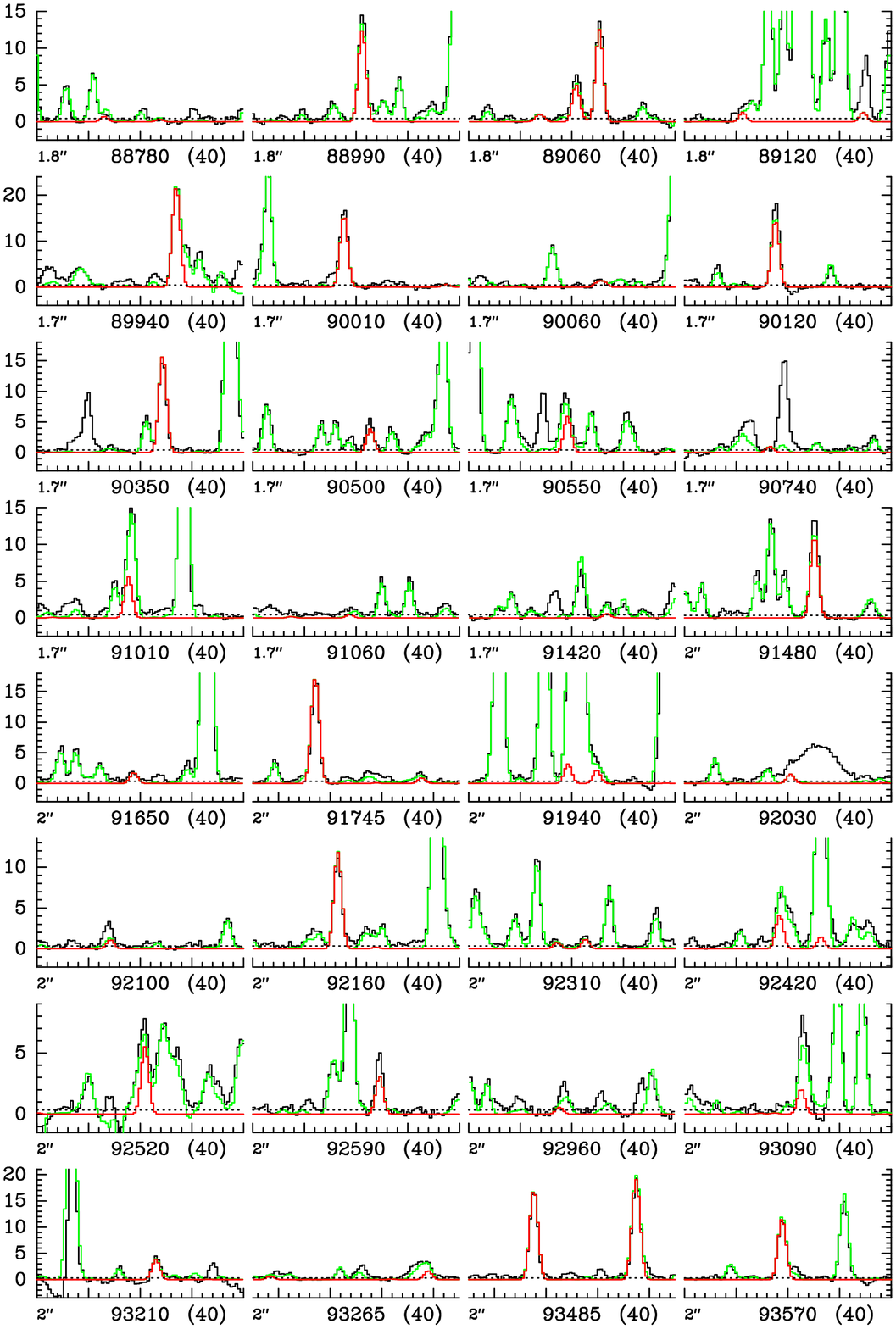}}}
\caption{continued.}
\end{figure*}
}
\addtocounter{figure}{-1}

\onlfig{
\begin{figure*}
\addtocounter{figure}{-1}
\centerline{\resizebox{0.9\hsize}{!}{\includegraphics[angle=0]{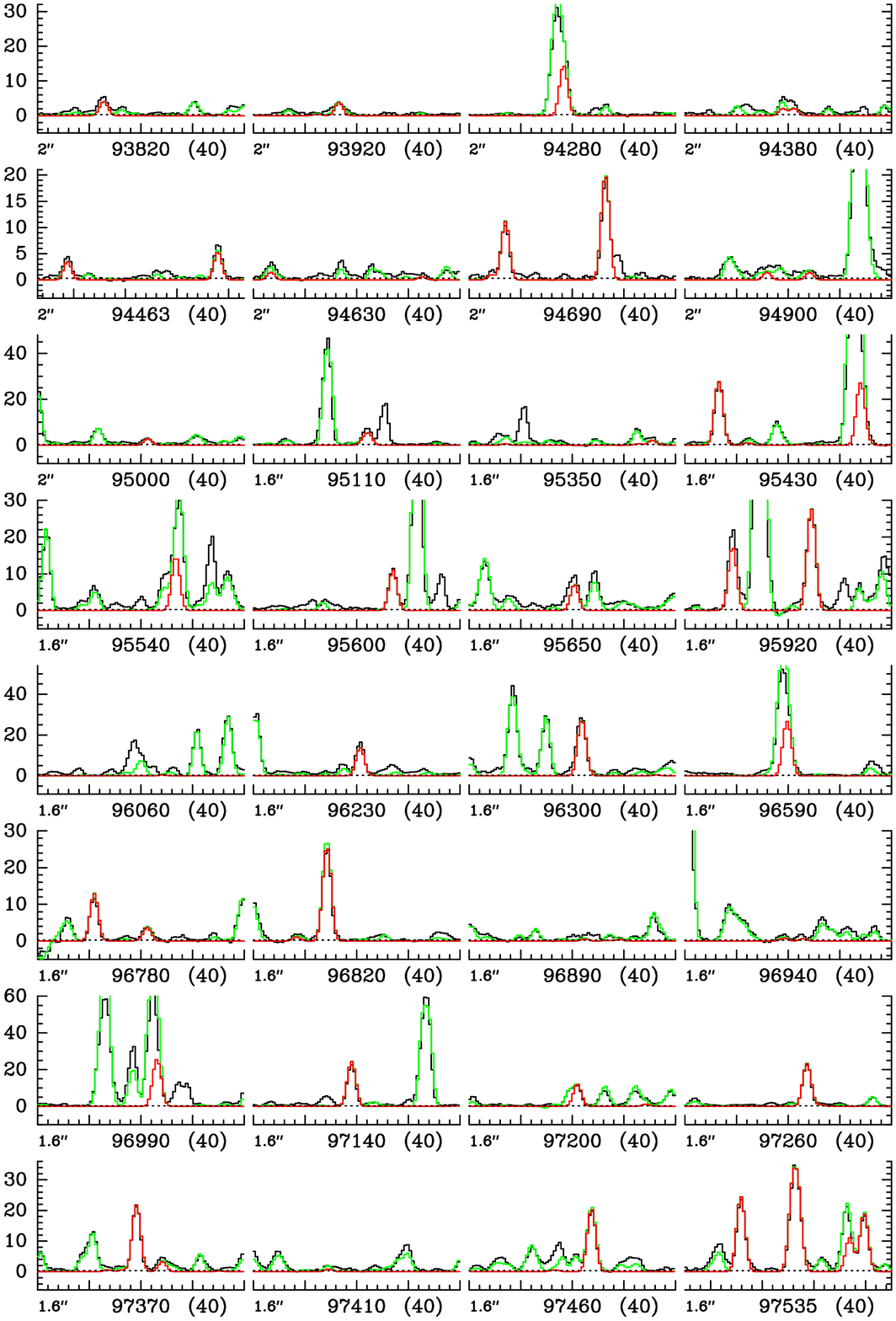}}}
\caption{continued.}
\end{figure*}
}
\addtocounter{figure}{-1}

\onlfig{
\begin{figure*}
\addtocounter{figure}{-1}
\centerline{\resizebox{0.9\hsize}{!}{\includegraphics[angle=0]{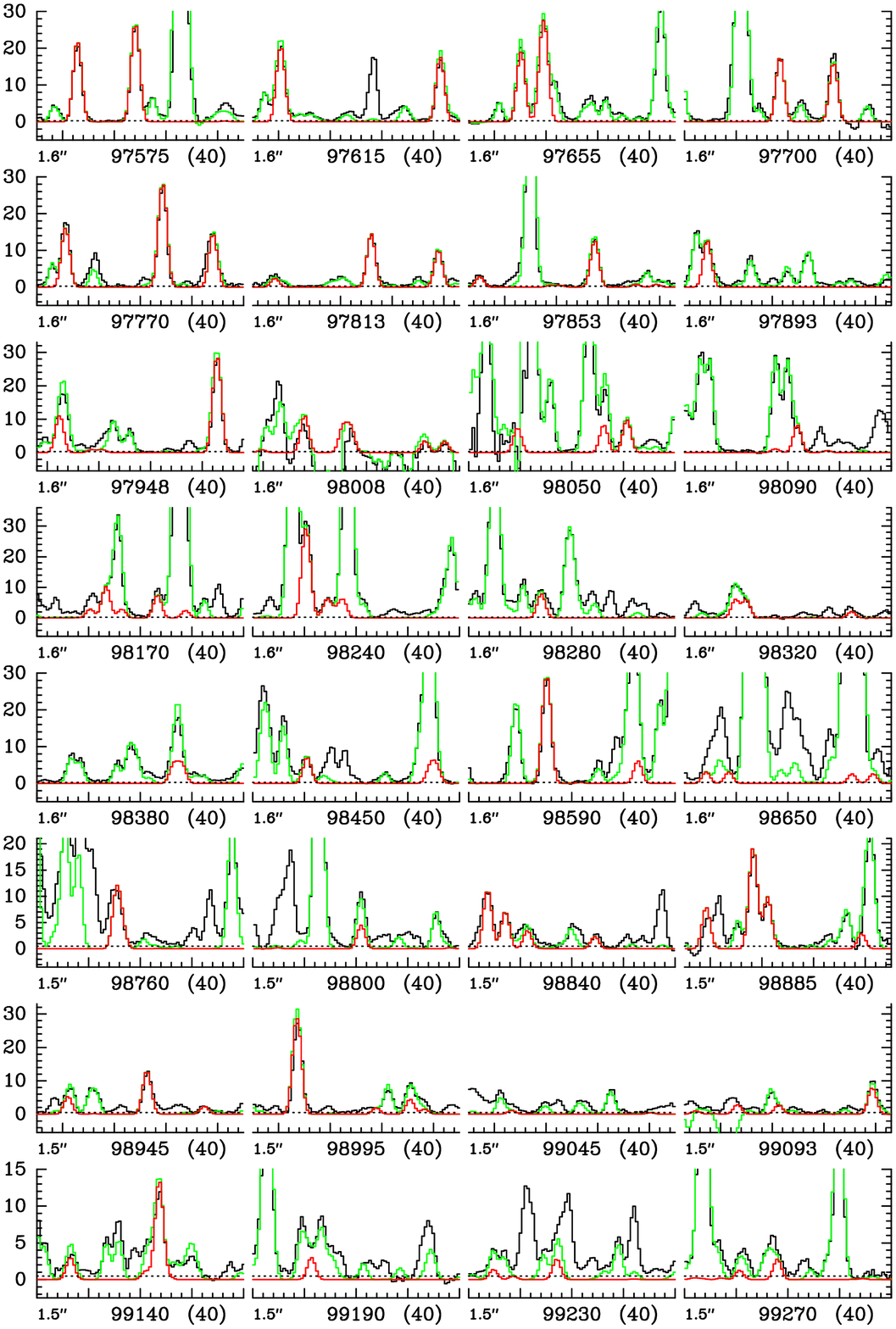}}}
\caption{continued.}
\end{figure*}
}
\addtocounter{figure}{-1}

\onlfig{
\begin{figure*}
\addtocounter{figure}{-1}
\centerline{\resizebox{0.9\hsize}{!}{\includegraphics[angle=0]{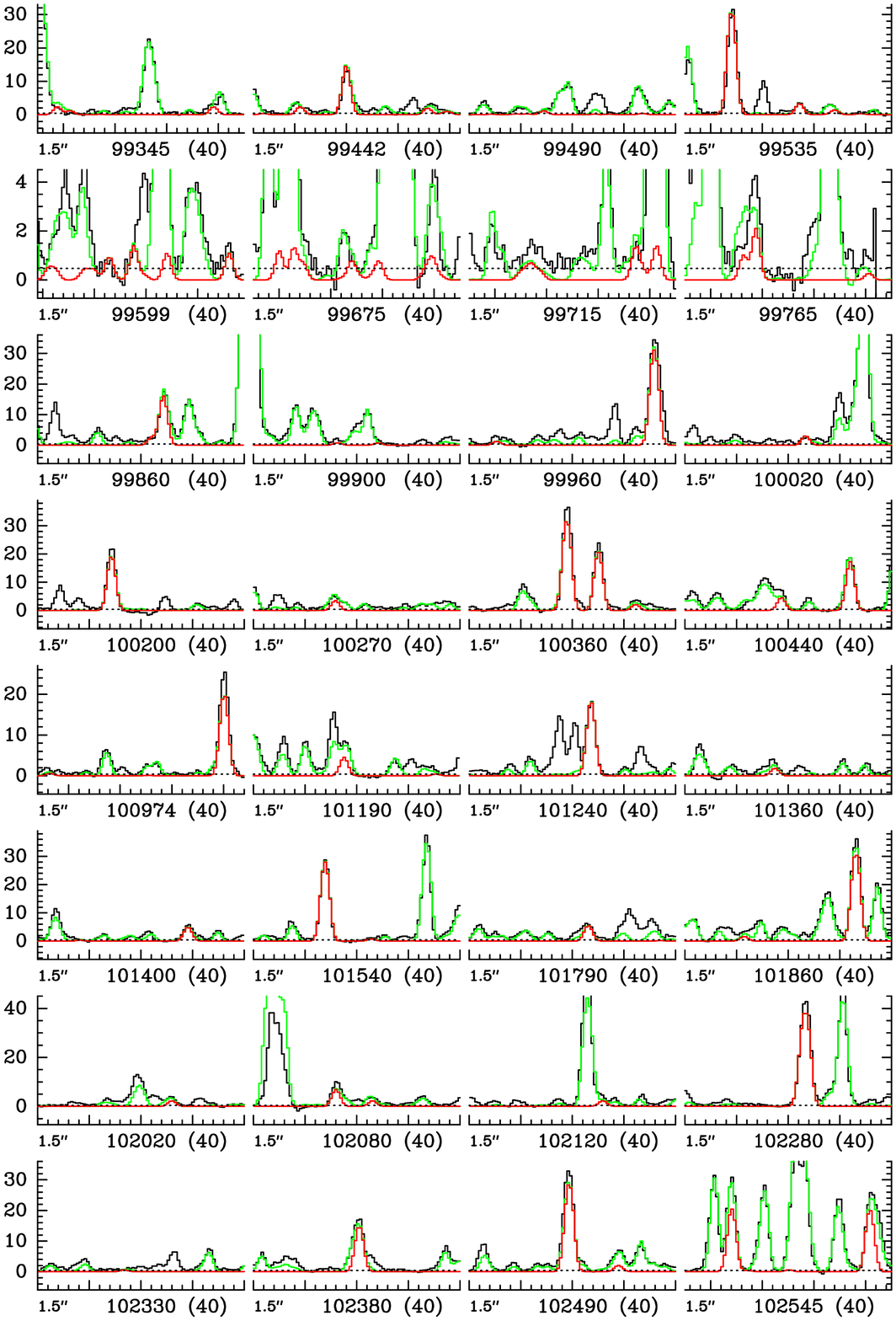}}}
\caption{continued.}
\end{figure*}
}
\addtocounter{figure}{-1}

\onlfig{
\begin{figure*}
\addtocounter{figure}{-1}
\centerline{\resizebox{0.9\hsize}{!}{\includegraphics[angle=0]{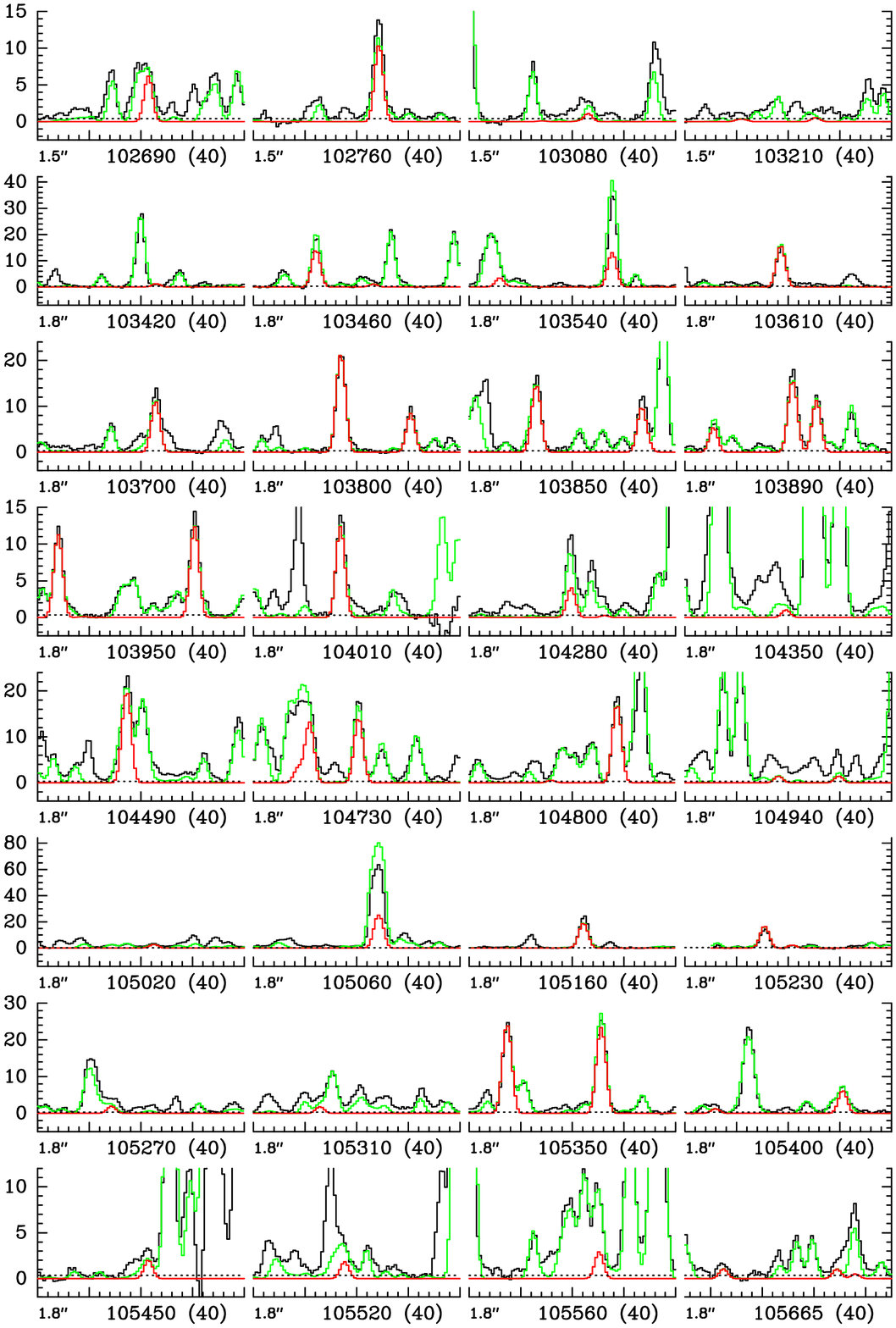}}}
\caption{continued.}
\end{figure*}
}
\addtocounter{figure}{-1}

\onlfig{
\begin{figure*}
\addtocounter{figure}{-1}
\centerline{\resizebox{0.9\hsize}{!}{\includegraphics[angle=0]{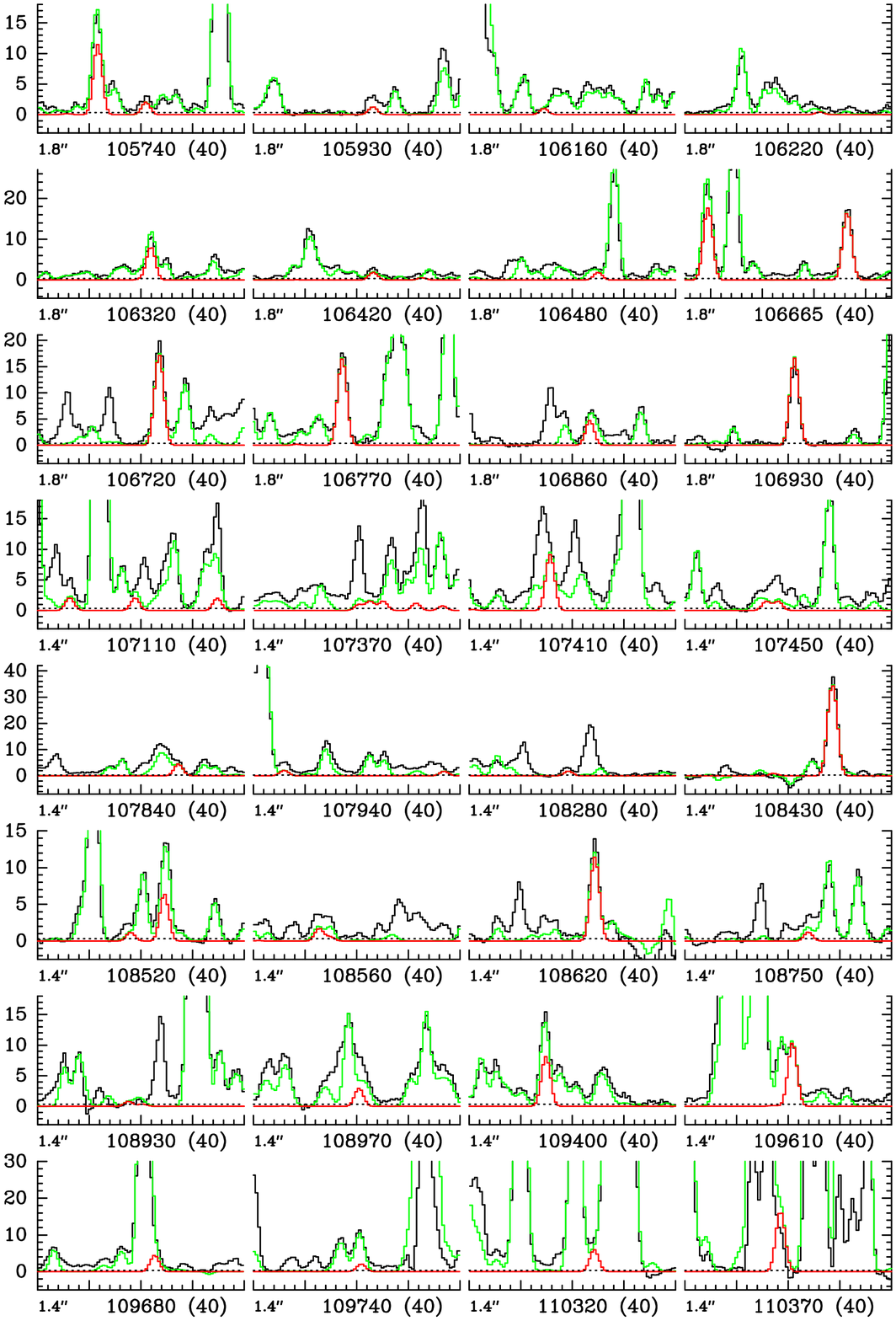}}}
\caption{continued.}
\end{figure*}
}
\addtocounter{figure}{-1}

\onlfig{
\begin{figure*}
\addtocounter{figure}{-1}
\centerline{\resizebox{0.9\hsize}{!}{\includegraphics[angle=0]{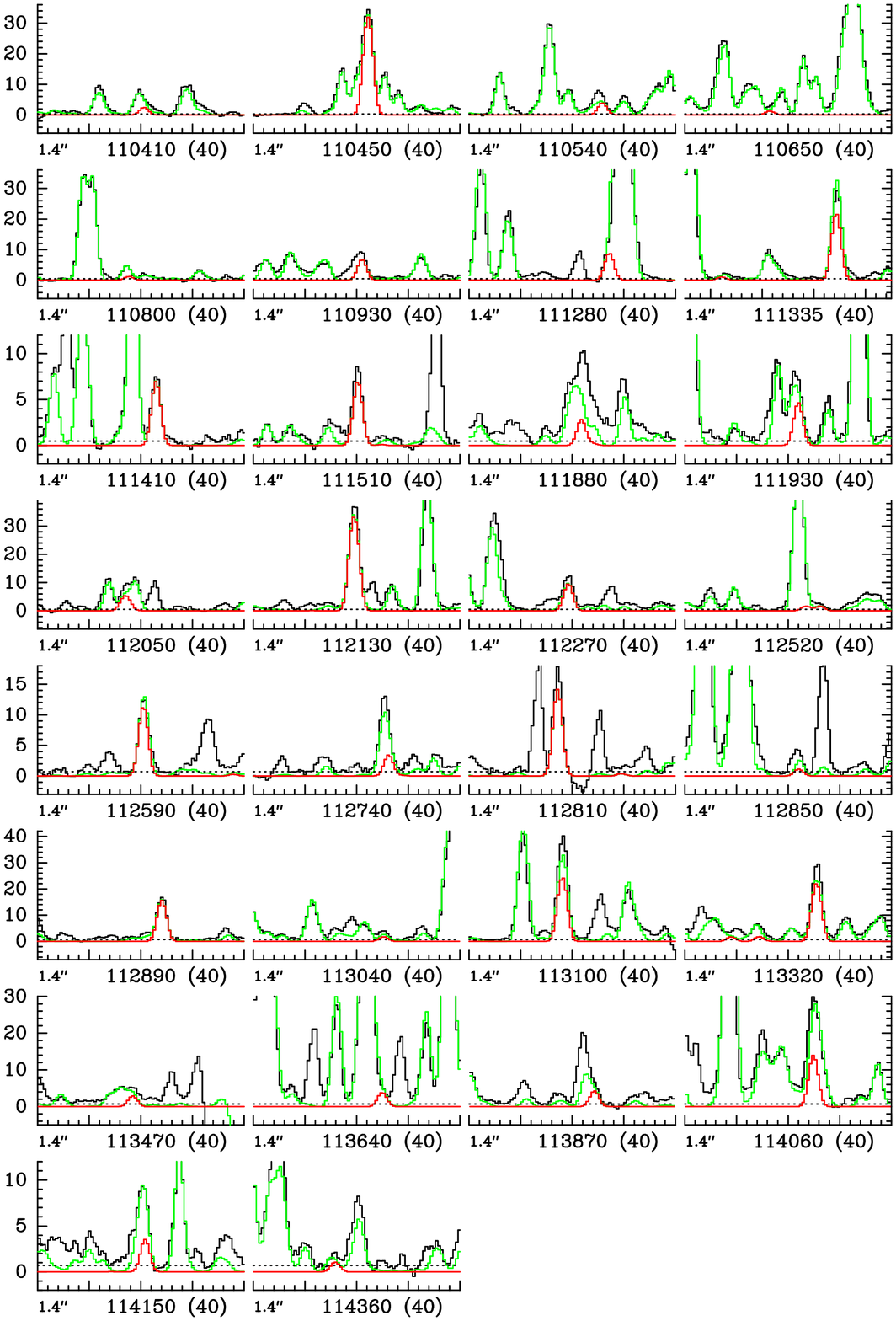}}}
\caption{continued.}
\end{figure*}
}
\addtocounter{figure}{-1}

\onlfig{
\clearpage
\begin{figure}
\centerline{\resizebox{1.0\hsize}{!}{\includegraphics[angle=0]{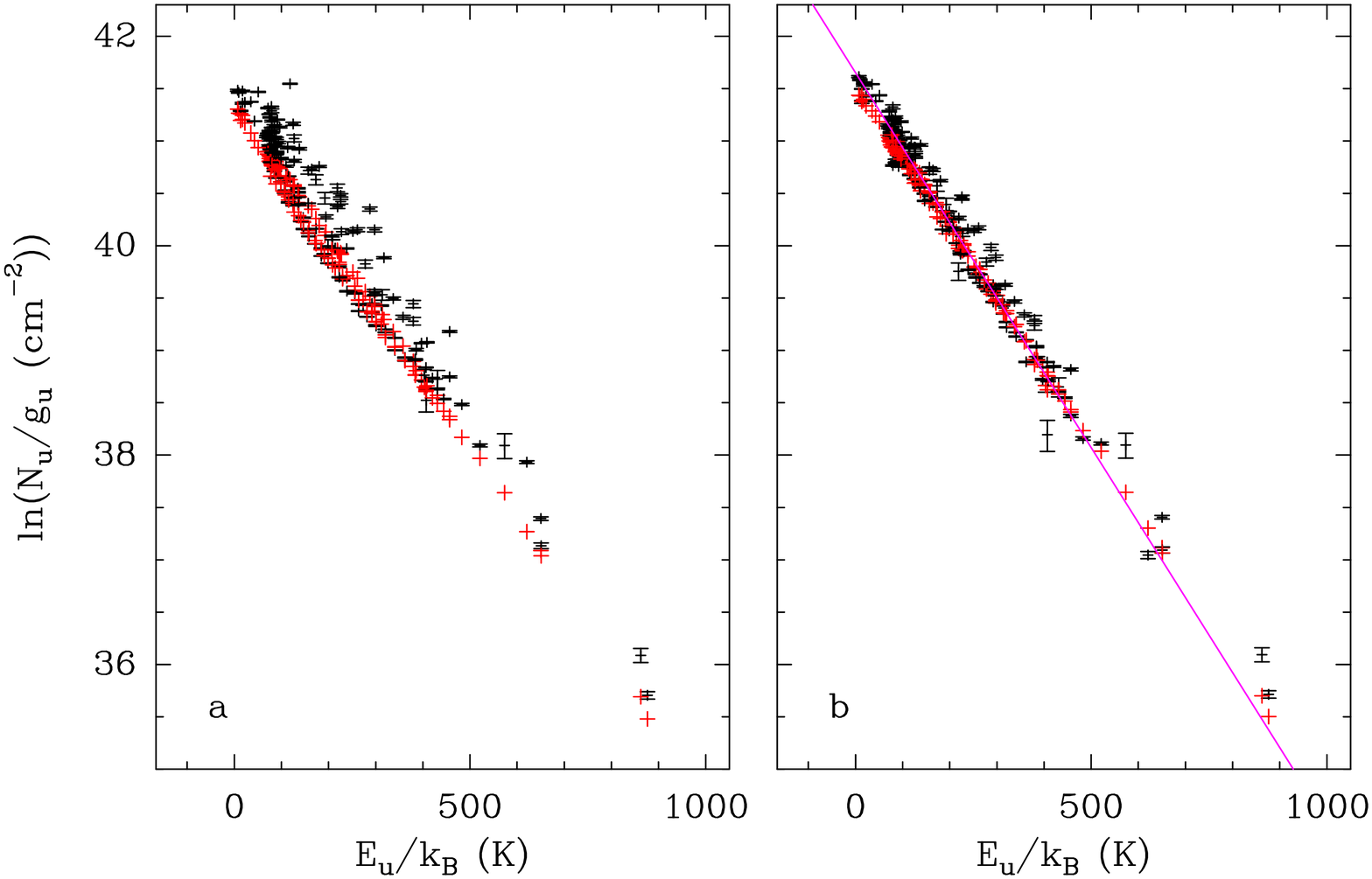}}}
\caption{Same as Fig.~\ref{f:popdiag_ch3oh} for C$_2$H$_5$OH, $\varv=0$.}
\label{f:popdiag_c2h5oh}
\end{figure}
}

Both $^{13}$C isotopologs of ethanol are detected in their \textit{anti} form with 
a few clearly detected lines each (Figs.~\ref{f:spec_c2h5oh_13c1_ve0} and 
\ref{f:spec_c2h5oh_13c2_ve0}). We derive a $^{12}$C/$^{13}$C isotopic ratio of $\sim$25 
for ethanol, consistent with the one derived for methanol (see Sect.~\ref{ss:ch3oh}). 

\onlfig{
\clearpage
\begin{figure*}
\centerline{\resizebox{0.9\hsize}{!}{\includegraphics[angle=0]{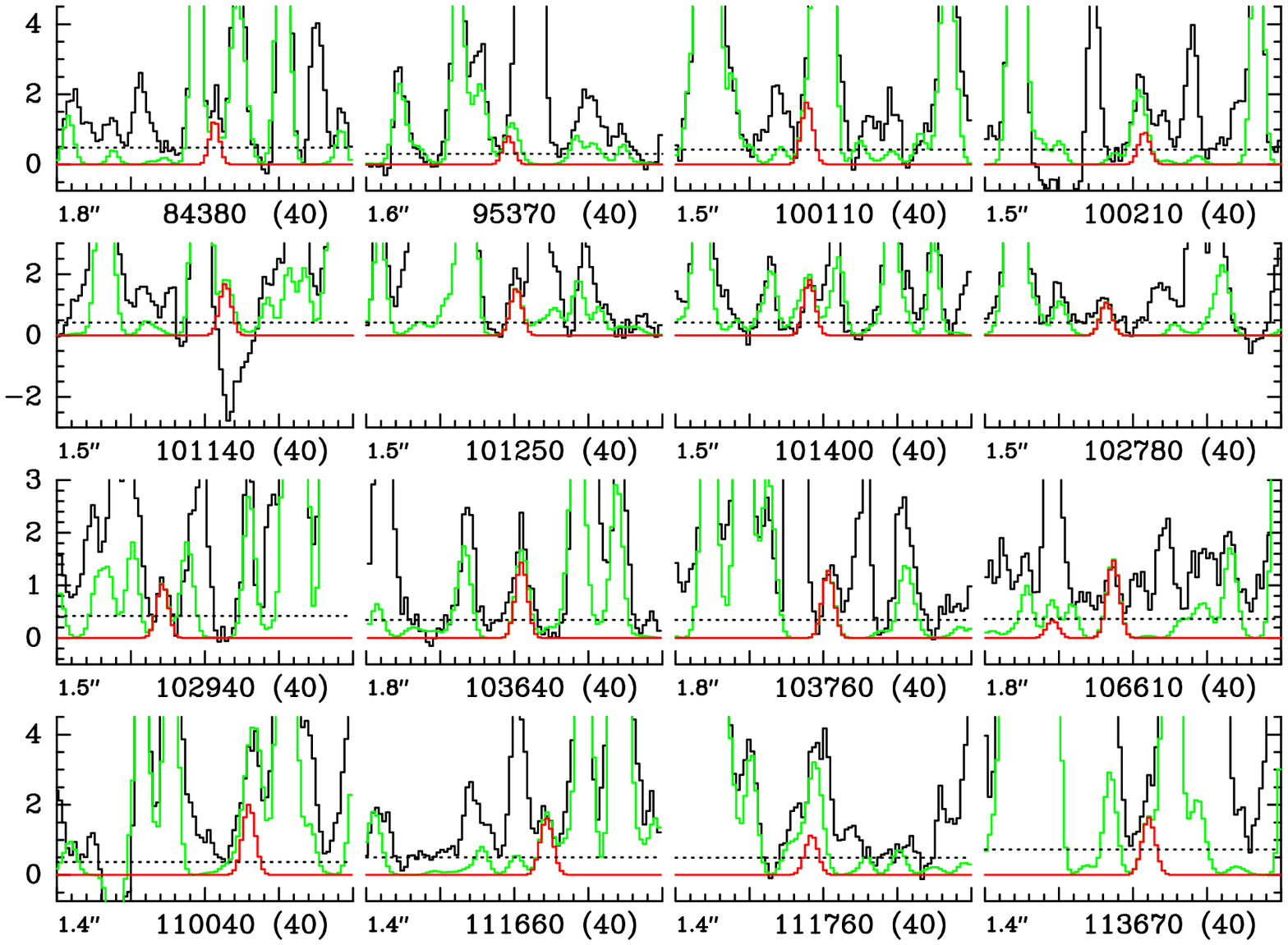}}}
\caption{Same as Fig.~\ref{f:spec_ch3oh_ve0} for $^{13}$CH$_3$CH$_2$OH, $\varv=0$.
}
\label{f:spec_c2h5oh_13c1_ve0}
\end{figure*}
}

\onlfig{
\begin{figure*}
\centerline{\resizebox{0.9\hsize}{!}{\includegraphics[angle=0]{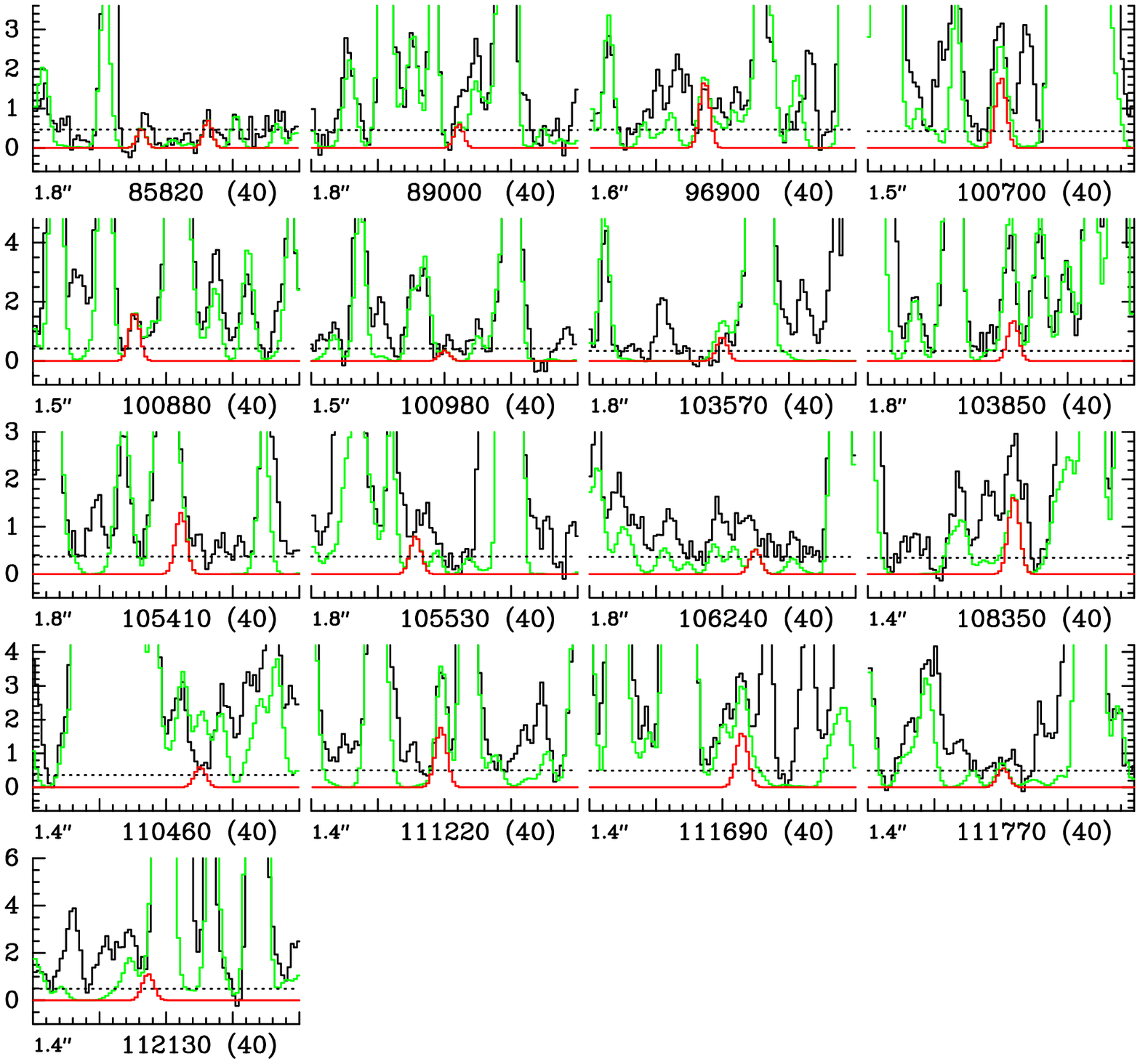}}}
\caption{Same as Fig.~\ref{f:spec_ch3oh_ve0} for CH$_3$$^{13}$CH$_2$OH, $\varv=0$.
}
\label{f:spec_c2h5oh_13c2_ve0}
\end{figure*}
}

\subsection{Propanol \textit{normal-}C$_3$H$_7$OH and \textit{iso-}C$_3$H$_7$OH}
\label{ss:c3h7oh}

Propanol is not detected toward Sgr~B2(N2) either in its straight chain form 
(\textit{normal}) or in its branched form (\textit{iso}). Assuming the same 
parameters as for ethanol, we derive upper limits to the column density of 
both forms (Table~\ref{t:coldens}).

\subsection{Methanethiol CH$_3$SH}
\label{ss:ch3sh}

With about 12 lines clearly detected in its vibrational ground state, methanethiol is 
securely identified toward Sgr~B2(N2) (Fig.~\ref{f:spec_ch3sh_ve0}). We also identified 
torsionally excited transitions ($\varv _{\rm t} = 1$) belonging to the $a$-type $J = 4 - 3$ 
branch, which are shown in Fig.~\ref{f:spec_ch3sh_ve1}. Even though they are slightly or 
heavily blended, two almost completely blended transitions, $k = +1$ of $E$ and 
$4_{2,3} - 3_{2,2}$ of $A$ torsional symmetry near 100806~MHz, account for essentially 
all of that emission feature. In addition, $k = +1$ near 100838~MHz and $4_{1,3} - 3_{1,2}$ 
of $A$ torsional symmetry near 101522~MHz account for emission features that are part 
of a blend. The $\varv _{\rm t} = 1$ state is not unambiguously detected; however, since 
the synthetic spectrum computed with the same parameters as for the ground state 
is fully consistent with the observed spectrum, we consider this state as tentatively 
detected and we include it in the full model (Fig.~\ref{f:spec_ch3sh_ve1}). 
The fits to the integrated intensity maps of the detected ground state lines yield 
a source size of $\sim$1.4$''$. The fit to the population diagram including both states 
is not well constrained ($T_{\rm rot} = 210 \pm 50$~K, Table~\ref{t:popfit} and 
Fig.~\ref{f:popdiag_ch3sh}). This fit is dominated by the more numerous lines in 
the vibrational ground state and tends to overestimate the intensities of the lines 
in the first vibrationally excited state. To avoid this overestimate when modeling 
the spectrum, we use a lower temperature of 180~K, still consistent with the 
fit result above. With this source size and temperature, the emission is optically 
thin ($\tau_{\rm max} = 0.3$). The apparent discrepancy between the synthetic and 
observed spectra at 108680 MHz is likely due to $^{13}$CN absorption produced by 
diffuse clouds along the line of sight, which we have not completely modeled so far. 
The small discrepancy at 110342~MHz may be due to an overestimate of the level of 
the baseline. This frequency corresponds to the data point significantly below the 
model at $E_{\rm u}/k_{\rm b} \approx 100$~K (in red) in the population diagram. 

\onlfig{
\clearpage
\begin{figure*}
\centerline{\resizebox{0.9\hsize}{!}{\includegraphics[angle=0]{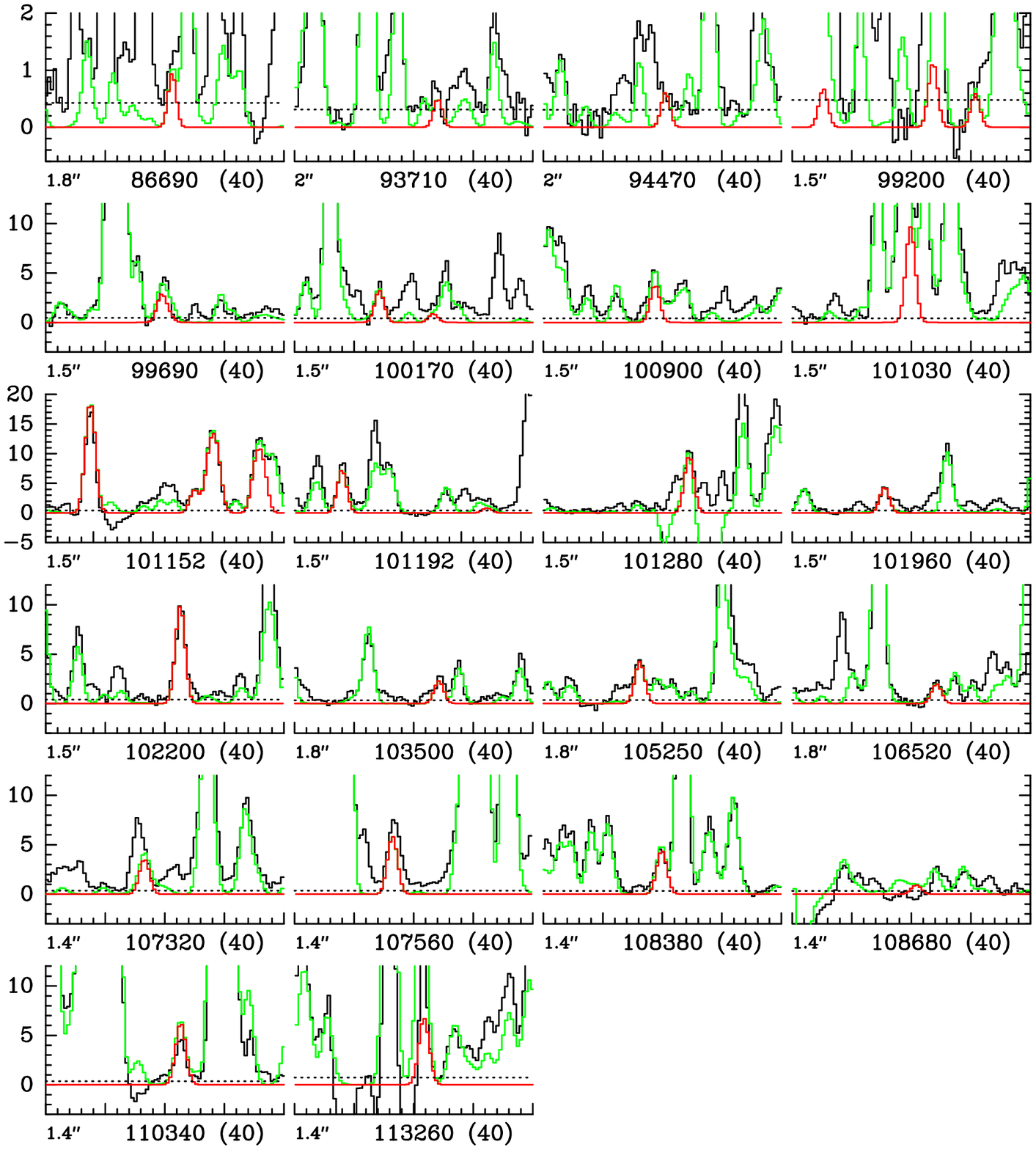}}}
\caption{Same as Fig.~\ref{f:spec_ch3oh_ve0} for CH$_3$SH, $\varv=0$.
}
\label{f:spec_ch3sh_ve0}
\end{figure*}
}

\onlfig{
\begin{figure*}
\centerline{\resizebox{0.9\hsize}{!}{\includegraphics[angle=0]{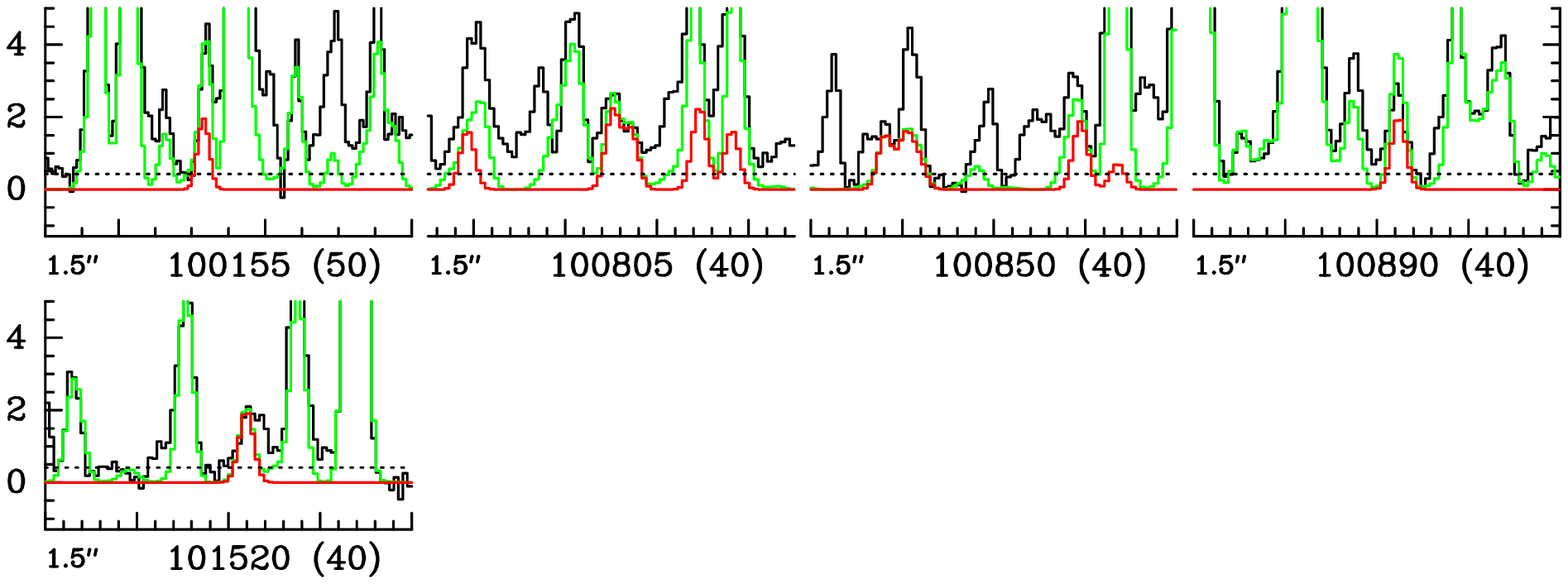}}}
\caption{Same as Fig.~\ref{f:spec_ch3oh_ve0} for CH$_3$SH, $\varv_{\rm t}=1$.
}
\label{f:spec_ch3sh_ve1}
\end{figure*}
}

\onlfig{
\begin{figure}
\centerline{\resizebox{1.0\hsize}{!}{\includegraphics[angle=0]{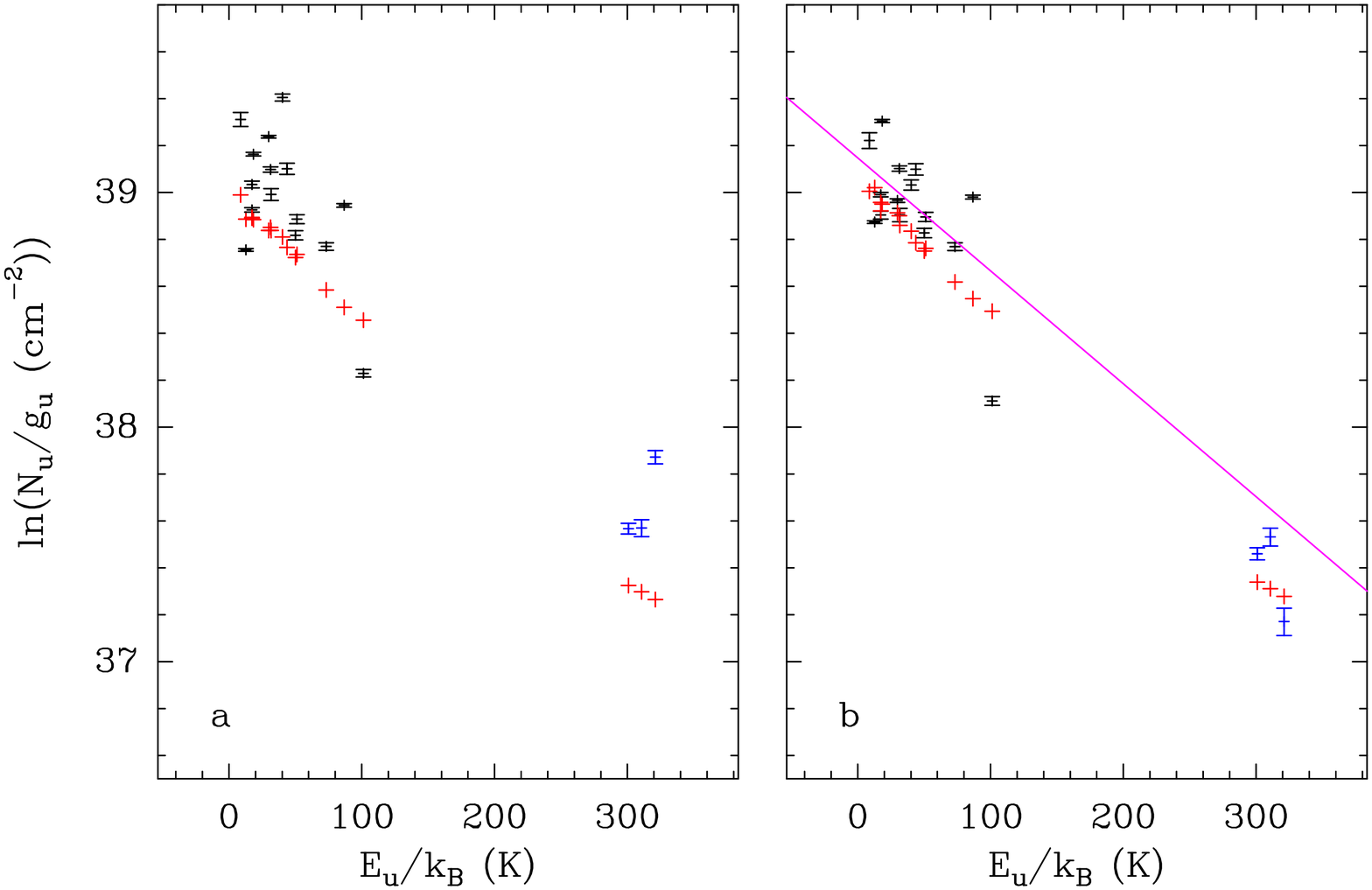}}}
\caption{Same as Fig.~\ref{f:popdiag_ch3oh} for CH$_3$SH, $\varv=0$ and
$\varv_{\rm t}=1$.}
\label{f:popdiag_ch3sh}
\end{figure}
}

\subsection{Ethanethiol C$_2$H$_5$SH}
\label{ss:c2h5sh}

Ethanethiol is not unambiguously detected toward Sgr~B2(N2). Under the assumption of 
the same temperature, source size, linewidth, and velocity offset as for methanethiol 
(Sect.~\ref{ss:ch3sh}), the synthetic spectrum shown in Fig.~\ref{f:spec_c2h5sh_ve0} is 
consistent with the observed spectrum, especially for the $11_4-10_4$ multiplet near 
111647~MHz, but the lack of clearly detected lines prevents a secure identification. 
We consider this model as a $3\sigma$ upper limit to the column density of ethanethiol 
toward Sgr~B2(N2) (Table~\ref{t:coldens}). 

\onlfig{
\clearpage
\begin{figure}
\centerline{\resizebox{0.9\hsize}{!}{\includegraphics[angle=0]{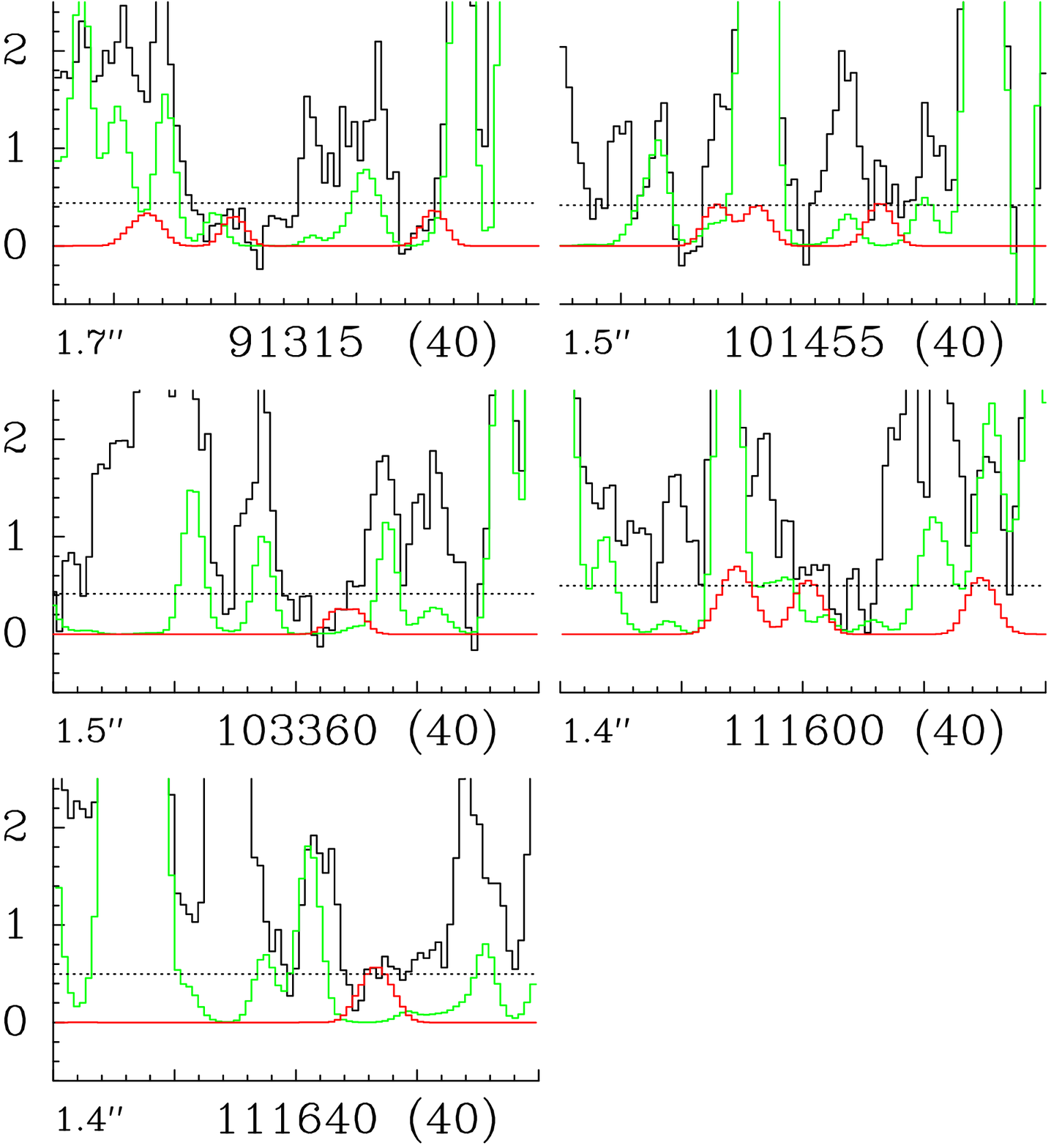}}}
\caption{Same as Fig.~\ref{f:spec_ch3oh_ve0} for \textit{gauche}-C$_2$H$_5$SH, $\varv=0$, 
but here, the green synthetic spectrum does not contain the contribution of  
\textit{gauche}-C$_2$H$_5$SH that is shown in red.
}
\label{f:spec_c2h5sh_ve0}
\end{figure}
}

\section{Astrochemical modeling}
\label{astro-chem}

In order to investigate the chemical formation and destruction mechanisms associated with 
CH$_3$SH and C$_2$H$_5$SH, we incorporate a new set of reactions into the network presented 
by \citet{i-PrCN_det_2014}. This expanded network is used to simulate the coupled gas-phase 
and grain-surface chemistry occurring in Sgr~B2(N2), using the kinetics model MAGICKAL 
\citep{3steps2heaven_2013}.

Current astrochemical networks include relatively few sulfur-bearing molecules; with reference 
to the production of alkanethiols, none of them (to the authors' knowledge) includes any 
treatment for the formation of sulfur-bearing molecules larger than H$_2$CS. Here, in analogy 
with the dominant formation routes for methanol and ethanol in hot cores, as predicted by the 
chemical models, the new network concentrates on the production of alkanethiols solely 
on the surfaces of the dust grains. However, gas-phase \textit{destruction} mechanisms for 
both molecules (as well as related intermediates) are included in the new network, the majority 
of which are ion-molecule processes or the subsequent dissociative recombination with electrons 
of the resultant molecular ions. Ion-molecule reactions are included for the major ionic species 
He$^+$, H$^+$, H$_3$$^+$, H$_3$O$^+$ and HCO$^+$. Estimates for the rates of photo-dissociation 
of new molecules, as caused by cosmic ray-induced and (where extinction allows) external UV photons, 
are also included (see \citet{modeling-warm-up_2008} and \citet{3steps2heaven_2013}).

For CH$_3$SH, we assume the same grain-surface desorption/binding energy as for methanol (5534~K), 
which is based on the laboratory study of \citet{MeOH-desorb_2004}. The weak hydrogen-bonding 
character of the thiol group is likely to result in a somewhat lower binding energy for methanethiol 
than for methanol in the amorphous water-dominated ices present on the dust grains; however, based 
on the greater observed rotational temperature for CH$_3$SH than for CH$_3$OH observed in our source 
(180~K versus 150~K), indicating a similar or greater desorption temperature for CH$_3$SH, the adoption 
of the methanol-based value appears justified in the absence of appropriate laboratory data. 
In the case of C$_2$H$_5$SH, a binding energy of 6230~K is used $-$ based on interpolation as detailed 
by \citet{3steps2heaven_2013} $-$ that is close to the value already adopted for ethanol (6259~K).

Methanethiol may form on grain surfaces through the addition of hydrogen atoms to H$_2$CS. 
Both gas-phase and grain-surface formation routes already exist in the network for CS, HCS, 
and H$_2$CS. The new network thus includes a chain of hydrogen addition reactions analogous 
to those for methanol, accounting for all hydrogenation states: CS, HCS, H$_2$CS, CH$_3$S/CH$_2$SH 
and CH$_3$SH. Methanethiol may also be formed via the addition of CH$_3$ and HS, and other species 
in the chain may also be formed via alternative atomic additions, e.g., CH$_2$ + S $\rightarrow$ 
H$_2$CS. Ethanethiol may be formed by a selection of radical-radical and hydrogen-addition 
reactions on the grains, including CH$_3$ + CH$_2$SH and C$_2$H$_5$ + HS.

The physical model follows that detailed in previous papers where a cold collapse phase to 
maximum density ($n_{\rm H} = 2 \times 10^{8}$~cm$^{-3}$) and minimum dust-grain temperature 
(8~K) is followed by a warm-up from 8$-$400~K; during this phase, the gas and dust temperatures 
are assumed to be well coupled.

The initial chemical compositions used in the model follow those of \citet{3steps2heaven_2013}. 
Crucially, a standard, depleted sulfur abundance of $8 \times 10^{-8}$ is used, which reproduces 
well the relative abundances of the sulfur-bearing molecules. The underlying reason for the 
anomalous depletion of atomic sulfur in dense regions (as compared to diffuse cloud values) is 
a long-standing problem (e.g., \citealt{S_in_dense-ISM_2013}), and will not be addressed here.

\subsection{Results of modeling}

\begin{figure}
\centering
\includegraphics[angle=0,width=8.8cm]{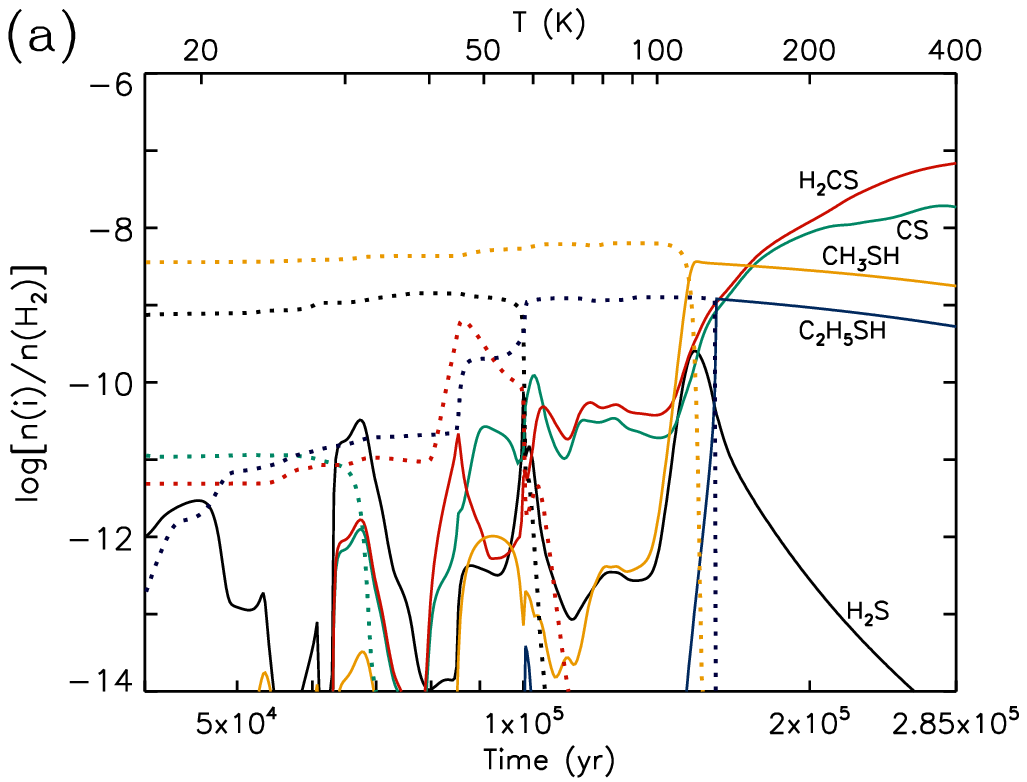}
\includegraphics[angle=0,width=8.8cm]{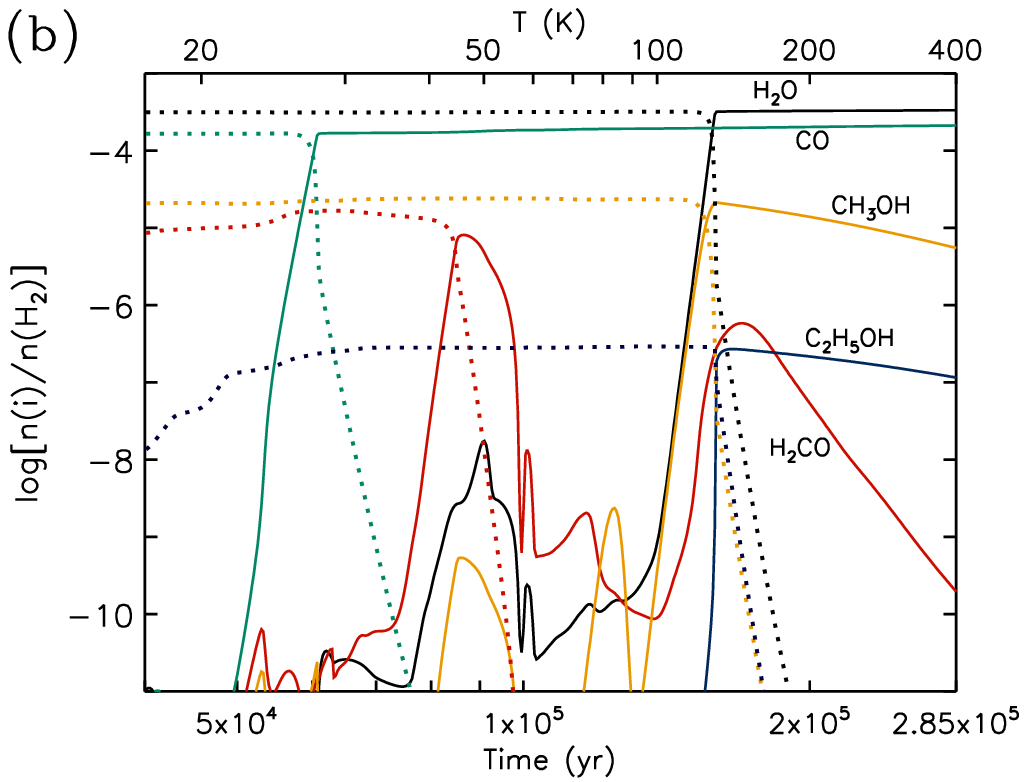}
  \caption{Chemical model abundances for a selection of sulfur-bearing molecules \textbf{(a)} 
   and their oxygen-bearing analogs \textbf{(b)}. Solid lines indicate gas-phase abundances; 
   dotted lines of the same color indicate the grain-surface/ice abundances of the same species.}
\label{model-fig}
\end{figure}

Figure~\ref{model-fig} shows the results of the models, for CH$_3$SH, C$_2$H$_5$SH, and 
a selection of related molecules (a), and their oxygen-substituted equivalents, each of which 
is a well-known hot-core molecule (b). The plots indicate the calculated time-dependent 
abundances of each molecule with respect to molecular hydrogen, during the warm-up phase 
of the chemistry. Solid lines indicate gas-phase abundances, while dotted lines of the same 
color indicate ice-mantle abundances for the same species.

Most of the CH$_3$SH is formed early, during the cold collapse phase, when the CS accreted 
from the gas-phase is efficiently converted to H$_2$CS via atomic hydrogen addition, and 
thence to CH$_3$SH. In common with the methanol-producing series, the addition of H to CS 
and H$_2$CS is expected in both cases to have an activation energy barrier. Previous networks 
have assumed a value of 1000~K for H + CS $\rightarrow$ HCS (e.g. \citealt{modeling_HHL_1992}), 
which is unaltered here. The hydrogenation of H$_2$CS is a newly added process in this network, 
to which a barrier of 1000~K is also assigned. This reaction is allowed three product branches, 
each of which are given equal weight: CH$_3$S, CH$_2$SH, and HCS + H$_2$ (corresponding to 
hydrogen abstraction). Most of the CS, that is accreted onto the grains during collapse, 
is ultimately converted into and stored as CH$_3$SH in the ices.

Desorption of CH$_3$SH from the dust grains becomes important above around 100~K; peak gas-phase 
abundance for this molecule occurs at a temperature of 119~K (see Table~\ref{chem-models-tab}). 
In spite of the identical binding energy adopted for methanethiol as for methanol, CH$_3$SH still 
reaches a gas-phase peak at a somewhat lower temperature than CH$_3$OH; the large total abundance 
of methanol on the grains causes its release to take longer and peak later than for methanethiol. 
The peak gas-phase abundance of CH$_3$SH is also lower than the peak value obtained on the grain 
surfaces; abstraction of hydrogen from methanethiol by surface radicals, including, most prominently, 
NH$_2$, causes destruction of this molecule just prior to its release into the gas phase, reducing 
the overall peak by a factor of $\sim$3. While similar H-abstraction processes also exist for methanol, 
the exothermicity of these reactions is low and results in larger activation energy barriers (and thus 
smaller rates). Barriers for abstraction reactions are based on experimental values where available, 
or calculated using the Evans-Polanyi relation (as detailed by \citealt{3steps2heaven_2013}); 
barriers involving methanethiol are obtained with this method.

Formation of C$_2$H$_5$SH is achieved through three main processes, occurring at $\sim$20$-$30, 
45, and 55~K. The first is the addition reaction CH$_3$ + CH$_2$SH $\rightarrow$ C$_2$H$_5$SH; 
this process is analogous to the dominant mechanism for ethanol formation in the model, and 
C$_2$H$_5$SH abundance in the ices shows similar behavior to that of ethanol in the 20$-$40~K 
temperature range.

At around 45~K, the addition of atomic H to the C$_2$H$_5$S radical becomes more important, 
contributing an abundance of a few 10$^{-10} \times n$(H$_2$). The C$_2$H$_5$S radical is 
itself formed by the addition of S atoms to C$_2$H$_5$, which is produced when acetylene in 
the bulk ice mantles becomes mobile, allowing it to be hydrogenated on the ice surface.

At 55~K, the main C$_2$H$_5$SH-producing event occurs, when H$_2$S becomes mobile, dredges up 
from the bulk ice mantle, and desorbs from the ice surface. While most H$_2$S is desorbed into 
the gas phase, some reacts with atomic H, which abstracts a hydrogen atom to give HS. 
This radical may then quickly react with C$_2$H$_5$ to produce ethanethiol. 
C$_2$H$_5$SH ultimately desorbs, reaching peak gas-phase abundance at a similar temperature 
to the alcohols. It should be noted that essentially all H$_2$S in the ices is converted 
to C$_2$H$_5$SH through this mechanism, as can be clearly seen in Fig.~\ref{model-fig}(a). 
However, there is no analogous conversion of H$_2$S into CH$_3$SH, as the CH$_3$ radical 
is no longer present in significant abundance when H$_2$S and HS become mobile on the grains. 
Neither does the addition of the OH radical to CH$_3$/C$_2$H$_5$ provide any significant 
formation route for methanol/ethanol, analogous to the HS-related mechanisms. While the 
photodissociation of water ice provides abundant OH, its mobility is relatively low 
at these temperatures.

By the end of the model run, at 400~K, CS and H$_2$CS comprise more than 50\,\% of the sulfur budget; 
the remainder resides mostly in SO and SO$_2$ (not shown).

\subsection{Discussion of modeling}

For the alcohols, the CH$_3$OH/C$_2$H$_5$OH ratio of 78 produced by the models, 
comparing peak abundances, is a reasonable match to the observed value of $\sim$20. 
In spite of the approximate nature of the overall sulfur abundance in the models due to 
the uncertainty in initial atomic sulfur budget, the modeled C$_2$H$_5$OH/C$_2$H$_5$SH 
ratio of 225 is in line with the observed lower limit of 125.

Peak abundance for CH$_3$SH produced by the model is somewhat lower than the observations demand. 
A modeled ratio of CH$_3$SH/C$_2$H$_5$SH = 3.1 is obtained, compared with the observed value 
of $>21$. Similarly, the ratio of methanol to methanethiol is around 50 times larger 
($\sim$5700 versus 120) than the observed ratio, suggesting that the low CH$_3$SH/C$_2$H$_5$SH ratio 
is caused more by an underprediction of CH$_3$SH relative to the other three molecules rather 
than by an overprediction of C$_2$H$_5$SH. However, it should be noted that CH$_3$SH is released 
from the grains at lower temperatures than either methanol or ethanethiol in the models. 
This would tend to produce a somewhat larger spatial extent for methanethiol under a three-dimensional 
treatment of the core, and would likely result in a greater average column density than is indicated 
when comparing peak abundances for each molecule. The differing sublimation characteristics 
of CH$_3$SH may therefore improve its column density ratio with other molecules. 
Direct laboratory determination of the methanethiol binding energy on a water/methanol surface 
would give greater confidence to the comparison with observed column density values.

The low initial abundance of sulfur assumed in this model allows room for the methanethiol abundance 
to be increased through this variable; however, alternative model runs indicate that ethanethiol 
abundances also scale in this way. Lower efficiency in the early-stage formation of H$_2$S on the 
grains could reduce the C$_2$H$_5$SH abundance, as could lower acetylene production. 
Acetylene abundances ultimately rely on the quantity of ice-mantle methane formed during the cold 
collapse stage, which is released into the gas phase at temperatures around 25~K, producing a range 
of hydrocarbons.


\begin{table}
\begin{center}
\caption{Results of chemical kinetics modeling of alkanethiols and alkanols toward Sgr~B2(N2).}
\label{chem-models-tab}
\renewcommand{\arraystretch}{1.10}
\begin{tabular}[t]{lcc}
\hline \hline
Species      &   Peak $n$[i]/$n$(H$_{2})$   & Temperature (K) at peak \\
\hline
CH$_3$SH              & $3.7 \times 10^{-9}$  & 119  \\
C$_2$H$_5$SH          & $1.2 \times 10^{-9}$  & 130  \\
CH$_3$OH              & $2.1 \times 10^{-5}$  & 130  \\
C$_2$H$_5$OH          & $2.7 \times 10^{-7}$  & 139  \\
\hline
\end{tabular}
\end{center}
\end{table}

\section{Discussion}
\label{discussion}

\begin{table}
\begin{center}
\caption{Column density ratios of alkanols and alkanethiols in different sources.}
\label{molecule-ratios}
\renewcommand{\arraystretch}{1.10}
\begin{tabular}[t]{lccccccc}
\hline \hline
                     & MeOH & & MeSH         & & MeOH & & EtOH          \\
                     \cline{2-2} \cline{4-4}   \cline{6-6} \cline{8-8} 
Source               & EtOH & & EtSH         & & MeSH & & EtSH          \\
\hline
Sgr~B2(N2)           &  20  & & $\gtrsim 21$ & & 118  & & $\gtrsim 125$ \\
Orion~KL$^{a,b}$     &  31  & &        5$^c$ & & 120  & &        20$^c$ \\
G327.3$-$0.6$^{a,d}$ &  82  & &              & &  58  & &               \\
\hline
\end{tabular}
\end{center}
\tablefoot{
$^a$ Methanol (MeOH) inferred from $^{13}$CH$_3$OH. EtOH, MeSH, and EtSH short for ethanol, 
     methanethiol, and ethanethiol, respectively. 
$^b$ Orion~KL data from \citet{EtSH_lab_Orion-KL_2014}. 
$^c$ Only for \textit{gauche}-EtSH; ratio would be smaller by about a factor of 1.25 with 
     \textit{anti}-EtSH assuming LTE. 
$^d$ G327.3$-$0.6 data from \citet{CH3SH_etc_G327_2000}. 
}
\end{table}


In light of the extensive number of emission features detected for CH$_3$OH, C$_2$H$_5$OH, 
and CH$_3$SH in our sensitive ALMA data toward Sgr~B2(N2), it is noteworthy that we detect 
only a few, rather weak, and partially or completely blended lines that can be attributed 
to C$_2$H$_5$SH. Consequently, we are unable to claim even a tentative detection of the 
molecule. We consider the low upper limit to the column density of ethanethiol 
(Table~\ref{t:coldens}) to be meaningful. In fact, comparing the Sgr~B2(N2) column density 
ratios of alkanols and alkanethiols (or their lower limits) in Table~\ref{molecule-ratios}, 
it is conceivable that the column density of C$_2$H$_5$SH is close to the upper limit. 
With so many potential ethanethiol lines partially or strongly blended in our ALMA data, 
additional observations at longer wavelengths may be more promising than more sensitive 
observations at 3~mm or observations at shorter wavelengths.

The non-detection of ethanethiol in Sgr~B2(N2) is rather surprising in the context of 
the purported detection of the molecule in Orion~KL at a column density that yields 
CH$_3$SH/C$_2$H$_5$SH and C$_2$H$_5$OH/C$_2$H$_5$SH ratios that are much lower than our 
lower limits. The ratios (Table~\ref{molecule-ratios}) are 5 and 20, respectively, ignoring 
the presence of the \textit{anti}-conformer, or 4 and 16 if we assume a thermal (200~K) 
population of this conformer. Our numbers are $\gtrsim 21$ and $\gtrsim 125$, respectively. 
In contrast, the CH$_3$OH/CH$_3$SH ratios are identical, and the CH$_3$OH/C$_2$H$_5$OH
values are quite similar, 20 and 31, respectively. Inspection of the features reported in 
\citet{EtSH_lab_Orion-KL_2014} as detected shows one isolated emission feature 
(near 243.58~GHz) assigned to ethanethiol. A small number of fairly isolated lines 
were assigned in part to the molecule. The majority of the potential C$_2$H$_5$SH lines, 
however, are strongly blended. More importantly, the model of all species excluding 
C$_2$H$_5$SH leaves no room for any ethanethiol contribution in the isolated emission 
feature near 263.90~GHz. The high abundance of C$_2$H$_5$SH relative to CH$_3$SH and the 
alkanols in Orion~KL compared to Sgr~B2(N2), and the issues we pointed out concerning 
the identification of C$_2$H$_5$SH in the Orion~KL spectrum suggest that the presence 
of ethanethiol in Orion~KL is uncertain. Unfortunately, our model calculations are not 
accurate enough, in particular for CH$_3$SH, to support or dismiss the reported 
C$_2$H$_5$SH detection in Orion~KL.

The column density of CH$_3$SH is relatively high in G327.3$-$0.6, yielding a 
CH$_3$OH/CH$_3$SH ratio of 58 \citep{CH3SH_etc_G327_2000}, half the value in Sgr~B2(N2) 
and Orion~KL. On the other hand, the CH$_3$OH/C$_2$H$_5$OH ratio is 82, higher than in 
Orion~KL, and even higher than in Sgr~B2(N2). On the basis of column density ratios alone, 
it is difficult to judge how promising a search for ethanethiol in G327.3$-$0.6 would be. 
We point out that we did not use the quoted column density of CH$_3$OH from 
\citet{CH3SH_etc_G327_2000}. Instead, we used the $^{13}$CH$_3$OH column density and the 
quoted $^{12}$C/$^{13}$C ratio of 40; the reported CH$_3$OH/$^{13}$CH$_3$OH ratio was 
actually 10, suggesting substantial CH$_3$OH opacities.

In their 1.3~mm line survey toward three positions in Sgr~B2, 
\citet{SgrB2_Nummelin_1998,SgrB2_Nummelin_2000} detected CH$_3$OH, C$_2$H$_5$OH, and 
CH$_3$SH toward not only Sgr~B2(N), but also toward Sgr~B2(M). Judging from their 
$^{12}$C/$^{13}$C ratios for methanol, these column densities are affected by substantial 
opacities. Emission of C$_2$H$_5$OH and CH$_3$SH may be less prone to opacity biases. 
Their ratios are 3.8 and 1.4 for Sgr~B2(N), whose emission is likely dominated by Sgr~B2(N1), 
and Sgr~B2(M), respectively. Both values are smaller than 5.9 for Sgr~B2(N2), though not 
by much for Sgr~B2(N). The Orion~KL value is essentially identical to the one from 
Sgr~B2(N), whereas the G327.3$-$0.6 value of 0.7 is even more extreme than the one 
from Sgr~B2(M).

Our upper limit to the column density of \textit{normal}-propanol is not very constraining 
in comparison to our CH$_3$OH/C$_2$H$_5$OH ratio of 20, as it translates to a 
C$_2$H$_5$OH/$n$-C$_3$H$_7$OH ratio of $\gtrsim 7.7$. In the case of the Galactic Center 
molecular cloud MC G+0.693$-$0.027, the column density ratio between CH$_3$OH and C$_2$H$_5$OH 
is $\sim$14 \citep{cold-GC-clouds_GBT_2008}, lower than the Sgr~B2(N2) value, but the lower 
limit to the C$_2$H$_5$OH to $n$-C$_3$H$_7$OH ratio is only $\sim$4.5, which was derived 
from the CH$_3$OH to C$_2$H$_5$OH and the CH$_3$OH to $n$-C$_3$H$_7$OH ratios 
\citep{cold-GC-clouds_GBT_2008}. In their paper on the tentative detection of ethyl methyl 
ether, \citet{search_EME_n-PrOH_2015} mention a ground state column density ratio of $\gtrsim 60$. 
Taking column density corrections by vibrational states for both molecules and by other conformers 
for \textit{n}-propanol into account (see Table~\ref{t:coldens}) the column density ratio becomes 
$\gtrsim 14.3$, slightly more constraining than ours. However, should our 
C$_2$H$_5$CN/$n$-C$_3$H$_7$CN ratio of $\sim$55 (\citealt{deuterated_SgrB2N2_2015} and 
rescaled for the same source size of $1.2''$ in the case of $n$-C$_3$H$_7$CN, 
\citealt{i-PrCN_det_2014}) be more appropriate for the estimation of the column density of 
$n$-C$_3$H$_7$OH, then all available observational results are not so meaningful. 
Our upper limit to the column density of \textit{iso}-propanol translates to a 
C$_2$H$_5$OH/$i$-C$_3$H$_7$OH ratio of $\gtrsim 21.5$, which may be meaningful if the 
$i$-C$_3$H$_7$OH/$n$-C$_3$H$_7$OH ratio is considerably larger than the 
$i$-C$_3$H$_7$CN/$n$-C$_3$H$_7$CN ratio of $\sim$0.4 in Sgr~B2(N2) \citep{i-PrCN_det_2014}.

Our tentative detection of torsionally excited methanethiol is the first such account; 
previously only ground state transitions had been identified. On the other hand, 
observations of CH$_3$OH, $\varv_{\rm t} = 2$ and $^{13}$CH$_3$OH, $\varv_{\rm t} = 1$ 
had been reported before, e.g., in Sgr~B2(N) by \citet{SgrB2_Nummelin_1998} and in 
Orion~KL by \citet{Orion-KL_607-725_2001}.

We determined an isotopic ratio of $\sim$25 for CH$_3$OH/$^{13}$CH$_3$OH and for the ratio 
of parent ethanol to both of its isotopologs with one $^{13}$C. This value is slightly larger 
than our ratio determined for the main isotopic species of vinyl cyanide to all three of its 
singly $^{13}$C substituted species of $21 \pm 1$ \citep{13C-VyCN_2008}. These ratios are 
in good to reasonable agreement with a value of $\sim$20 in Galactic Center sources  
\citep{isotopic_abundances_1994,12C-13C_gradient_2005} or the range of $\sim$20 to 25 
derived by \citet{H2CO_12-13_GC_1985}.

Our CH$_3$OH/CH$_3^{18}$OH and $^{13}$CH$_3$OH/CH$_3^{18}$OH ratios of $\sim$180 and $\sim$7.3 
are in very good agreement with respective Sgr~B2 values of $210 \pm 40$ and $7.5 \pm 1.0$ 
\citep{CH3O-18-H_det_1989} or with $\sim$200 and $8.6 \pm 1.2$ for respective H$_2$CO 
values in Sgr~B2 and the Galactic Center source M$-$0.13$-$0.18 \citep{H2CO_12-13_GC_1985}. 
\citet{isotopic_abundances_1994} summarized isotopic ratios of $\sim$250 and $\sim$12.5 in 
the Galactic Center ISM, in reasonable agreement with our values.

The isotopic ratios in the Galactic Center ISM are rather different from the solar system values 
of about 89 and 490 for the $^{12}$C/$^{13}$C and $^{16}$O/$^{18}$O ratios, respectively. 
These changes reflect gradients in the isotopic ratios from the center of the Milky Way to 
its outer regions which, in addition, change with time. This aspect explains small, but 
non-negligible differences between the solar system values and those determined in the 
local ISM (e.g., \citealt{H2CO_12-13_GC_1985,isotopic_abundances_1994,12C-13C_gradient_2005}).

\section{Conclusion}
\label{conclusion}

Our sensitive ALMA data permitted clear detections of methanethiol with the first, albeit 
tentative, detection of lines in its first excited torsional state, ethanol, with the first 
unambiguous detection of the singly $^{13}$C substituted isotopomers, and three isotopologs 
of methanol; the main species was detected in several torsional states, including some 
evidence of its third excited torsional state. However, they were not sensitive enough 
for a clear detection of ethanethiol, though a comparison to CH$_3$SH and alkanols suggests 
that its column density could be close to the upper limit derived here. We also obtained upper 
limits to the column densities of \textit{normal}- and \textit{iso}-propanol. Our values for 
the $^{12}$C/$^{13}$C ratio obtained for methanol and ethanol and for the $^{16}$O/$^{18}$O 
ratio obtained for methanol are in line with the values known for the ISM in the Galactic 
Center region.

The column density ratios involving methanol, ethanol, and methanethiol in Sgr~B2(N2) are 
similar to values reported for Orion~KL, but those involving ethanethiol are significantly 
different and suggest that the detection of ethanethiol reported toward Orion~KL is rather 
uncertain.

The chemical models presented here include new reactions for the formation and destruction 
of methanethiol and ethanethiol. This represents a step forward in the complexity of sulfur 
chemistry that can be treated with astrochemical models. The results indicate that adequate 
abundances of both alkanethiols may be formed in hot cores through dust-grain surface/ice-mantle 
chemistry. Methanethiol is formed at the low temperatures achieved during cold core collapse 
(on the order 10~K). Ethanethiol is produced on the grains at elevated temperatures 
($\sim$45$-$55~K), with most of the surface H$_2$S at those temperatures being converted 
almost entirely to ethanethiol. The modeled gas-phase ratio CH$_3$SH/C$_2$H$_5$SH $\simeq$~3 
is substantially lower than the observed value of $>21$. The strongest influence on this 
ratio is likely the underprediction of CH$_3$SH (in comparison to CH$_3$OH), although the 
efficient formation of C$_2$H$_5$SH from H$_2$S on the grains is also a source of uncertainty.

In addition, more sensitive observations of Sgr~B2(N), Orion~KL, or other sources with ALMA 
or other interferometers are required to establish whether and possibly in what quantities 
ethanethiol, \textit{normal}-propanol, and \textit{iso}-propanol are present in the ISM. 
Such observations may be more promising at wavelengths longer than 3~mm for the first two molecules.


\begin{acknowledgements}
This paper makes use of the following ALMA data: 
ADS/JAO.ALMA\#2011.0.00017.S, ADS/JAO.ALMA\#2012.1.00012.S. 
ALMA is a partnership of ESO (representing its member states), NSF (USA) and NINS (Japan), 
together with NRC (Canada), NSC and ASIAA (Taiwan), and KASI (Republic of Korea), 
in cooperation with the Republic of Chile. The Joint ALMA Observatory is operated by ESO, 
AUI/NRAO and NAOJ. The interferometric data are available in the ALMA archive at
https://almascience.eso.org/aq/. 
This work has been supported by the Deutsche Forschungsgemeinschaft (DFG) through the 
collaborative research grant SFB 956 ``Conditions and Impact of Star Formation'', 
project area B3. We thank Dr. Monika Koerber for making an ethanol spectrum available 
for inspection of the relative intensities. L.H.X. and R.M.L. acknowledge financial 
support from the Natural Sciences and Engineering Research Council (NSERC) of Canada. 
R.T.G. acknowledges support from the NASA Astrophysics Theory Program through grant 
NNX11AC38G. Our research benefited from NASA's Astrophysics Data System (ADS).
\end{acknowledgements}



\end{document}

%% file: abb/tab_r-sh_popfit.tex
\begin{table}
 {\centering
 \caption{
 Rotational temperatures derived from population diagrams of alkanols and alkanethiols toward Sgr~B2(N2).
}
 \label{t:popfit}
 \vspace*{0.0ex}
 \begin{tabular}{lll}
 \hline\hline
 \multicolumn{1}{c}{Molecule} & \multicolumn{1}{c}{States\tablefootmark{a}} & \multicolumn{1}{c}{$T_{\rm fit}$\tablefootmark{b}} \\ 
  & & \multicolumn{1}{c}{\scriptsize (K)} \\ 
 \hline
CH$_3$OH & $\varv=0$, $\varv_{\rm t}=1$, $\varv_{\rm t}=2$ & 150.8 (1.7) \\ 
$^{13}$CH$_3$OH & $\varv=0$, $\varv_{\rm t}=1$ & 160.6 (6.8) \\ 
CH$_3$$^{18}$OH & $\varv=0$ &   143 (12) \\ 
\hline 
C$_2$H$_5$OH & $\varv=0$ & 139.6 (1.6) \\ 
\hline 
CH$_3$SH & $\varv=0$, $\varv_{\rm t}=1$ &   208 (46) \\ 
\hline 
 \end{tabular}
 }\\[1ex] 
 \tablefoot{
 \tablefoottext{a}{Vibrational states that were taken into account to fit the population diagram.}
 \tablefoottext{b}{The standard deviation of the fit is given in parentheses. As explained in 
    \citet{deuterated_SgrB2N2_2015}, these uncertainties should be viewed with caution. They may be underestimated.}
 }
 \end{table}

%% file: abb/tab_r-sh_weedsmodel.tex
\begin{table*}[!ht]
 {\centering
 \caption{
 Parameters of our best-fit LTE model (or upper limit) of alkanols and alkanethiols toward Sgr~B2(N2).
}
 \label{t:coldens}
 \vspace*{0.0ex}
 \begin{tabular}{lcrccccccr}
 \hline\hline
 \multicolumn{1}{c}{Molecule} & \multicolumn{1}{c}{Status\tablefootmark{a}} & \multicolumn{1}{c}{$N_{\rm det}$\tablefootmark{b}} & \multicolumn{1}{c}{Size\tablefootmark{c}} & \multicolumn{1}{c}{$T_{\mathrm{rot}}$\tablefootmark{d}} & \multicolumn{1}{c}{$N$\tablefootmark{e}} & \multicolumn{1}{c}{$C$\tablefootmark{f}} & \multicolumn{1}{c}{$\Delta V$\tablefootmark{g}} & \multicolumn{1}{c}{$V_{\mathrm{off}}$\tablefootmark{h}} & \multicolumn{1}{c}{$\frac{N_{\rm ref}}{N}$\tablefootmark{i}} \\ 
  & & & \multicolumn{1}{c}{\scriptsize ($''$)} & \multicolumn{1}{c}{\scriptsize (K)} & \multicolumn{1}{c}{\scriptsize (cm$^{-2}$)} & & \multicolumn{1}{c}{\scriptsize (km~s$^{-1}$)} & \multicolumn{1}{c}{\scriptsize (km~s$^{-1}$)} & \\ 
 \hline
 CH$_3$OH, $\varv=0$$^\star$ & d & 41 & 1.4 &  160 &  4.0 (19) & 1.00 & 5.4 & $-$0.5 &       1 \\ 
 \hspace*{8ex} $\varv_{\rm t}=1$ & d & 16 & 1.4 &  160 &  4.0 (19) & 1.00 & 5.4 & $-$0.2 &       1 \\ 
 \hspace*{8ex} $\varv_{\rm t}=2$ & d & 3 & 1.4 &  160 &  4.0 (19) & 1.00 & 5.4 & $-$0.5 &       1 \\ 
 \hspace*{8ex} $\varv_{\rm t}=3$ & t & 0 & 1.4 &  160 &  4.0 (19) & 1.00 & 5.4 & $-$0.5 &       1 \\ 
 $^{13}$CH$_3$OH, $\varv=0$ & d & 19 & 1.4 &  160 &  1.6 (18) & 1.00 & 5.4 & $-$0.2 &      25 \\ 
 \hspace*{9.5ex} $\varv_{\rm t}=1$ & d & 7 & 1.4 &  160 &  1.6 (18) & 1.00 & 5.4 & $-$0.2 &      25 \\ 
 CH$_3$$^{18}$OH, $\varv=0$ & d & 8 & 1.4 &  160 &  2.2 (17) & 1.00 & 5.4 & $-$0.0 &     182 \\ 
 \hspace*{9.5ex} $\varv_{\rm t}=1$ & t & 0 & 1.4 &  160 &  2.2 (17) & 1.00 & 5.4 & $-$0.0 &     182 \\ 
\hline 
 C$_2$H$_5$OH$^\star$ & d & 168 & 1.3 &  150 &  2.0 (18) & 1.24 & 5.4 & 0.0 &       1 \\ 
 \textit{anti}-$^{13}$CH$_3$CH$_2$OH & d & 4 & 1.3 &  150 &  8.0 (16) & 2.96 & 5.4 & 0.0 &      25 \\ 
 \textit{anti}-CH$_3$$^{13}$CH$_2$OH & d & 3 & 1.3 &  150 &  8.0 (16) & 2.96 & 5.4 & 0.0 &      25 \\ 
\hline 
 \textit{Ga-normal}-C$_3$H$_7$OH & n & 0 & 1.3 &  150 & $<$  2.6 (17) & 5.20 & 5.4 & 0.0 & $-$ \\ 
\hline 
 \textit{iso}-C$_3$H$_7$OH & n & 0 & 1.3 &  150 & $<$  9.3 (16) & 1.86 & 5.4 & 0.0 & $-$ \\ 
\hline 
 CH$_3$SH, $\varv=0$$^\star$ & d & 12 & 1.4 &  180 &  3.4 (17) & 1.00 & 5.4 & $-$0.5 &       1 \\ 
 \hspace*{7.3ex} $\varv_{\rm t} = 1$ & t & 1 & 1.4 &  180 &  3.4 (17) & 1.00 & 5.4 & $-$0.5 &       1 \\ 
\hline 
 \textit{gauche}-C$_2$H$_5$SH & n & 0 & 1.4 &  180 & $<$  1.6 (16) & 1.95 & 5.4 & $-$0.5 & $-$ \\ 
\hline 
 \end{tabular}
 }\\[1ex] 
 \tablefoot{
 \tablefoottext{a}{d: detection, t: tentative detection, n: non-detection.}
 \tablefoottext{b}{Number of detected lines (conservative estimate, see \citealt{deuterated_SgrB2N2_2015}). Here, one line of a given species may mean a group of transitions of that species that are blended together.}
 \tablefoottext{c}{Source diameter (\textit{FWHM}).}
 \tablefoottext{d}{Rotational temperature.}
 \tablefoottext{e}{Total column density of the molecule. $X$ ($Y$) means $X \times 10^Y$.}
 \tablefoottext{f}{Correction factor that was applied to the column density to account for the contribution of vibrationally or torsionally excited states or other conformers, in the cases where this contribution was not included in the partition function of the spectroscopic predictions.}
 \tablefoottext{g}{Linewidth (\textit{FWHM}).}
 \tablefoottext{h}{Velocity offset with respect to the assumed systemic velocity of Sgr~B2(N2) $V_{\mathrm{lsr}} = 74$ km~s$^{-1}$.}
 \tablefoottext{i}{Column density ratio, with $N_{\rm ref}$ the column density of the previous reference species marked with a $\star$.}
 }
 \end{table*}